\documentclass{aa}  
\usepackage{xcolor}

\usepackage{graphicx}
\usepackage{txfonts}
\usepackage{hyperref}
\usepackage{multirow}
\usepackage{todonotes}

\newcommand{\BD}{BD$-$11$^{\rm{o}}$162}
\newcommand{\F}{Feige\,80}

\renewcommand{\sun}{$_{\odot}$}
\newcommand{\aref}{EZZ74T}

\begin{document} 

   \title{Orbital and atmospheric parameters of two wide O-type subdwarf binaries: \\ \BD\ and \F}
    \titlerunning{Orbital and atmospheric parameters of two wide O-type subdwarf binaries}
    \authorrunning{F.\ Molina et al.}

   \author{Francisco Molina \inst{1,2,3}
          \and
          Joris Vos \inst{1}
          \and
          P\'eter N\'emeth \inst{4,5}
          \and
          Roy \O{}stensen \inst{6}
          \and
          Maja Vu\u{c}kovi\'{c} \inst{7}
          \and
          Andrew Tkachenko \inst{8}
          \and
          Hans van Winckel \inst{8} 
          }

              \institute{Institut f\"{u}r Physik und Astronomie, Universit\"{a}t Potsdam, Karl-Liebknecht-Str. 24/25, 14476, Golm, Germany
              \and 
              Laboratorio LCH, Centro de Investigación en Química Sostenible (CIQSO), Universidad de Huelva, 21007, Spain
              \and
              Universidad Internacional de Valencia, Spain
              \and
              Astronomical Institute of the Czech Academy of Sciences, 25165, Ond\v{r}ejov, Czech Republic
              \and 
              Astroserver.org, F\H{o} t\'er 1, 8533 Malomsok, Hungary
              \and
              Department of Physics, Astronomy, and Materials Science, Missouri State University, Springfield, MO 65804, USA
              \and
              Instituto de Física y Astronomía,
              Facultad de Ciencias,
              Universidad de Valparaíso,
              Av. Gran Bretaña 1111, Playa Ancha, 
              Valparaíso 2360102, Chile 
              \and 
              Instituut voor Sterrenkunde, KU Leuven, Celestijnenlaan 200D, 3001, Leuven, Belgium
             }

   \date{Received October 7, 2021; accepted October 14, 2021}

 
  \abstract
  {There are 23 long-period binary systems discovered to date that contain a B-type hot subdwarf (sdB) whose orbital parameters have been fully solved. 
   They evolve into O-type subdwarfs (sdO) once the helium burning transitions from the core to the He shell. Their study will help constraint parameters on the formation and evolution of these binaries and explain some of their puzzling features.}
   {In this study, we aim to solve orbital and atmospheric parameters of two long-period sdO binaries and, for the first time, investigate the chemical composition of their main-sequence (MS) companions.}
   {HERMES high-resolution spectra are used to obtain radial velocities and solve their orbits. The GSSP code is used to derive the atmospheric parameters and photospheric chemical abundances of the MS companions. Stellar evolution models (MIST) are fitted to the companion atmospheric parameters to derive masses.}
   {The orbital and atmospheric parameters have been fully solved. Masses of the companions and the sdOs were obtained. The photospheric chemical abundances of the MS stars for elements with available lines in the optical range have been derived. They match general trends expected from Galactic chemical evolution but show a depletion of yttrium in both systems and an enrichment of carbon in the \BD\  MS.}
   {In the bimodal period-eccentricity diagram, the orbital parameters indicate that \F\  matches the same correlation as the majority of the systems. The analysis suggests that \F\  has a canonical subdwarf mass and followed a standard formation channel. However, \BD\  is an exceptional system with a lower mass. It also shows a C overabundance, which could be caused by a higher progenitor mass. The Y depletion in the MS companions could indicate the existence of a circumbinary disk in these systems' pasts. Nevertheless, a chemical analysis of a larger sample is necessary to draw strong conclusions.
    }
    
   \maketitle
    
\section{Introduction}
At least half of stars reside in binary or multiple systems.
Therefore, binary interactions rule the late-stage stellar evolution textbook scenarios for a high percentage of systems \citep{2010ApJS..190....1R,Hilditch2001}.
Binary interaction and mass transfer give rise to many new stellar objects that populate the extreme horizontal branch (EHB), including hot subdwarf stars.

Most B-type hot subdwarfs (sdBs) are core-helium-burning stars with a very thin hydrogen envelope (<0.02 M\sun) and a mass close to the core-helium-flash mass of 0.47 M\sun \ \citep{1994ApJ...432..351S,2001ApJ...563.1013B}.
To ignite He burning in the core while not having sufficient hydrogen remaining to sustain H-shell burning, the sdB progenitor needs to lose most of its hydrogen envelope near the tip of the red-giant branch \citep[RGB; e.g.][]{Heber2016}.

Unlike with other theories postulated in the past, there is currently a consensus that binary interactions are responsible for the formation of sdB stars, as was first proposed by \citet{1975BAAS....7Q.256M}. 
The three formation channels that contribute to the sdB population are the common-envelope (CE) ejection channel \citep{1976IAUS...73...75P,Han2002}, the stable Roche-lobe-overflow (RLOF) channel \citep{2000MNRAS.319..215H,Han2002}, and the formation of a single sdB star as a result of a merger between a binary and a white dwarf  \citep[WD;][]{1984ApJ...277..355W}.

Long-orbital-period systems are formed through the RLOF channel, producing binaries composed of an sdB and a main-sequence (MS) star, with orbital periods of up to 1600 days when atmospheric RLOF is taken into account \citep{Chen2013}.
\citet{Green2001} predicted their existence, but the first long-period systems were not published until 2012 \citep{Ostensen2012, Deca2012, Vos2012}. 
This led to the realisation that a new model was needed \citep{Chen2013}, as the orbital periods of some systems exceeded the predictions of binary-population-synthesis (BPS) studies  at that time \citep{Han2002, Han2003}.

These works, \citet{Han2002, Han2003} and \citet{Chen2013}, provide the theoretical framework for the long-period systems. 
However, the increasing number of solved systems show new puzzling features not explained by the models. 
This is the case for the eccentricity at long orbital periods: Where orbital circularisation is expected due to tidal forces, it is found that the eccentricity of the orbits increases at longer orbital periods.
Although efforts have been made to provide a theoretical framework \citep{Vos2015} for this question, it remains unsolved.
The study of these evolved O-type subdwarfs (sdOs) will increase the small family of solved long-period hot subdwarf systems and might provide additional constraints on the subject.

The evolution of an sdB star is determined by the remaining mass of the RGB star once the mass loss ends, after the RLOF. Whether the He flash is early or late will depend on this factor. 
But, in general terms, the sdB core burns helium through the triple-alpha reaction  for $\sim$100 Myr, producing $^{12}$C and $^{16}$O as a side product.

Once the core exhausts its helium, the star continues burning helium in a shell that surrounds the C/O core \citep{Heber2009}, expanding, heating, and evolving off the EHB to a more luminous and hotter area of the Hertzsprung-Russell diagram, and in turn giving rise to an sdO star. 
Once the helium shell is depleted and fusion ceases, the subdwarf star will start to contract as it proceeds on its way to the WD graveyard.

The described mechanism corresponds to the standard post-EHB channel. 
However, other possible scenarios might lead to the formation of an sdO as the post-RGB or the post-asymptotic giant branch (AGB) channels.
The post-RGB channel implies an enhanced mass loss, leading to objects whose remaining low mass prevents their core-He flash. 
They are progenitors of He WDs with masses from 0.2 to 0.32 M\sun\, that, on their cooling evolution, cross the EHB: the so-called pre-extremely-low-mass WDs \citep[e.g.][]{Istrate2016}.
\citet{Otoole2008ASPC..392...67O} proposed them as the forming components of the lower helium sequence in the bimodal behaviour shown by the $T_{\rm eff}-{\rm He}$ abundance correlation. 

Early spectroscopic studies of the \BD\  and \F\  systems classified both subdwarfs as  sdO spectral class stars \citep{Zwicky1957moas.book.....Z,Berger1980, Ulla1998A&AS..132....1U}. However, no clear agreement about the class of the MS companions was attained.
More recently, \citet{Ostensen2012} indicated that the spectroscopic study of the companion of \BD\  is compatible with an MS G star and that of \F\  with an early G- or late F-type star.
Nevertheless, the more remarkable results come from the study of their preliminary orbital parameters. 
\BD\  shows a preliminary orbital period in agreement with the distribution predicted by the \citet{Han2003} model (< 500 days).
However, \F\  seriously exceeds it, similar to the other long-period systems that are included in the study. 
This was key in driving the need for a new theoretical model, which was later developed by \citet{Chen2013}.

\BD\  and \F\  are the first long-period sdO binaries with solved orbits. 
By increasing the number of solved systems, one can use the added new information and ask whether sdO stars show different features with respect to sdBs.
Solved systems can also contribute to the discussion on the puzzling eccentricity issue, as they show very different orbital parameters.
Furthermore, we provide the first look into the atmospheric chemical composition of the MS companions.
The abundances of the $\alpha$, CNO (carbon, nitrogen, and oxygen), and slow neutron capture process (s-process) elements of the MS stars could provide information about the surrounding medium in which they were born \citep{Duong2018}, the mass-exchange process \citep{Dervisoglu2018}, the evolutionary stage of the sdB's progenitor \citep{Frost1996}, or, as mentioned above, possible evolutionary paths to be taken into account other than the post-EHB scenario.

\section{Spectroscopic observations}

Respectively, 56 and 67 spectra of \BD\  and \F\  were obtained as part of a long-term observing campaign from July 2009 to April 2017 \citep{Gorlova2013}. 
The observations were performed at the Roque de los Muchachos Observatory (La Palma, Islas Canarias, Spain) with the 
High Efficiency and Resolution {\it Mercator} Echelle Spectrograph (HERMES) connected by optical fibres to
the 1.2 m Mercator Telescope. 
HERMES has a spectral resolution of R = 85000, and covers a spectral range from 3770 to 9000 \AA\,\citep{Raskin2011}. 
It is managed in conditions of high wavelength stability within a temperature controlled enclosure.

The high-resolution mode of HERMES was used, and Th-Ar-Ne exposures were made at the beginning, middle and end of the night. 
The exposure time was calculated to reach a signal-to-noise ratio (S/N) of 25 or higher in the V band.
The reduction of the spectra was performed using the fifth version of the HERMES pipeline \footnote{http://www.mercator.iac.es/instruments/hermes/drs/}, which includes barycentric correction. 

\section{Orbital solutions}
\subsection{Radial velocities}
\label{RV_sect}
The determination of the radial velocities (RVs) of the MS star is straightforward, as they have many lines visible in the spectra. 
The ranges used in the cross-correlation are the following: 4725-4820 \AA, 4895-5385 \AA, 5440-5870 \AA, 5880-6270 \AA, 6310-6540 \AA. 
Lines from the sdO component and telluric lines are avoided. These intervals cover orders 55-75 of the HERMES spectrograph and offers the best compromise between maximum S/N, absence of telluric lines and low sdO flux contribution.

The MS lines in these ranges are cross-correlated \citep{Zverko2007} with respect to a local thermodynamic equilibrium (LTE) high-resolution synthetic template ($\log{g}=4.5$, $T_{\rm eff}=\,6000$ K, $[{\rm M}/{\rm H}]=0.0$\,dex and ${ v_{\rm rot}\sin{i}}$ = 1.0 km s$^{-1}$).
The rotational velocity affects the broadening of the spectral lines, and the cross-correlation of spectra with different rotational velocities might introduce systematic errors. 
As a first step, the synthetic template is convolved to match the rotational velocity, after which the cross-correlation is performed and the preliminary orbital parameters are obtained. 
The quality parameters: reduced $\chi^2$, the Akaike information criterion \citep[AIC;][]{akaike1974}, the Bayesian information criterion \citep[BIC;][]{schwarz1978}, and errors from the orbital fitting are assessed to determine the best value of the rotational velocity. 
The AIC and BIC penalise the over-fitting that might be produced by an excess of parameterisation.

The RV errors are obtained by performing a Monte Carlo (MC) simulation for each spectrum. 
Firstly, Gaussian noise is added to the spectrum, determined from the noise level of the continuum in the wavelength ranges used for the cross-correlation.
This synthetic spectrum is used in a new cross-correlation, and the final error is calculated from the standard deviation of the RV results from 100 simulations.

The orbital parameters are obtained by fitting a Keplerian orbit to the RVs \citep{Hilditch2001} with six free parameters: orbital period ($P$), time of periastron ($T_0$), eccentricity ($e$), angle of periastron ($\omega$), semi-amplitude ($K$), and systemic velocity ($\gamma$). The parameters are iterated to obtain the set of best-fitting parameters to the RVs.

The derivation of the RVs of the sdO components is more challenging. 
The available sdO lines are scarce, broad, and the only non-blended lines are the \ion{He}{ii} $\lambda$ 4686 \AA\, and \ion{He}{i} $\lambda$ 5876 \AA. 
In the case of the hotter sdO star in \BD, the low S/N of the \ion{He}{i} line prevented its use.
To increase the S/N, spectra taken at similar phases (up to 0.05 range in orbital phase) are co-added together. Using a range of 5\% in orbital phase produces negligible smearing or broadening of the He lines \citep{Vos2017}.

Due to the lack of a suitable synthetic high-resolution sdO spectrum from the grids, a custom template spectrum is calculated using the {\sc XTgrid} code \citep{Nemeth12,Nemeth2019}, based on parameters obtained by fitting the best S/N phased-binned sdO spectrum (see Sect. \ref{s:XTgrid}). 
The RV errors are determined using the MC simulations as described above.

\subsection{Orbital parameters}
\label{OrbPar}
To obtain the final orbital parameters, the phase binned spectra are used for both the sdO and the MS components.
Firstly, the orbital period is derived using the RVs of the MS component from the individual spectra. 
Then all spectra with similar phase (interval of 0.05) are summed, and new RVs for both the sdO and the MS components are determined by cross-correlation. 
A Keplerian binary orbit is fitted to the RVs, with common parameters as $e$, $\omega$, and individual ones as $K$ and $\gamma$ . The systemic velocities ($\gamma$) of both components are considered as separate parameters, as the difference in the surface gravity between the components can cause a line shift to the red (gravitational redshift; see e.g. \citealt{Vos2012, Vos2013}). 

The starting tentative parameter intervals are those obtained from the individual companion orbits. The errors on the parameters are determined using an MC algorithm in which, with each iteration, random Gaussian noise within the RV error intervals is added to the RVs.

The reduced $\chi^2$ metric is used to determine the best fit.
Based on the orbital parameters, additional related ones are derived \citep{Hilditch2001}:
the mass ratio, q, from the semi-amplitude relationship,
    \begin{equation}
     \label{q}
        q = \frac{M_{sdO}}{M_{MS}} =\frac{K_{MS}}{K_{sdO}}
    ,\end{equation}
    the minimum sdO and MS star masses (M\sun \:units),
    \begin{equation}
    \label{massproj}
        M_1\sin{i}^3=1.0361.10^7(1-e^2)^{3/2}(K_1+K_2)^2K_2P
    ,\end{equation}
    the projected semi-major axis (R\sun \:units),
    \begin{equation}
    \label{axproj}
        a_1\sin{i}=1.9758.10^{-2}(1-e^2)^{1/2}K_1P
    ,\end{equation}
    and the projected separation between stars (R\sun \:units),
    \begin{equation}
    \label{sepproj}
        a\sin{i}=a_1\sin{i}+a_2\sin{i}.
    \end{equation}

\begin{figure}
   \centering
   \includegraphics[width=9cm]{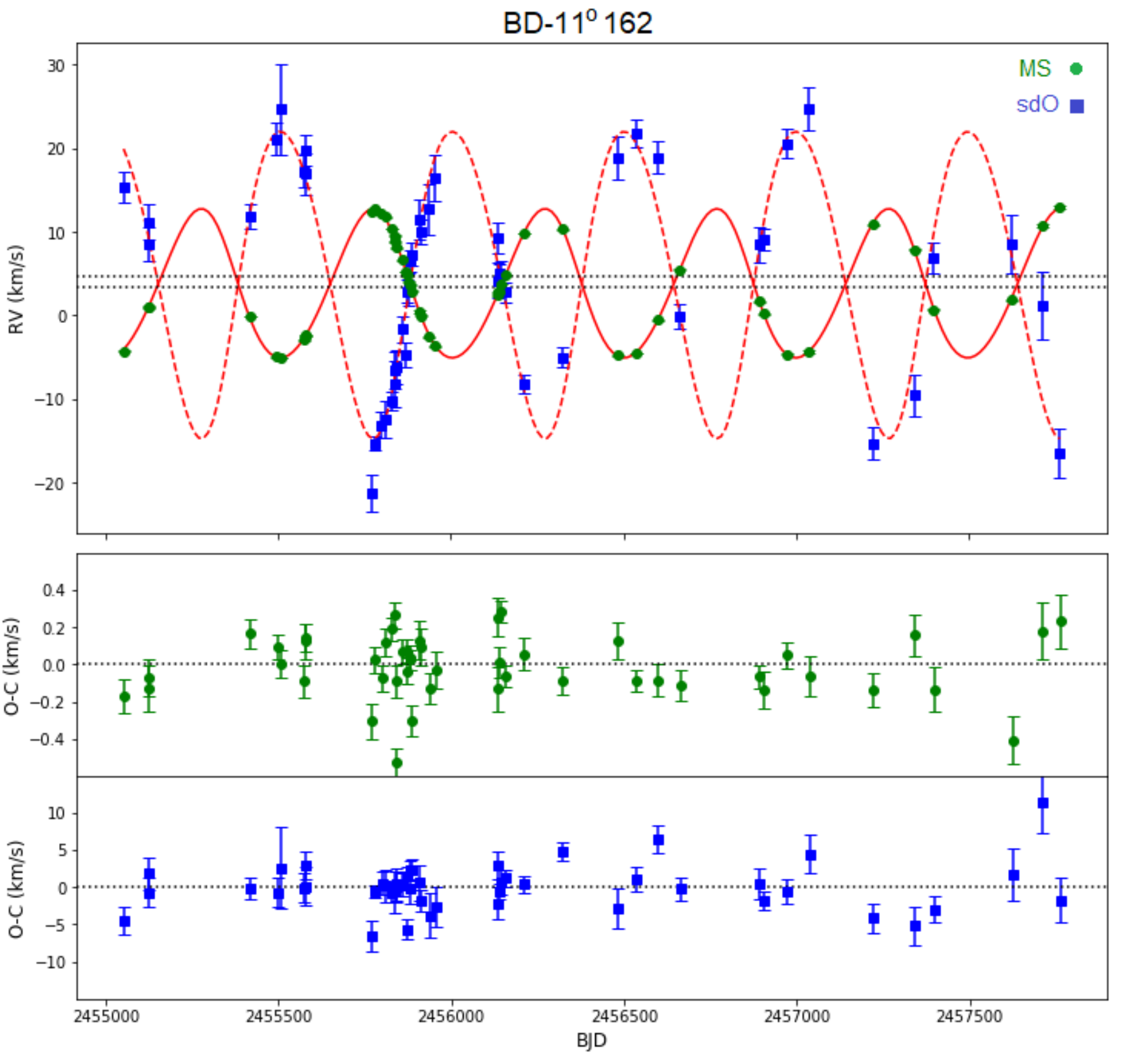}
    \includegraphics[width=9cm]{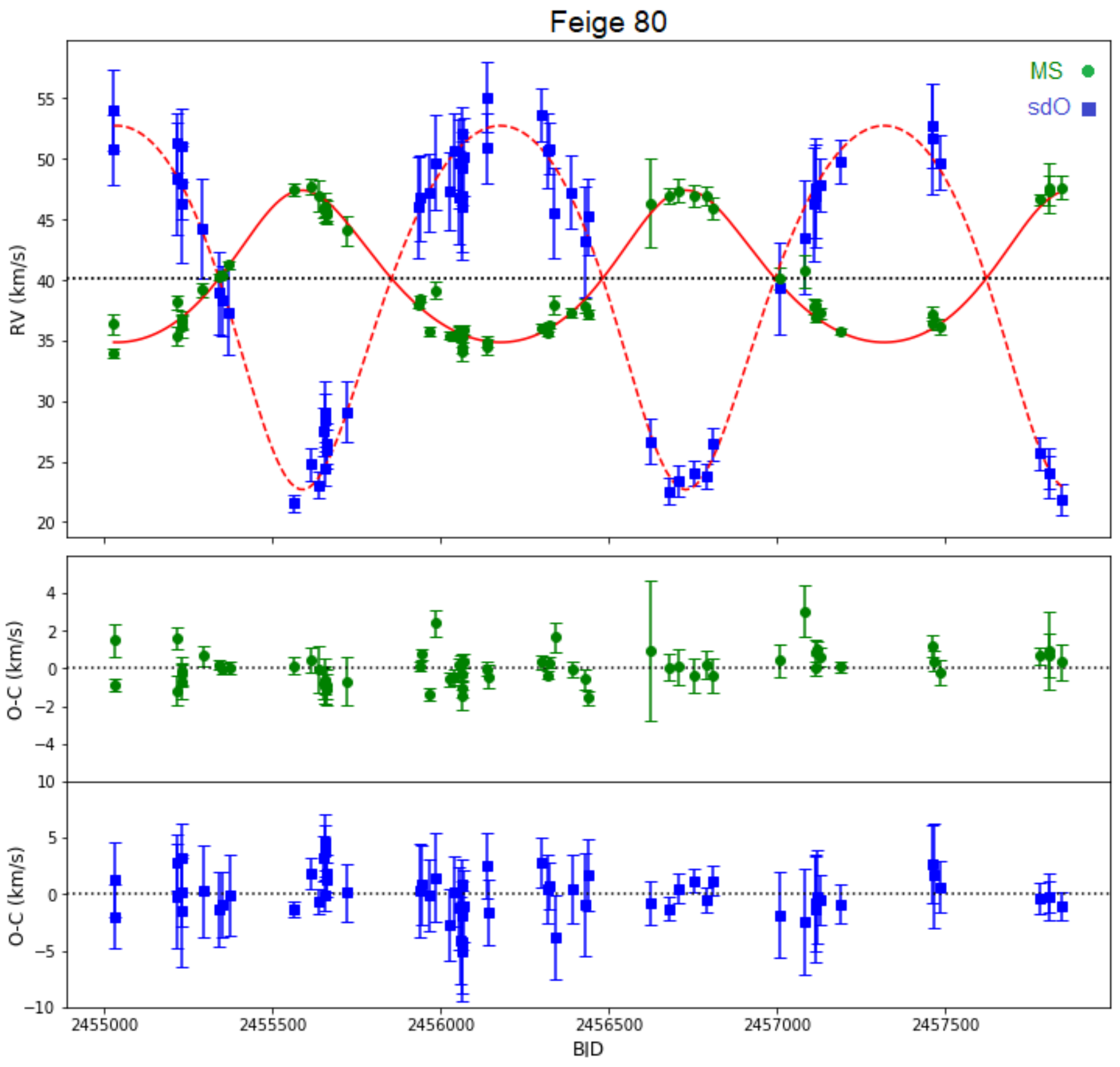}
      \caption{RV curves and residuals (O-C) from individual observations of \BD\  (upper) and \F\  (bottom). The RVs of the cool companions are plotted with green filled circles, and those of the sdO are shown as blue squares.  RV error bars ($1\sigma$) obtained by MC simulations, as described in Sect. \ref{RV_sect}, are also included. The best Keplerian orbits simultaneously fitted (MS+sdO) are shown as solid red lines for the cool companions and dashed red lines for the sdOs.}
         \label{KP}
   \end{figure}

\begin{figure}
   \centering
   \includegraphics[width=9cm]{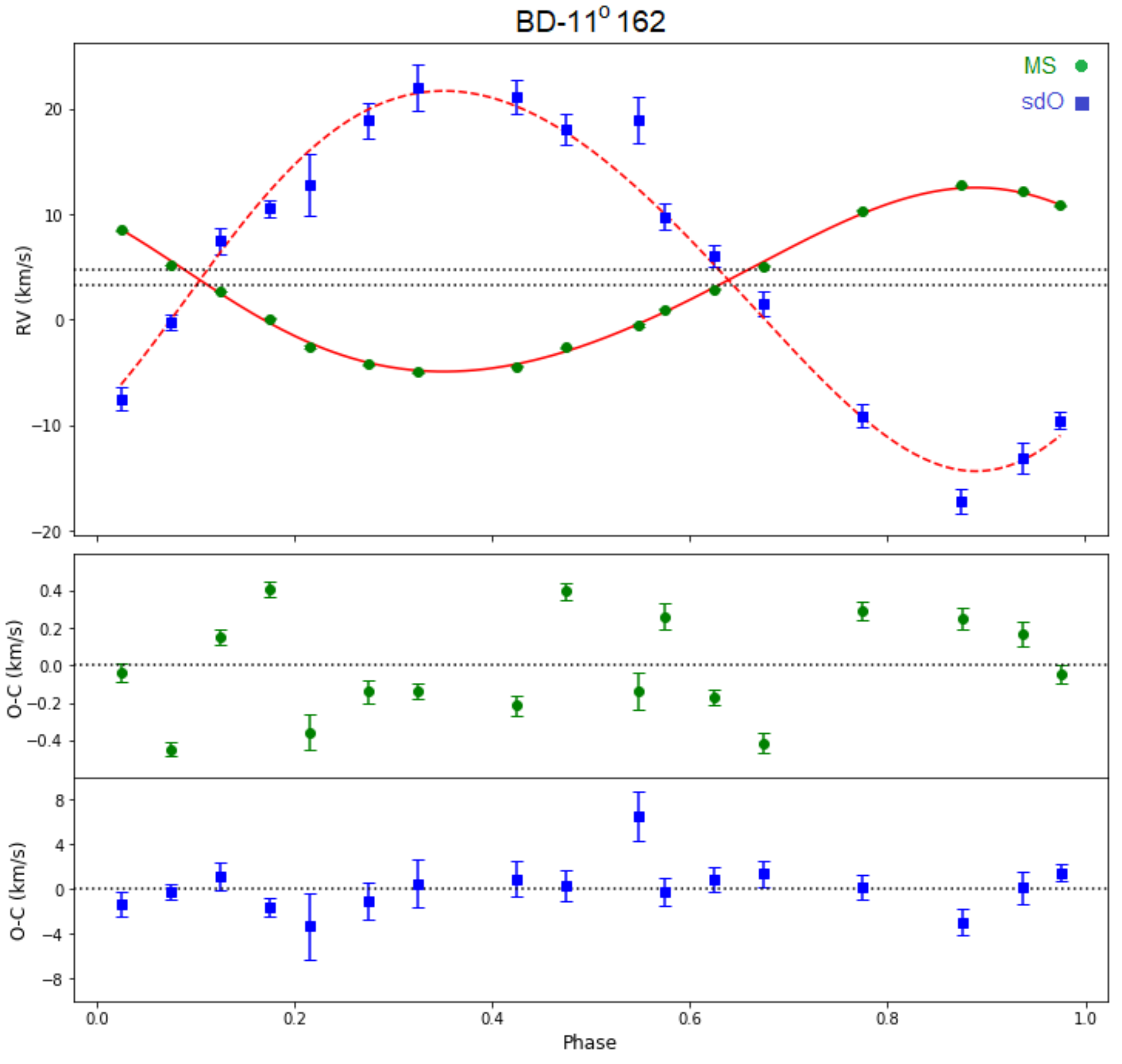}
    \includegraphics[width=9cm]{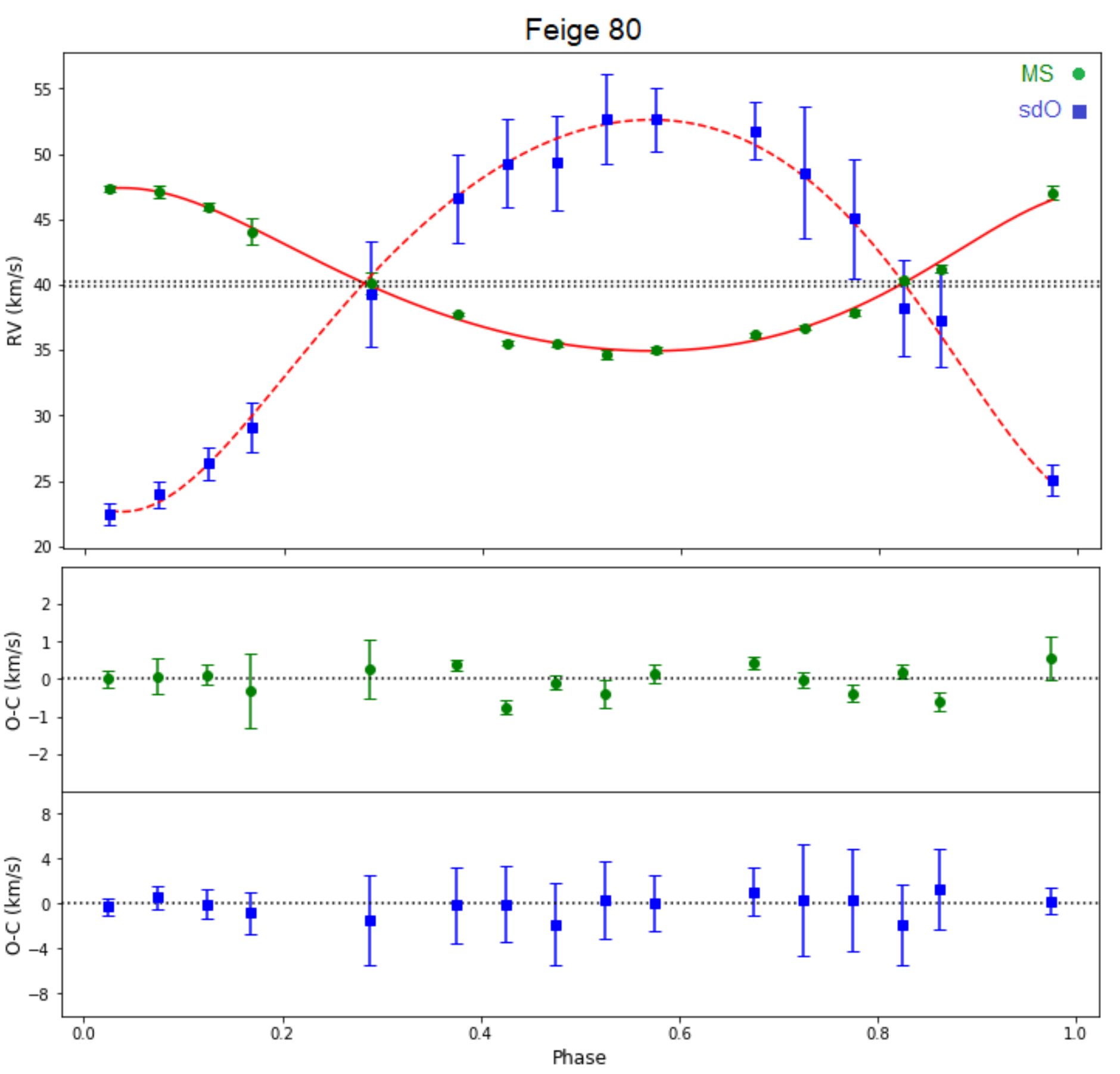}
      \caption{RV curves and residuals (O-C) from phase-binned spectra of \BD\  (upper) and \F\  (bottom). The RVs of the cool companions are plotted with green filled circles, and those of the sdO are shown as blue squares. RV error bars ($1\sigma$) obtained by MC simulations, as described in Sect. \ref{RV_sect}, are also included. The best Keplerian orbits simultaneously fitted (MS+sdO) are shown as solid red lines for the cool companions and dashed red lines for the sdOs.}
         \label{KP_BIN}
   \end{figure}

The RVs from individual spectra and the best fitting Keplerian orbits are shown in Fig.\,\ref{KP}. The RVs derived from the phase merged spectra are shown in Fig.\, \ref{KP_BIN} jointly with the final best fitting Keplerian orbits. The final orbital parameters and their errors are given in Table \ref{OP Table}.

\begin{table}
\caption{Spectroscopic orbital solutions for the MS and sdO components.}
\label{OP Table}
\centering
\renewcommand{\arraystretch}{1.4}
\begin{tabular}{lcc}
\hline\hline
\noalign{\smallskip}
Parameter    &   \BD    &   \F   \\
\hline
\noalign{\smallskip}
P(d)                                & 497.1 $\pm$\, 0.2         & 1140.4 $\pm$\, 5.0   \\
T$_0$\,(d)                          & 2454837 $\pm$\, 3         & 2454430 $\pm$\, 7 \\
e                                   & 0.082 $\pm$\, 0.002       & 0.16 $\pm$\, 0.02   \\
$\omega$                                & 0.808 $\pm$\, 0.002   & 5.96 $\pm$\, 0.02   \\
q$\, (\frac{M_{\rm sdO}}{M_{\rm MS}})$  & 0.48 $\pm$\, 0.01     & 0.42 $\pm$\, 0.02 \\
K$_{\rm MS}$\, (km s$^{-1}$)            & 8.71 $\pm$\, 0.02     & 6.2 $\pm$\, 0.1 \\
$\gamma_{\rm MS}$\, (km s$^{-1}$)       & 3.29 $\pm$\, 0.01     & 40.21 $\pm$\, 0.07 \\
M$_{\rm MS}sin^3\:i$\, (M\sun)          & 0.65 $\pm$\, 0.04     & 0.76 $\pm$\, 0.08 \\
a$_{\rm MS}sin\:i$\, (R\sun)            & 85 $\pm$\,  < 1       & 139 $\pm$\, 2 \\
K$_{\rm sdO}$\, (km s$^{-1}$)           & 18.0 $\pm$\, 0.5      & 15.0 $\pm$\, 0.7 \\
$\gamma_{\rm sdO}$\, (km s$^{-1}$)      & 4.7 $\pm$\, 0.3       & 39.9 $\pm$\, 0.7 \\
M$_{\rm sdO}sin^3\:i$\, (M\sun)         & 0.32 $\pm$\, 0.01     & 0.32 $\pm$\, 0.02 \\
a$_{\rm sdO}sin\:i$\, (R\sun)           & 176 $\pm$ 5           & 333 $\pm$ 15   \\
\hline
\end{tabular}
\tablefoot{$P$ and $T_0$ obtained from the Keplerian orbit fitting to the RV of individual observations. The rest of the orbital parameters obtained from the Keplerian orbit fitting to the phase-binned data. Errors ($1\sigma$) obtained by MC simulations from RV error intervals as described in Sect. \ref{OrbPar}.}
\end{table}

\subsection{Results}
\citet{Ostensen2012} showed preliminary results from Mercator observations of \BD\ and \F. We present the orbital parameters obtained from the complete set of observations:

\subsubsection*{\BD}
From the 56 observed spectra, 8 are dismissed due to their poor S/N or being heavily affected by cosmic rays. The companion in this system (see Sect. \ref{AP}) is a G-type MS star as noted by \citet{Berger1980}.
The average MS RV error from individual observations is 0.09 km s$^{-1}$ (0.06 for 5\% phase-binned spectra). 
The fit of a Keplerian orbit to the MS RVs shows a reassuring match.
The standard deviation of the O-C residues from the individual observations is 0.17 km s$^{-1}$. The average of the O-C residues, considered in absolute values, is 0.13 km s$^{-1}$. The quality of these data is due to the great number of available sharp metal lines in the MS spectra.
The only line available for cross-correlation in the hot sdO is the \ion{He}{ii} $\lambda$ 4686 \AA\, line.
However, this line is broad and noisy due to blending with some weak MS lines (see Table \ref{tb:XTgrid_param}). 
Therefore, the quality of the derived sdO RVs from individual observations is low.

To increase the S/N, spectra within 5\% of phase, previously determined from the MS Keplerian orbit fitting, are binned. The nominal 0.175 phase-binned spectrum shows the best S/N, and it is chosen to derive the disentangled sdO template at rest position by {\sc XTgrid}, which is used in the cross-correlation to determine the sdO RVs of the binned spectra. The average sdO RV error is 1.4 km s$^{-1}$.

The top panel of Fig.\,\ref{KP} corresponds to the sdO in \BD. 
The RVs from individual observations (blue squares) show one outlier. 
Its RV is obtained from the spectrum 759480 (see Table A1 in the appendix, BJD-2455053 = ca. 2658), which shows the second highest sdO RV error of the sample. 
The O-C residuals around the fitted Keplerian orbit of the sdO is equivalent to +4.3$\sigma$.
Even though the \ion{He}{ii} line is not affected by other factors (such as artefacts or cosmic rays), it provides poor quality RVs.

We were able to increase the S/N of the \ion{He}{ii} line in the phase-binned (0.75-0.80) merged spectrum to 46.1. 
The RV from the merged spectrum falls closer to the Keplerian orbit (see Fig.\,\ref{KP_BIN}, top panel, blue squares, point at phase 0.775). The sample of individual observations is large, and the best-fit Keplerian orbits simultaneously fitted to the binary components (MS+sdO) use weights to the RV points inversely proportional to their errors. 
Taking all considerations into account, the RV data obtained from the 759480 spectrum are included in the sample.

A second outlier can be observed in the phase-binned RV diagram of the sdO (see Fig.\,\ref{KP_BIN}, top panel, blue squares, 0.50-0.55 phase). 
The O-C residuals around the fitted Keplerian orbit of the sdO is equivalent to +4.5$\sigma$ (after excluding this point from the sample).
By chance, the interval contains a unique spectrum, the 505540 (see Table A.1, BDJ-2455053= ca. 1547). 
The interval retains the commented poor quality of the \ion{He}{ii} line in the individual observation spectra. But otherwise the 505540 spectrum yields a standard behaviour, either in the MS or sdO individual observation diagrams or in the corresponding phase-binned interval of the MS diagram. Similar reasons as previously developed led to its acceptation.

\BD\ shows an orbital period of 497.1 $\pm$ 0.2 days and an eccentricity of 0.082 $\pm$ 0.002. The small errors on the RVs of the MS companion allow us to obtain the period and eccentricity with high precision.
The higher errors from the RVs of the sdO component contribute to a higher error in the amplitude, systemic velocity and derived parameters of the hot sdO star.

\subsubsection*{\F}
From the 67 HERMES spectra, 4 are rejected due to their poor S/N or being heavily affected by cosmic rays. The companion in this system (see Sect. \ref{AP}) is an F-type MS star. 

The average MS RV error from individual observations is 0.70 km s$^{-1}$ (0.35 for 5\% phase-binned spectra). 
As was noted from preliminary observations \citep{Ostensen2012}, the O-C residuals of the MS RVs around the fitted Keplerian orbit are higher than for the MS star in  \BD. 
The standard deviation of the O-C residues from the individual observations is 0.88 km s$^{-1}$. The average of the O-C residues, considered in absolute values, is 0.65 km s$^{-1}$.
The higher rotational and thermal broadening in the metal lines of the cool companion are the main reason for the overall larger errors.

The sdO in \F\  is cooler than the one in \BD\  and the \ion{He}{i} $\lambda$ 5876 \AA\ line is stronger. 
This line is used together with the \ion{He}{ii} $\lambda$ 4686 \AA\ in the cross-correlation method. To increase the S/N, spectra within 5\% of orbit phase, previously determined from the MS Keplerian orbit fitting, are binned. The nominal 0.375 phase-binned spectrum shows the best S/N, and it is chosen to derive the disentangled sdO template at rest position by {\sc XTgrid}, being used later to cross-correlate the binned spectra. The average sdO RV error in the binned spectra is 2.8 km s$^{-1}$.

\F\  shows an orbital period of 1140 $\pm$ 5 days and an eccentricity of 0.160 $\pm$ 0.017. The larger errors on the RVs of the MS and sdO companions decrease the precision of the best-fit parameters, although they are comparable to the typical errors in other long-period sdB binaries \citep{Vos2019}.

\section{The mass ratio--eccentricity distribution} \label{q-EccDist}
\begin{figure}
\centering
\includegraphics[width=9cm]{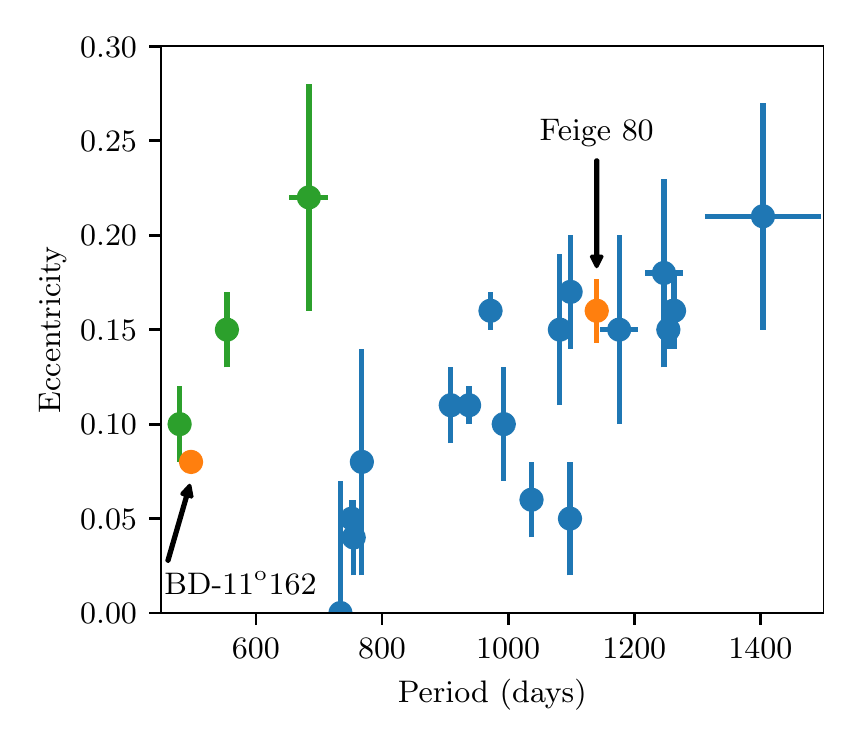}
  \caption{Eccentricity-orbital period distribution to date. Group II is shown in green. Error bars ($1\sigma$) are included.
  Orbital parameters are obtained from \citet{Vos2019}.}
     \label{Ecc-P}
\end{figure}

The eccentricity of evolved long-period binaries, including hot subdwarf binaries, has become a puzzling question in the last years \citep[e.g.][]{Vos2017, Murphy2018, Escorza2019}.
Even at the observed long periods, interaction models predict the orbits to circularise due to tidal forces long before the interaction phase. 
However, observations show the opposite case (see Fig. \ref{Ecc-P}). 

A Modules for Experiments
in Stellar Astrophysics (MESA) study published by \citet{Vos2015} supports the idea of the formation of a circumbinary disk (CB) during the mass-loss phase as an eccentricity pumping mechanism.
The existence of dusty disks around eccentric post-AGB binaries \citep{deRuyter2006, Hillen2014} favours this idea, as well as the detection of dust around the post-CE WD binary NN Serpentis \citep{Hardy2016} and the discovery of dusty post-RGB systems \citep{Kamath2015}. 
But to date, no direct evidence for the existence of dust structures has been found in sdB systems. 
An observing programme with the ALMA radio-telescope focusing on two long-period hot subdwarf binaries: \BD\ and the sdB + subgiant (SG) binary BD-7$^{\rm o}$5977, aimed at detecting the remnants of a CB disk. 
However, the results are negative (see Sect.\,\ref{sec:alma_observations} in the appendix for a description of the observations and their results). 
A recent study by \citet{Oomen2020} shows that the same disk - binary interaction that could work in sdB binaries, cannot explain the observed eccentric orbits in post-AGB binaries. Clearly, the eccentricity evolution during binary interactions is not well understood yet.

\begin{figure}[h]
    \centering
    \includegraphics[width=9cm]{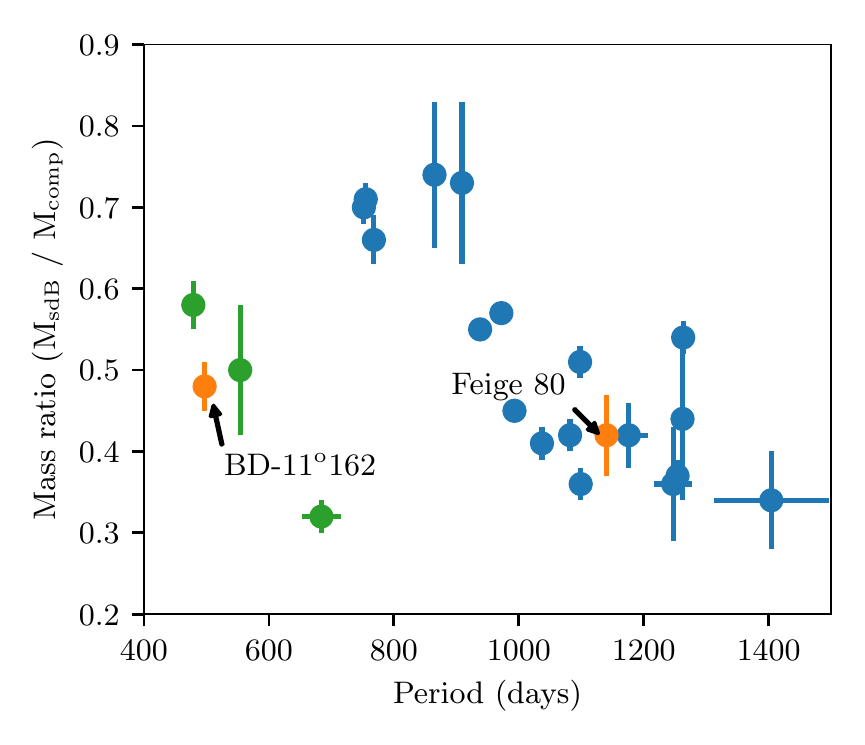}
    \caption{Mass  ratio  as  a  function  of  orbital  period. Group II is shown in green. Error  bars ($1\sigma$)  are  included. Orbital parameters are obtained from \citet{Vos2019}.}
    \label{qP}
\end{figure}

The eccentricity question has got even more complex with the increasing number of sdB+MS systems with well known orbital properties: the eccentricity-orbital period relation ($e$-$P$) shows a bimodal behaviour. 
A similar bimodal behaviour is also present in the mass ratio - orbital period relation ($P$-$q$) as shown in Fig. \ref{qP}. 
The orbital parameters of \BD\ and \F\ corroborate these trends. \F\ belongs to the main group in the $e$-$P$ relation, while \BD\ is a new member of the minority group II (see Fig. \ref{Ecc-P}).

\section{Spectroscopic analysis of the main-sequence companions}
\label{GSSP}

The spectroscopic analysis of the MS companions is performed using the Grid Search in Stellar Parameters (GSSP) code \citep{Tkachenko2015}.
GSSP uses the method of atmosphere models and spectrum synthesis, comparing the observations with each theoretical spectrum of the grid. The synthetic spectra are obtained using the SynthV LTE-based radiative transfer code \citep{Tsymbal1996} and a grid of LTE atmosphere models pre-computed with the {\sc LLmodels} code \citep{Shulyak2004}. 

GSSP optimises up to 7 stellar parameters at a time:
surface temperature (T$_{\rm eff}$), surface gravity (log$g$), metallicity ([M/H]), micro-turbulent velocity ($v_{\rm mic}$), macro-turbulent velocity ($v_{\rm mac}$), projected rotational velocity (${\bf v_{\rm rot}\sin{i}}$) and the dilution factor (F$_{\rm MS}$/F$_{\rm Total}$) of the MS star flux.
The grid of theoretical spectra is built from all possible combinations of these parameters, and GSSP compares each grid element with respect to the observed normalised spectrum.
The $\chi^2$ merit function judges the match between the observed and synthetic spectra, obtaining the set of best-fit parameters.

\subsection{Atmospheric parameters}
\label{AP}
Master spectra are obtained by shifting the observed spectra from \BD\, and \F\ to the rest velocity of the MS stars. The wavelength interval of 5910-6510 \AA \, is normalised before using it as input for with GSSP to obtain the atmospheric parameters. 
The selected spectral range in the red is a compromise between high S/N and high contribution of the cool companion. The range is fairly short, avoiding problems with the wavelength-dependent dilution factor.
The spectral region (6270-6330 \AA) that contains telluric lines is avoided.

The initial setup of the parameter search covers wide ranges with large step sizes in order to assure that the global minimum is found and prevent an excessive computational time.
A second coarse grid search is processed, which also updates the $v_{\rm mic}$ and $v_{\rm mac}$ values as described below.
Eventually, a third smaller step-size grid search is set up around the parameters found in the second run. 
With the third run, GSSP obtains the final atmospheric parameters.

Disentangling macro-turbulent velocity from rotational velocity is difficult, and it requires spectra of very high resolution and a high S/N to solve the degeneracy between these parameters \citep{Doyle2014}. 
In order to avoid this degeneracy and to establish a standard procedure for studying spectra of lower resolution and S/N, the $v_{\rm mic}$ and $v_{\rm mac}$ are set using the \citet{Bruntt2010} and \citet{Doyle2014} relations. 
These relations were obtained from the respective studies on the spectroscopic and asteroseismic data of MS stars. They provide standard values of $v_{\rm mic}$ (as a function of T$_{\rm eff}$) and $v_{\rm mac}$ (as a function of T$_{\rm eff}$ and $\log{g}$) from a calibrated sample of MS dwarfs.

Therefore, $v_{\rm mic}$ and $v_{\rm mac}$ are not freely refined in the successive code runs, but their values are fixed for the next run based on the standard relations and the currently computed T$_{\rm eff}$ and $\log{g}$.
Nevertheless, the change of their values after the second run is hardly noticeable. 
The final values are derived from the T$_{\rm eff}$ and $\log{g}$ obtained in the last run of the code.

Errors for the atmospheric parameters are derived by fitting a polynomial function to the reduced $\chi^2$ coefficients, and the error interval is obtained by determining where the polynomial reaches the $\chi^2$ level corresponding to 1$\sigma$ errors. 
Errors for $v_{\rm mic}$ and $v_{\rm mac}$ are obtained from the root-mean-square (RMS) scatter of the relations found by \citet{Bruntt2010} and \citet{Doyle2014}, respectively, adding the standard deviation of the velocities when developing an MC simulation as a function of the surface temperature and gravity error intervals. 
The set of best-fit parameters are summarised in Table \ref{AP_Tab}. They point to a G1V-type star for \BD\  and F7IV/V-type for \F.

\begin{table}
\caption{Atmospheric parameters.}
\label{AP_Tab}
\centering
\renewcommand{\arraystretch}{1.4}
\begin{tabular}{lcccc}
\hline\hline
\noalign{\smallskip}
System  & \multicolumn{2}{c}{\BD} & \multicolumn{2}{c}{\F} \\
 \hline
Parameter & GSSP  &  Error   & GSSP    & Error        \\
\hline
\noalign{\smallskip}
T$_{\rm eff}$ (K)                           &   5630   &        $^{+230}_{-170}$    &  6200   &  $^{+320}_{-290}$ \\
log $g$ (dex)                           &   4.3    &   $^{+0.5}_{-0.5}$     &  3.9    & $^{+0.6}_{-0.6}$   \\
${ v\sin{i}}$ ($\rm km \ s^{-1}$) &   12     &   $^{+1}_{-1}$         &  29     & $^{+2}_{-2}$   \\
\big[M/H\big] (dex)                     &   -0.6   &   $^{+0.2}_{-0.2}$     &  -0.5   & $^{+0.2}_{-0.2}$   \\
Dilution factor                         &   0.52   &   $^{+0.06}_{-0.08}$   &  0.55   & $^{+0.11}_{-0.09}$ \\
$v_{\rm mic}$ ($\rm km \ s^{-1}$)       &   1.0    &    $^{+0.2}_{-0.2}$         &  1.3    & $^{+0.3}_{-0.3}$  \\
$v_{\rm mac}$ ($\rm km \ s^{-1}$)       &   3.1    &    $^{+1.5}_{-1.5}$         &  5.5    & $^{+2.1}_{-2.1}$    \\
\hline
\end{tabular}
\tablefoot{Obtained from the 5910-6510 \AA\,interval, excluding 6270-6330 \AA. Errors at 1$\sigma$ of reduced $\chi^2$. }
\end{table}

\subsection{Abundances}
A study of individual chemical abundances is developed, including the CNO elements.
Their relation is a key to understand stellar evolution of stars that have evolved past the MS, either on to the RGB or further along the AGB \citep{Frost1996}. For example, the C/O fraction is the criteria to classify AGB stars \citep{Ramstedt2014,Abia2003}.
Due to convective episodes known as dredge-ups (DUs), different mixing events happen along the RGB and AGB ascent. 
They transport newly synthesised material \citep{DeSmedt2016} to the surface from deep shells or the core, altering the atmospheric chemical composition. 
Therefore, the surface abundance relations are connected with the evolutionary stages. The different mixing episodes are: the first dredge-up (FDU) during the RGB evolution and the second and third dredge-up (SDU and TDU) during the AGB evolution.

The companions that we have studied here are still on the MS. 
However, they have likely undergone a mass-transfer event when the sdB progenitor started losing its envelope. 
Studying the atmospheric abundances of the MS components might reveal interesting facts on the sdB progenitor and the mass transfer event.

The C/N ratio is a key parameter to understand mass exchange processes \citep{Dervisoglu2018,Sarna1996}. The mass-loss rate of the donor is related to the level of exposition of their inner layers, which contain processed material from the CNO cycle where the C/N ratio is reduced. On the other hand, the accreted mass fraction by the MS, after the initial impact on their outer layers, will lead to thermohaline mixing \citep{Dervisoglu2018,Kippenhahn1980}, decreasing the C/N ratio on the surface.
Therefore, both processes might leave their fingerprints on the MS's C/N ratio of the hot sdB binary systems, leading to an enhancement of the $^{14}$N and depletion of the $^{12}$C abundances.

Other valuable information on the evolutionary stage of the donor might be obtained from the abundance of the s-process elements (the light Sr, Y, and Zr from the first peak of stability and the heavy Ba as the prototype element of the second peak). 
These elements are produced in evolved AGB stars and mixed on the surface by the TDU.
The moderate poor metallicities of these binary systems lead us to expect an optimised production of heavy second peak s-process elements as Ba.
At solar metallicity, the AGB production is especially enhanced in the first peak s-process elements as Sr, Y, and Zr, at the neutron magic number of N=50.
At lower [Fe/H], more neutrons per Fe seeds are available and the Fe nuclei do not dominate the neutron capture so efficiently. 
Hence, more neutrons are available to push the $s$-process reactions ahead, beyond the first peak of $s$-process elements, overcoming the stability bottleneck at N=50. 
This process feeds elements to the second $s$-process peak, the Ba-peak, at N=82, with a maximum production yield at [Fe/H]$\sim$-0.6 \citep{Travaglio2004ApJ...601..864T}. 
The metallicities of our systems are close to the sweet spot production of second peak s-process elements in case that the sdB's progenitors would have evolved to the AGB stage. 
If the sdB progenitor had evolved on the AGB before the mass transfer episode, we might expect some impact in Sr, Y, Zr or Ba abundances of the MS due to the mass-exchange process.

For abundance calculations, the 5910-6510 \AA\, interval is used as standard for all elements. This is the same interval used for the atmospheric parameters study. As commented in the preceding section, this interval gives the best compromise between a high S/N and a high contribution of the cool companion, and there are no spectral lines of the sdOs components visible. This range is short enough for considering the dilution as a wavelength-independent factor, and hence representative of the interval \citep{Vos2018a}.

Flux dilution is a wavelength-dependent parameter.
In \BD\  the dilution factor goes from 0.14 (ca. 4000 \AA) to 0.76 (ca. 8700 \AA) and \F\  from 0.15 to 0.74, respectively. 
This is due to the difference in $T_{\rm eff}$ between the sdO and the companion stars (see Tables \ref{AP_Tab} and \ref{tb:XTgrid_param}). 
Therefore, the use of larger intervals is avoided in order not to introduce distortion in the obtained abundances via the dilution factor.

Although this interval is rich in MS lines, it is only possible to derive abundances from a few elements with frequent and strong absorption lines (e.g. Fe, Si, and Ni). Additional intervals and specific line studies for each element of interest have to be performed along the spectrum. The intervals and line information used for each element are shown in Tables \ref{BDInt}, \ref{FGInt}, and \ref{Lines} of the appendix.

Firstly, the procedure is to cut and normalise an interval, selected from the literature, containing the line(s) of the element under study. This larger interval is used to determine the dilution factor. It must contain iron lines for this assessment.

Secondly, a final abundance calculation is performed on the specific line while keeping fixed the previously derived dilution factor.
Derived abundances are summarised in Table \ref{GSSPabund}.

\begin{table}
\caption{Element fractions and [X/Fe] abundance ratios of \BD\  and \F.}
\centering
\renewcommand{\arraystretch}{1.4}
\begin{tabular}{lcccccc}
\hline
\hline
 & \multicolumn{3}{c}{\BD} & \multicolumn{3}{c}{\F} \\
\hline
\small{Elem.} & \small{log $Z$}  &  \small{Err.}  & \small{[X/Fe]}   & \small{log $Z$}    & \small{Err.} & \small{[X/Fe]} \\
\hline
\noalign{\smallskip}

Na          &    -6.35      &   $^{+0.06}_{-0.12}$  &    +0.12 &   -6.18    &    $^{+0.12}_{-0.12}$   &   +0.19     \\
Mg          &    -4.97      &   $^{+0.00}_{-0.00}$  &    +0.14 &   -4.79    &   $^{+0.05}_{-0.02}$   &   +0.22     \\
Al          &    -5.87      &   $^{+0.06}_{-0.06}$  &    +0.40 &   -5.93    &   $^{+0.06}_{-0.12}$   &   +0.24      \\
Si      &    -4.90      &       $^{+0.08}_{-0.06}$  &    +0.23 &   -4.87    &    $^{+0.10}_{-0.05}$   &   +0.16      \\
S      &     -5.18      &       $^{+0.12}_{-0.31}$  &    +0.32 &   -5.12    &    $^{+0.06}_{-0.18}$   &   +0.28      \\
K      &     -7.02      &       $^{+0.00}_{-0.00}$  &    +0.54 &   -7.06    &    $^{+0.12}_{-0.00}$   &   +0.40     \\
Ca      &    -6.16      &       $^{+0.07}_{-0.12}$  &    +0.17 &   -5.99    &    $^{+0.12}_{-0.17}$   &   +0.24      \\
Sc      &    -9.38      &       $^{+0.06}_{-0.12}$  &    +0.21 &      **    &    $^{}_{}$   &      **      \\
Ti      &    -7.67      &       $^{+0.12}_{-0.12}$  &    +0.07 &   -7.49    &    $^{+0.12}_{-0.25}$   &   +0.15     \\
V      &     -8.45      &       $^{+0.37}_{-0.37}$  &    +0.19 &   -8.67    &    $^{+0.37}_{-0.61}$   &   -0.13     \\
Cr      &    -6.94      &       $^{+0.06}_{-0.12}$  &    +0.06 &   -6.74    &    $^{+0.06}_{-0.18}$   &   +0.16     \\
Mn      &    -7.27      &       $^{+0.20}_{-0.24}$  &    -0.02 &   -7.18    &    $^{+0.29}_{-0.28}$   &   -0.03     \\
Co      &    -7.67      &       $^{+0.24}_{-0.61}$  &    +0.05 &   -7.55    &    $^{+0.61}_{-0.86}$   &   +0.07     \\
Ni      &    -6.42      &   $^{+0.08}_{-0.08}$  &    -0.01 &   -6.18    &   $^{+0.06}_{-0.18}$   &   +0.13      \\
Cu      &    -8.39      &   $^{+0.00}_{-0.16}$  &    +0.04 &   -8.18    &   $^{+0.12}_{-0.12}$   &   +0.15      \\
Zn      &    -8.00      &   $^{+0.06}_{-0.00}$  &    +0.04 &   -8.00    &   $^{+0.12}_{-0.00}$   &   -0.06       \\
Sr      &    -9.69      &   $^{+0.12}_{-0.00}$  &    +0.03 &   -9.57    &   $^{+0.12}_{-0.06}$   &   +0.05       \\
Y       &   -10.67      &   $^{+0.00}_{-0.06}$  &    -0.24 &  -10.74   &   $^{+0.12}_{-0.18}$   &   -0.41      \\
Zr      &    -9.82      &   $^{+0.24}_{-0.24}$  &    +0.23 &  -9.51    &   $^{+0.35}_{-0.35}$   &   +0.44       \\
Ba      &   -10.43      &   $^{+0.06}_{-0.06}$  &    +0.04 &  -10.31   &   $^{+0.12}_{-0.12}$   &   +0.06      \\
La      &   -11.17      &  $^{+0.31}_{-0.37}$   &    +0.34 &   **      &   $^{}_{}$             &   **      \\
Ce      &   -11.00      &  $^{+0.18}_{-0.31}$   &    +0.06 &   **      &   $^{}_{}$             &   **      \\
\hline
\end{tabular}
\label{GSSPabund}
\tablefoot{Element fractions $Z=n{\rm X}/n_{\rm {tot}}$. Units in dex.}
\end{table}

\begin{figure*}
  \centering
  \includegraphics[width=18.2cm]{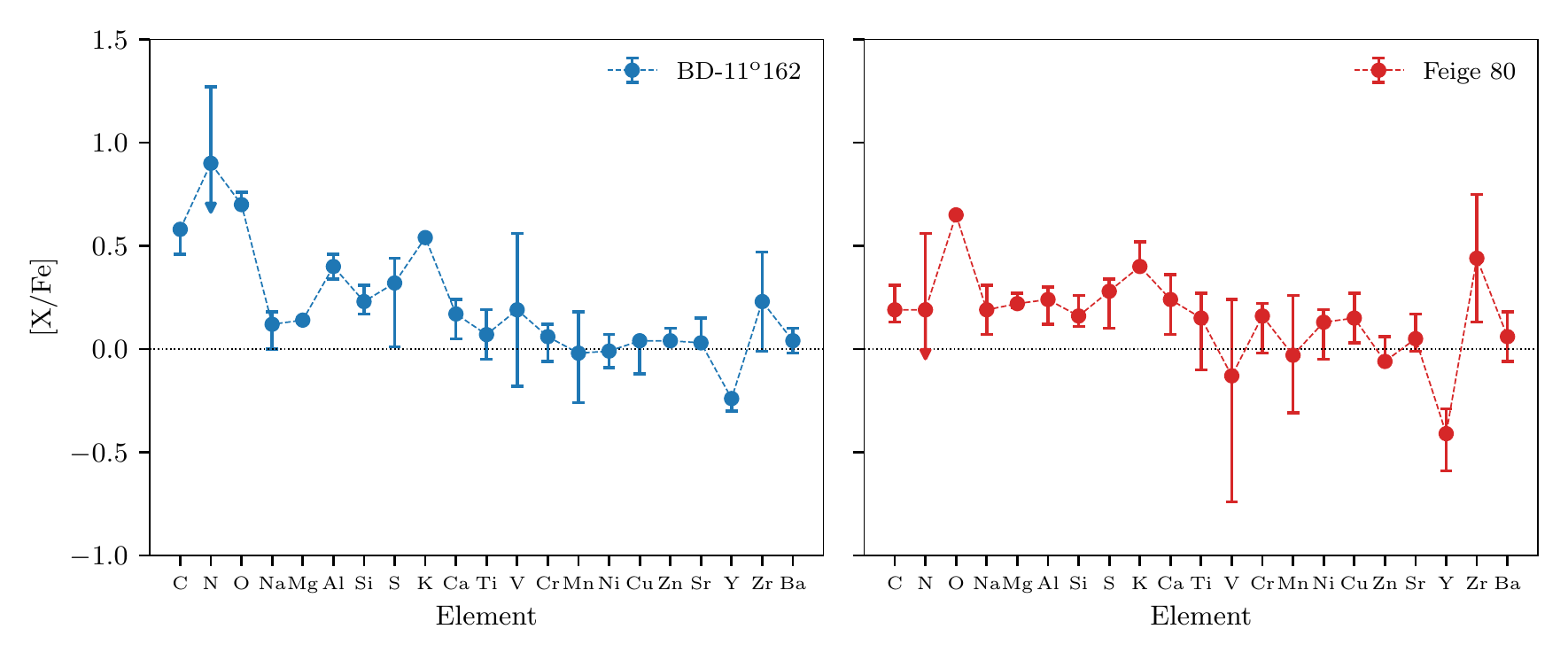}
      \caption{[X/Fe] ratio of the derived elemental abundances in common for \BD\  and \F. The lower error limit of nitrogen abundances, not constrained, is shown. Error  bars ($1\sigma$)  are  included. Units are in dex.}
\label{Diff}
\end{figure*}

\begin{figure}
  \centering
  \includegraphics[width=9.1cm]{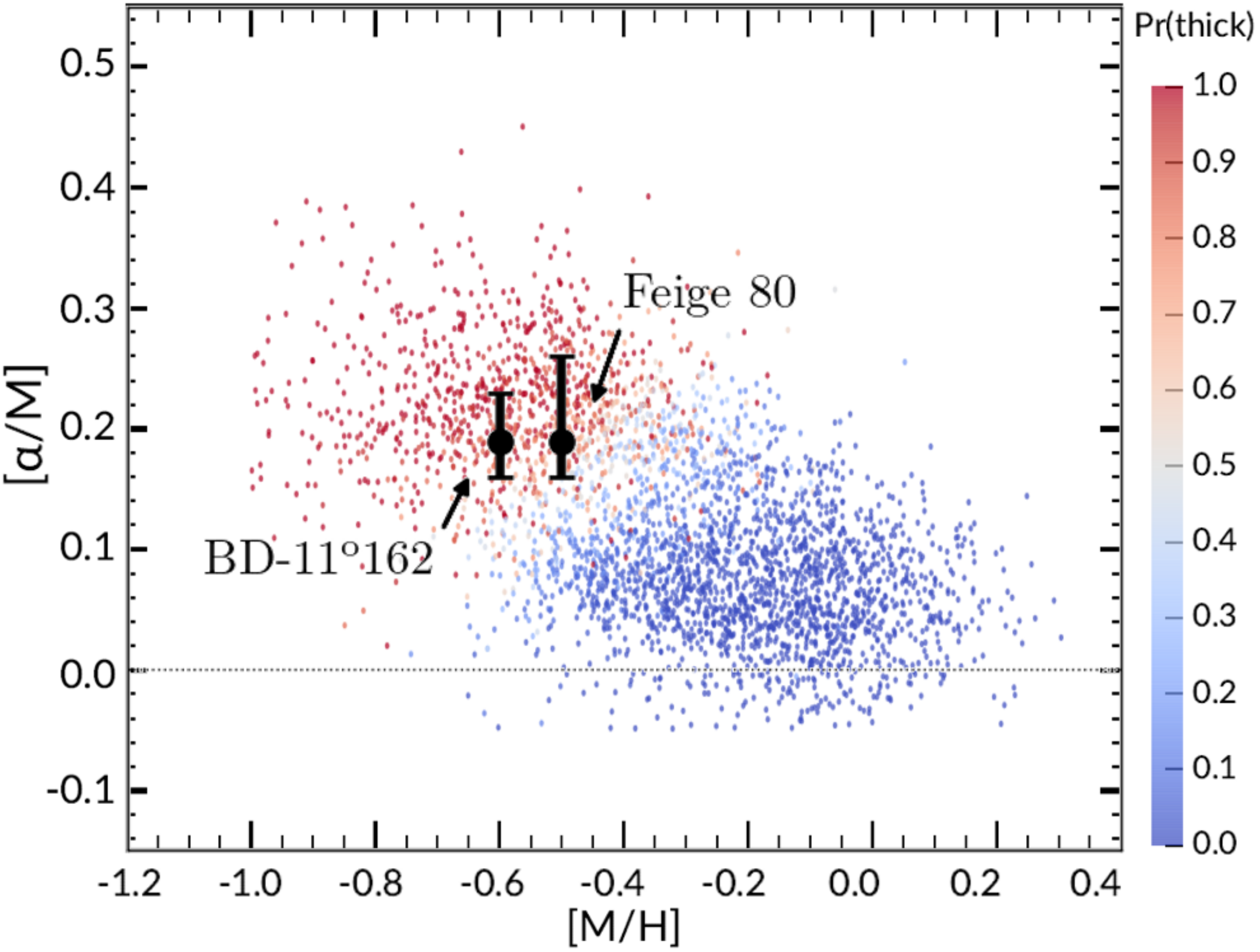}
      \caption{Study on the probability of solar neighbourhood stars belonging   to the thin or thick disk using kinematics and chemical data by \citet{Duong2018}: projection along the [$\alpha$/M]-[M/H] plane used in their paper and adapted to include \BD\  and \F. Error  bars ($1\sigma$)  are  also included. Units are in dex.
             }
\label{alphaDuong}
\end{figure}   

The $\alpha$ elements are produced in the later burning stages of massive stars (>8 M\sun) previous to core-collapse supernovae (SNeII) and their distributions are due to the Galactic chemical evolution (GCE). The early universe was enriched in these elements because the SNeII occurred on a much faster timescale than the Type 1a supernovae (SNeIa) that produce more iron than $\alpha$ elements \citep{Edvardsson1993A&A...275..101E}.  
Likewise, the thin and thick disks were formed in different evolutionary moments and hence the stars from both disks show a different chemical and kinematic nature.

The `two-infall' chemical evolution model for the Milky Way halo and disks, originally developed by \citet{Chiappini1997ApJ...477..765C,Chiappini2001ApJ...554.1044C} has been widely used in literature as a basic formalism and understanding of the Galactic formation in order to build GCE models.
This model assumes the inner halo and thick disk to form fast, in less than 1 Gyr, out of a first episode of accretion of virgin gas. During these earlier evolutionary stages, the star formation proceeds very efficiently and turns a large fraction of the available gas into stars. Eventually, the critical threshold for gas density below which the star formation halts is reached.
The second infall episode, which is delayed by 1 Gyr with respect to the previous one, starts replenishing the disk with fresh gas and star formation is reignited \citep{Romano2020A&A...639A..37R}. 

The thin disk forms on longer timescales that are functions of the Galactocentric distance \citep{Matteucci1989MNRAS.239..885M,Romano2000ApJ...539..235R}.
Hence, both factors are contributing to the typical profile of the $\alpha$-element abundances with respect to metallicity, showing an increasing enrichment over solar abundance to metal-poorer stars, and eventually reaching a plateau at a metallicity of ca -1.0 dex.

The $\alpha$-element abundances in these systems show moderate enhancements, as shown in Table \ref{GSSPabund} and Fig. \ref{Diff}.
The +0.19 dex enrichment of the $\alpha$-element index from the weighted average of abundances of Mg, Si, and Ti are in concordance with expected by GCE for metal-poor stars belonging to the thick disk \citep{Duong2018}, as observed in Fig.\,\ref{alphaDuong}. Or at least with the chemical behaviour of edge stars between the thin and thick disks \citep{Adibekyan2013}.

Most of the elemental abundances match with the expected Galactic chemical trends, but there are some intriguing exceptions. One of them is the light first peak s-process element, yttrium (Y). Its abundance has been determined from the \ion{Y}{II} lines at 4883.68 \AA\, and 4900.12 \AA.
Recent studies of atmospheric abundances in FGK \citep{Delgado2017} and FG dwarf stars of the solar neighbourhood \citep{Bensby2014} indicate a slight under-solar abundance (ca -0.1 dex), at the same range of metallicities as \BD\  and \F. However, the MSs of \BD\  and \F\ yield moderate depleted abundances, respectively, ca -0.2 and -0.4 dex, values under the Galactic chemical trend.

Yttrium is an element showing a high condensation temperature at 1622 K. Therefore, it is an element that easily gets locked in dust grains in the early stages of a circumstellar or CB disk formation. The radiation pressure prevents re-accretion on the star, leading to a depletion of its abundance on the photospheric layer, as observed in some reported stars by \citet{Maas2005}.
As above commented in the previous section, the formation of a CB disk during the mass-loss phase might help to explain, from a theoretical point of view, some features observed in wide hot subdwarf systems as the eccentricity.
While a disk formed during the mass-loss phase would eventually be ejected from the system, part of the material might be accreted by the MS companion. Those elements showing high condensation temperatures, as is the case of yttrium, would be removed in early stages, depleting their abundances on the MS star surface.

Another interesting element is barium (Ba), as a heavy second peak s-process prototype. It shows nearby scaled solar abundances, reinforcing the donor evolutionary stage as an RGB not an AGB star.

\subsubsection*{CNO abundances}

The lines used for determining CNO abundances have been taken from the literature (see Tables \ref{BDInt}, \ref{FGInt}, and \ref{Lines} in the appendix). The results are summarised in Tables \ref{CNO} and \ref{CNO ratios}.

Oxygen is an $\alpha$ element produced by the triple-$\alpha$ reaction as a by-product ($^{16}$O), delimiting the end-point of He burning in AGB stars, but mainly in later fusion stages of massive stars (>8 M\sun) preceding to the SNeII \citep{Meyer2008}. Especially stars more massive than 15 M\sun, contributed to the ISM enrichment at very early stages of the universe \citep{Delgado2017}. This element shows the typical $\alpha$-element profile in the [X/Fe]-[Fe/H] relation.

The abundances obtained from the \ion{O}{I} triplet at 7771.94, 7774.15, and 7775.39 \AA\, must be taken with precaution as these lines are known to be affected by strong non-LTE effects \citep{Amarsi2015,Amarsi2016}, which cause an overestimation of the abundances obtained under LTE assumptions. This is the main reason why the oxygen is not included in the $\alpha$-element index \citep{Duong2018}.

Non-LTE corrections were derived by interpolating the \BD\  and \F\  dwarf atmospheric parameters into the STAGGER grid of 3D stellar atmospheres \citep{Collet2011,Magic2013} obtained by \citet{Amarsi2015}. By applying these, we can compare our results with a non-LTE corrected compilation from previous oxygen abundance studies included in their work.
\BD\ is enhanced (ca +0.2 dex) over the expected Galactic chemical trend, while \F\  is fully coincident with it. Nevertheless, the poor constraint of the surface gravity introduces high uncertainty into the interpolations on the grid and the obtained non-LTE correction. Hence, these results must be considered as a rough approximation.
In comparison with preceding studies as \citet{Bensby2014,Takeda2005}, both dwarf abundances indicate enhancements over the average trends by ca +0.2 dex.

Unfortunately, alternative lines in the optical range, as the forbidden [\ion{O}{I}] at 6300 or 6363 \AA, which are negligibly affected by non-LTE effects, are not available as they are gravity-dependent and extremely weak or not present in metal-poor dwarf stars \citep{Fulbright2003}.

\begin{table}
\caption{CNO abundances against scaled solar abundance. }
\label{CNO}
\centering
\renewcommand{\arraystretch}{1.4}
\begin{tabular}{lcccccc}
\hline
\hline
    & \multicolumn{3}{c}{\BD} & \multicolumn{3}{c}{Feige 80} \\
\hline
\small{Elem.} & \small{log $Z$}  &  \small{Err.}  & \small{[X/Fe]}   & \small{log $Z$}    & \small{Err.} & \small{[X/Fe]} \\
\hline
\noalign{\smallskip}
C        &  -3.67   &   $^{+0.00}_{-0.12}$  &  +0.58  &  -3.96    &  $^{+0.12}_{-0.06}$   & +0.19    \\
N    &  -3.96   &       $^{+0.37}_{-\;n.c.}$ &  +0.90  &  -4.57    &  $^{+0.37}_{-\;n.c.}$     & +0.19     \\
O    &  -3.28   &   $^{+0.06}_{-0.00}$  &  +0.70  &  -3.23    &  $^{+0.00}_{-0.00}$   & +0.65      \\
\hline
\end{tabular}
\tablefoot{Element fractions $Z=n{\rm X}/n_{\rm {tot}}$. Units in dex.}
\end{table}

\begin{table}
\caption{C/O against the scaled solar ratio.}
\label{CNO ratios}
\centering
\renewcommand{\arraystretch}{1.6}
\begin{tabular}{ccccc}
\hline
\hline
    \multicolumn{2}{c}{\BD} & \multicolumn{2}{c}{Feige 80} &  \\
\hline

C/O &  \small{Err.}    & C/O     &  \small{Err.} & C/O\sun \\
\hline
\noalign{\smallskip}
0.41    &   $^{+0.00}_{-0.13}$    &     0.19    & $^{+0.05}_{-0.03}$    & 0.54  \\

\hline
\end{tabular}
\end{table}

The nitrogen abundances were derived from the high-excitation \ion{N}{I} 8683.40 \AA\,line. Although unblended, it is a weak line only useful for higher metallicities \citep[$> -0.4$ dex;][]{Takeda2005}. Unfortunately, it is the only possibility in the optical range as other lines of similar strength are strongly blended.
The CN bands in the near-UV in the range of 3850-3890 \AA, are heavily impacted by UV flux from the sdO subdwarf and the NH bands in the range of 3360-3370 \AA\, \citep{Israelian2004} lie outside the HERMES spectral range.

The nitrogen abundance on the surface of the sdB's progenitors is expected to be strongly enhanced after the RGB bump \citep{Gratton2000}.
In addition, during RLOF, the inner layers enriched in $^{14}$N are exposed, and both factors might produce a nitrogen enrichment of the companion atmosphere due to mass transfer and thermohaline mixing. 

However, nitrogen analysis of \BD\ and \F\ is limited by the low S/N and quality of the \ion{N}{I} 8683 \AA\,line. As a consequence, only upper limits of nitrogen abundances could be derived. The obtained upper limit of \BD\ is remarkably higher than the Galactic chemical abundance trend. However, the lack of constraint in the lower abundance limit prevents the validation of the nitrogen overabundance. The upper limit of the \F\  MS constrains its abundance as not noticeably enhanced with respect to the Galactic chemical trend. 

\begin{figure}
  \centering
  \includegraphics[width=9.1cm]{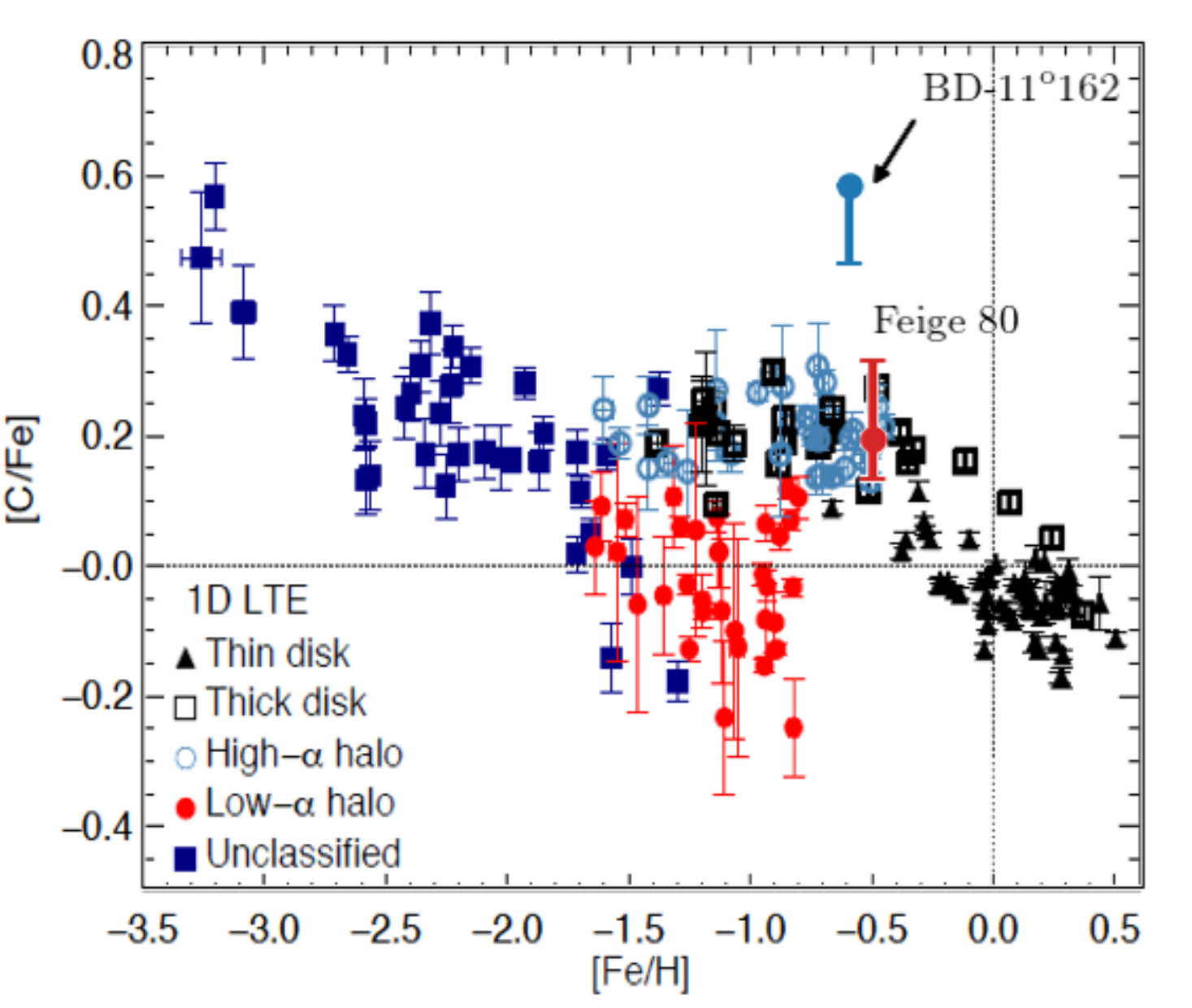}
      \caption{Carbon-to-iron abundance ratios. The sample consists of three different datasets of F and G dwarfs used by \citet{Amarsi2019A&A...630A.104A}: the 67 disk stars (mainly of the thin disk, and including the Sun) in the HARPS-FEROS sample of \citet{Nissen2014A&A...568A..25N}; the 85 thick-disk and halo stars in the UVES-FIES sample of
\citet{Nissen2014A&A...568A..25N}; and the 40 halo stars in the VLT/UVES sample
of \citet{Nissen2007A&A...469..319N}, which were recently reanalysed by \citet{Amarsi2019A&A...622L...4A}. This plot, originally used in \citet{Amarsi2019A&A...630A.104A}, has been adapted to include \BD\ (blue circle) and \F\ (dark red circle) and their error bars ($1\sigma$). Units are in dex.}
\label{Ccompilation}
\end{figure}

The carbon (C) enrichment of \BD\ MS over Galactic chemical trend is a surprising result (see Fig. \ref{Ccompilation}). While the \F\  MS star is matching the general trend, \BD\ yields an unexpected overabundance.
This cannot be explained by RLOF binary interactions, as the sdB progenitor at the moment of mass transfer is expected to have an under abundance of C due to the FDU \citep{Gratton2000, Karakas2014}.

Nevertheless, the C abundances show a striking bimodal behaviour in the sample of He-rich sdO stars (He-sdO) from the European Southern Observatory (ESO) SNeIa Progenitor Survey \citep[SPY;][]{Heber2016}. Striking C overabundance in hot subdwarfs are found at the hotter end of the distribution (T$_{eff}>$\,43900 K). 
They are related with late hot flasher scenarios in which H-rich envelope is engulfed and burnt, leading to an enrichment of He, C and N \citep{Lanz2004,Mocak2010A&A...520A.114M}. However, the hot flash takes place after the RLOF phase and the MS companion should not be impacted by the chemical evolution of the sdB once the accretion phase is over.

Overabundance of C might suggest a different evolutionary stage of the \BD\ sdO progenitor than an RGB star at the moment of the accretion.
However, no hint of enhancement from the second peak s-process elements such as Ba is found, as should be expected \citep{Travaglio2004ApJ...601..864T} in case of a post-AGB scenario. Although since s-process elements have a high T$_{cond}$, depletion processes might have taken place and possible s-process overabundance could be diluted in case of a CB disk formation \citep{Maas2005}.

Regarding the lack of overabundance from the s-process elements and the C peculiarity, other evolutionary paths than the standard channel are on the table. The C anomaly could indicate that the \BD\  subdwarf’s progenitor might have been a star with larger mass than 2 M\sun, and ignited He in a non-degenerate core. On the other hand, the post-RGB scenario can be considered for \BD, given both, the derived sdO's low mass and its atmospheric parameters as will be shown in the following sections.

\section{Analysis of the composite spectra} \label{s:XTgrid}
Beyond the GSSP analysis of the cool companions, we performed a composite spectrum analysis of the available ESO New Technology Telescope (NTT), Large Sky Area Multi-Object Fibre Spectroscopic Telescope (LAMOST), and HERMES spectra to derive the atmospheric parameters of both components in the two binaries. 
We employed a wavelength space direct spectral decomposition \citep{Simon94} with the {\sc XTgrid} code \citep{Nemeth12}. 
The procedure renders synthetic composite spectra from a linear combination of non-LTE ({\sc Tlusty}, \citealt{Hubeny17}) and LTE ({\sc Atlas}, \citealt{Bohlin17}) atmosphere models and iteratively fits all observations simultaneously. 
The fitted parameters include the $T_{\rm eff}$, $\log{g}$, $v_{\rm rot}\sin{i}$, individual abundances of He, C, N, O, Mg, Al, Si, S, Ar, and Fe for the hot subdwarf component, and  $T_{\rm eff}$, $\log{g}$, $v_{\rm rot}\sin{i}$ and the scaled solar metallicity for the cool companion. 
Synthetic composite spectra were built from these components by adjusting a monochromatic flux contribution, which is iterated along with the atmospheric parameters until the fit converges on the best solution. 
The flux contribution of the binary members (dilution factor) shows a considerable degeneracy with surface parameters. 
In particular, a strong anti-correlation between the metallicity of the cool companion and the dilution factor can be observed, but all other atmospheric parameters are affected. 
To overcome these in general, a wide spectral coverage and very high S/N spectra are necessary, which are not always available. 
Therefore, we updated our procedure with ultraviolet and infrared photometric data to constrain the dilution factor across the spectral energy distribution (SED). 
This also allows the spectroscopic solution from {\sc XTgrid} to be compared with an independent SED analysis, as in Sect. \ref{sec:sed}.

Another important update in the fitting procedure is the application of pixel weights for each atmospheric parameter of the model. 
The $\chi^2$ contributions are recorded for each data point in the spectrum (pixel) for each model parameter. 
These weight masks allow the same sensitivity to be achieved for each element in the model, regardless of the number of lines actually seen in the spectrum. 
The most important feature of this approach is to increase the sensitivity for Fe, which has only a very few lines in the optical spectrum of hot stars, while it usually dominates the ultraviolet spectrum. 
Thus, the ultraviolet blocking due to Fe has an effect on the temperature structure of the atmosphere and a measurable contribution to the Balmer line profiles. 
The new procedure aims at increasing the sensitivity for individual elements by applying pixel weights across the optical spectrum. 

{\sc XTgrid} provides a clear spectroscopic solution to the observations and does not apply constrains for the distances, luminosities, and radii of the components. 
These parameters are implicitly included in the value of the dilution factor and the normalisation factor of the composite model to the flux calibrated observation. 
The procedure is tuned to fit hot stars and treats the companion as a background source. 
While this approach requires fitting both components in composite binary spectra, the parameter errors are not evaluated for the cool companion.

To independently confirm the binary parameters, we made an attempt to fit the components in both binaries with {\sc XTgrid}. 
While the solutions were able to reproduce the composite spectra and confirm the temperature and flux contributions of the companions found by GSSP, the $\log{g}$ was systematically off, higher for the cool companions and lower for the sdO stars.
Therefore, we used the results from Sect.\,\ref{GSSP} as starting parameters in the fit. 

\begin{figure*}
  \centering
  \includegraphics{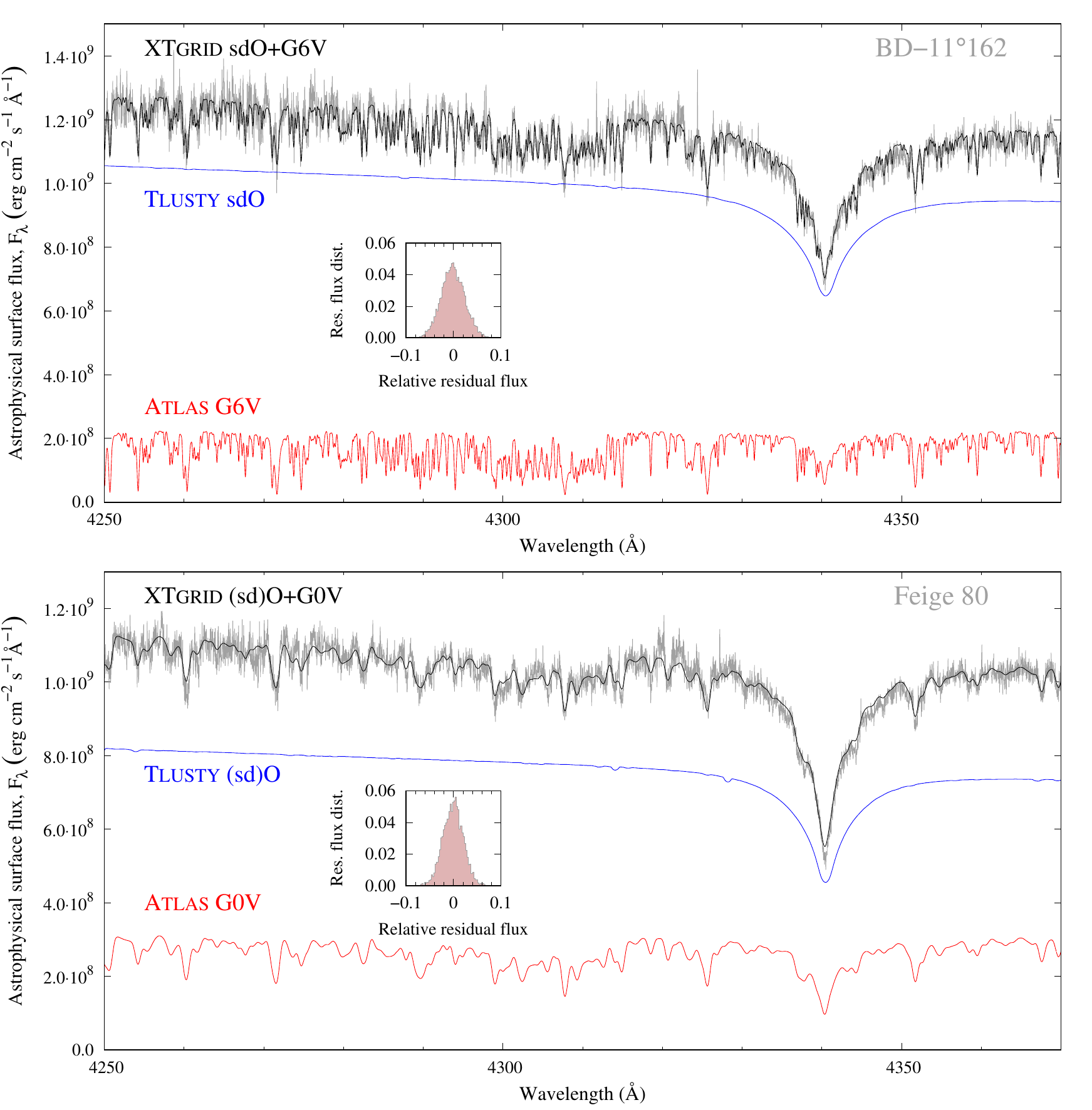}
      \caption{Spectral decomposition of \BD\  and \F. The sum of a {\sc Tlusty} model (blue) and an {\sc Atlas} model (red) fits the observations well. The residual flux distribution shows a symmetric profile, and in 68\% of the data points the models fit the observations better than 2.5\%. Both spectral models are available in their entirety, online at: \protect\url{https://astroserver.org/EZZ74T}}
\label{fig:XTgrid_fit}
\end{figure*}

Figure\,\ref{fig:XTgrid_fit} shows the spectral decomposition in the 4250-4370 \AA\ region, and the parameters from our spectral analysis are listed in Table \ref{tb:XTgrid_param}. 
For the chemical composition of the hot subdwarf, we list the number abundances and the solar abundance fractions with respect to \citet{Asplund09} in Table \ref{tb:XTgrid_abund}.

The goodness of fit can be evaluated from the distribution of residuals in Fig.\,\ref{fig:XTgrid_fit}. 
The more symmetric and narrower the distribution, the better the fit is. 
In a good fit the relative residuals correspond to the S/N, assuming that all spectral properties are calibrated well (instrumental resolution, model completeness, etc.). 
It is fulfilled for both systems, in 68\% of the data points the observed flux is within 2.5\% of the final model and the S/N (4250-4370 \AA) of the HERMES spectra of \BD\  and \F\  are 45 and 56, respectively. 
The spectral models of the sdO components were used in the RV cross-correlation to resolve the few broad and heavily blended lines of the hot subdwarf, and therefore to improve the RV curves and the final mass ratios.

The {\sc XTgrid} analysis found that the absolute flux contributions of the components are equal at $7400\pm50$ \AA\ in \BD\  and at $7000\pm50$ \AA\ in \F. 
In both systems the sdO component dominates most of the optical spectrum. 

\begin{table}[]
    \centering
    \caption{{\sc XTgrid} results of \BD\  and \F.}
    \label{tb:XTgrid_results}
    \begin{tabular}{lr@{ $\pm$ }lr@{ $\pm$ }l}
      \hline\hline
      Parameter & \multicolumn{2}{c}{\BD}& \multicolumn{2}{c}{\F} \\\hline
      \noalign{\smallskip}
      T$_{\rm eff, MS}$ (K)              & 5830 & 250    & 6250  & 150  \\
      $\log{g}_{\rm MS}$ (cm\,s$^{-2}$)  & 4.80 & 0.30   & 4.49  & 0.30 \\
      $[{\rm Fe}/{\rm H}]_{\rm MS}$      & -0.42& 0.30   & -0.50 & 0.30 \\
      $v\sin{i}$ (km s$^{-1}$)           & \multicolumn{2}{c}{12} & \multicolumn{2}{c}{29} \\
      \noalign{\smallskip}
      T$_{\rm eff, sdO}$ (K)             & 52400 & 2570  & 39680 & 660  \\
      $\log{g}_{\rm sdO}$ (cm\,s$^{-2}$) & 5.45  & 0.04  & 4.70  & 0.05  \\
      $\log{(n{\rm He}/n{\rm H})}$       & -1.88 & 0.03  &-2.29  & 0.02 \\
      $v\sin{i}$ (km s$^{-1}$)           & 47    & 15      & 33  & 6     \\
      \hline
    \end{tabular}
\label{tb:XTgrid_param}
\tablefoot{The projected rotation velocities of the MS companions were adopted from the GSSP analysis.}
\end{table}

\begin{table*}
\caption{Element abundances for the sdO components of \BD\  and \F.}
\label{Abun}
\centering
\renewcommand{\arraystretch}{1.4}
\begin{tabular}{l|rlr|rlr}
\hline
\hline
 & \multicolumn{3}{c|}{\BD} & \multicolumn{3}{c}{\F} \\
\hline
\small{Elem.} & \small{log $Z$}  &  \small{Err.}  & \small{$\log(\epsilon/\epsilon_\odot)$}   & \small{log $Z$}    & \small{Err.} & \small{$\log(\epsilon/\epsilon_\odot)$} \\
\hline
He      &  -1.89 &   $\pm0.02$  &    -0.81 &   -2.30    &   $\pm0.02$   &   -1.22  \\
C       & <-5.12 &   $\pm0.07$  &   <-1.62 &   -5.11    &   $\pm0.37$   &   -1.54  \\
N       & <-4.69 &   $\pm0.02$  &   <-0.54 &  <-5.00    &   $\pm0.02$   &  <-0.84  \\
O   &  -4.71 &   $\pm0.23$  &    -1.40 &   -4.19    &   $\pm0.31$   &   -0.88  \\
Mg  &  -4.40 &   $\pm0.22$  &     0.00 &   -3.54    &   $\pm0.50$   &   +0.86  \\
Si  & <-5.14 &   $\pm0.04$  &   <-0.69 &  <-4.95    &   $\pm0.02$   &  <-0.48  \\
Fe  &  -3.56 &   $\pm0.12$  &    +0.94 &   -3.71    &   $\pm0.43$   &   +0.79  \\
\hline
\end{tabular}
\label{tb:XTgrid_abund}
\tablefoot{Element fractions $Z=n{\rm X}/n{\rm H}$. Abundance scale $log\,\epsilon({\rm X})=log\,(n{\rm X}/n{\rm H})+12$. Units in dex.}
\end{table*}

\begin{figure}
  \centering
  \includegraphics[width=\linewidth]{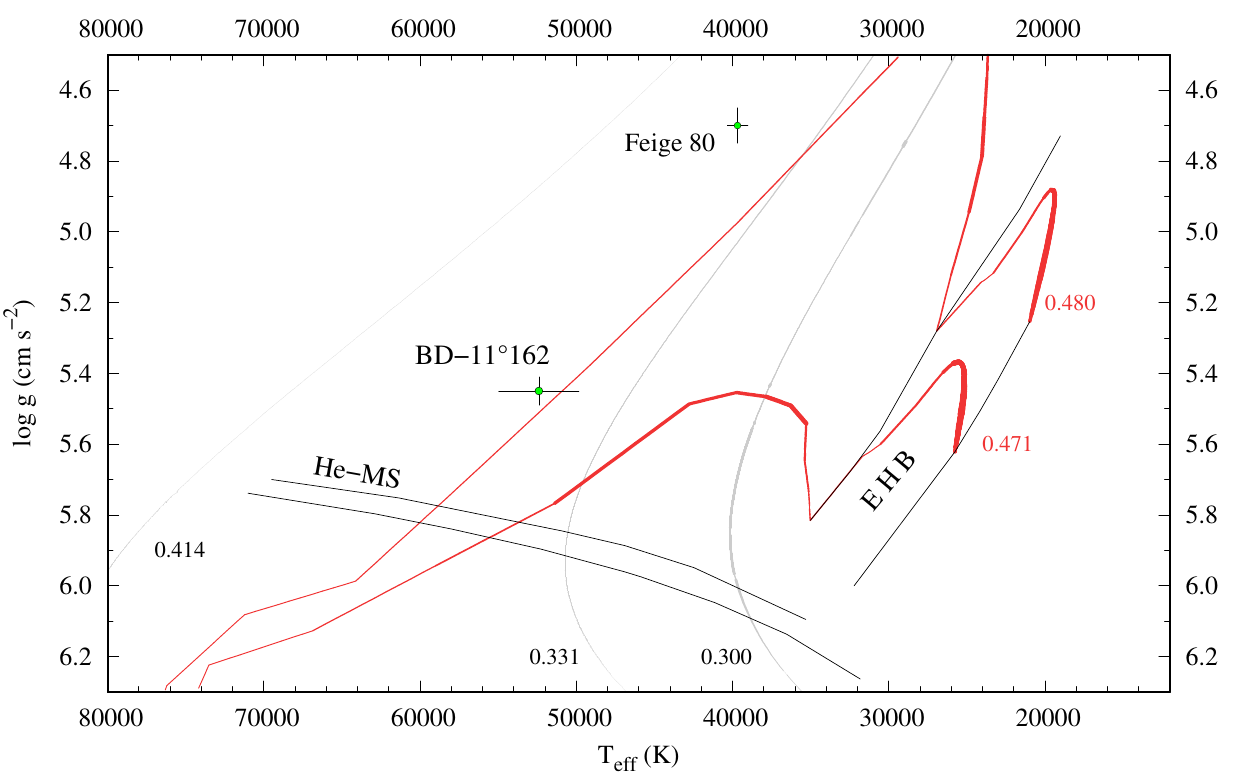}
      \caption{Hot subdwarf components of \BD\  and \F\  in the $T_{\rm eff}-\log{g}$ diagram. 
      Low-mass He-core WD evolutionary tracks for 0.300, 0.331, and 0.414 M$_\odot$ models from \cite{1998A&A...339..123D} are shown in grey and labelled with the stellar mass.
      Two evolutionary tracks for hot subdwarf stars with 0.471 and 0.480\,M$_\odot$ from \citet{dorman93} are marked with red. }
\label{fig:Tefflogg}
\end{figure}

\section{Spectral energy distribution}\label{sec:sed}
An independent set of atmospheric parameters can be derived by comparing the photometric SED with theoretical atmosphere models. However, since the main goal of this exercise is to obtain radii for both components and since the effective temperature of the sdO stars is hard to constrain without reliable UV photometry, we constrain the atmospheric parameters to those obtained from the spectroscopic analysis and only derive the radii and luminosity from the SED.

Literature photometry from Gaia Early Data Release 3 \citep[EDR3;][]{GaiaDR2, Riello2018, Evans2018}, AAVSO Photometric All Sky Survey Data Release 9 \citep[APASS DR9;][]{Henden2015}, Two Micron All-Sky Survey \citep[2MASS;][]{Skrutskie2006}, and Wide-field Infrared Survey Explorer W1 (3.4 $\mu$m) and W2 (4.6 $\mu$m) filters \citep[WISE W1 and W2;][]{Cutri2012} is used for both systems. Furthermore, there is Stromgren photometry available for \BD\  from the \citet{Paunzen2015} catalogue and for \F\  from \citet{Graham1970}. 
By using the Gaia EDR3 parallax \citep{Lindegren2018, Luri2018}, the radii and luminosity of both components can be derived. The distance to \BD\  and \F\  obtained from the Gaia parallax is, respectively, 326 $\pm$ 11 and 546 $\pm$ 18 pc. The reddening in the direction of \BD\ as determined from the dust maps of \citet{Lallement2019} is E(B-V) = 0.028 $\pm$ 0.018, while for \F\  it is 0.027 $\pm$ 0.019.  Both values are very low and will have little influence on the fit. The surface gravity cannot be constrained by an SED fit, and is kept fixed at the values determined from spectroscopy. The error on the surface gravity is propagated in the SED fit.

To fit the SED, models from the T\"ubingen non-LTE Model-Atmosphere Package (TMAP; \citealt{Werner2003}) are used for the sdO component, and Kurucz atmosphere models \citep{Kurucz1979} for the companion. A Markov chain Monte Carlo (MCMC) approach is used to find the global minimum and determine the error on the fit parameters. Errors on the constraints are propagated to the final results. A more detailed description of the fitting process is given in \citet{Vos2013, Vos2017} and \citet{Vos2018a}. The code to fit the SEDs is included in the {\sc speedyfit} package, which is available on github\footnote{https://github.com/vosjo/speedyfit}.

\begin{figure*}
  \centering
  \includegraphics{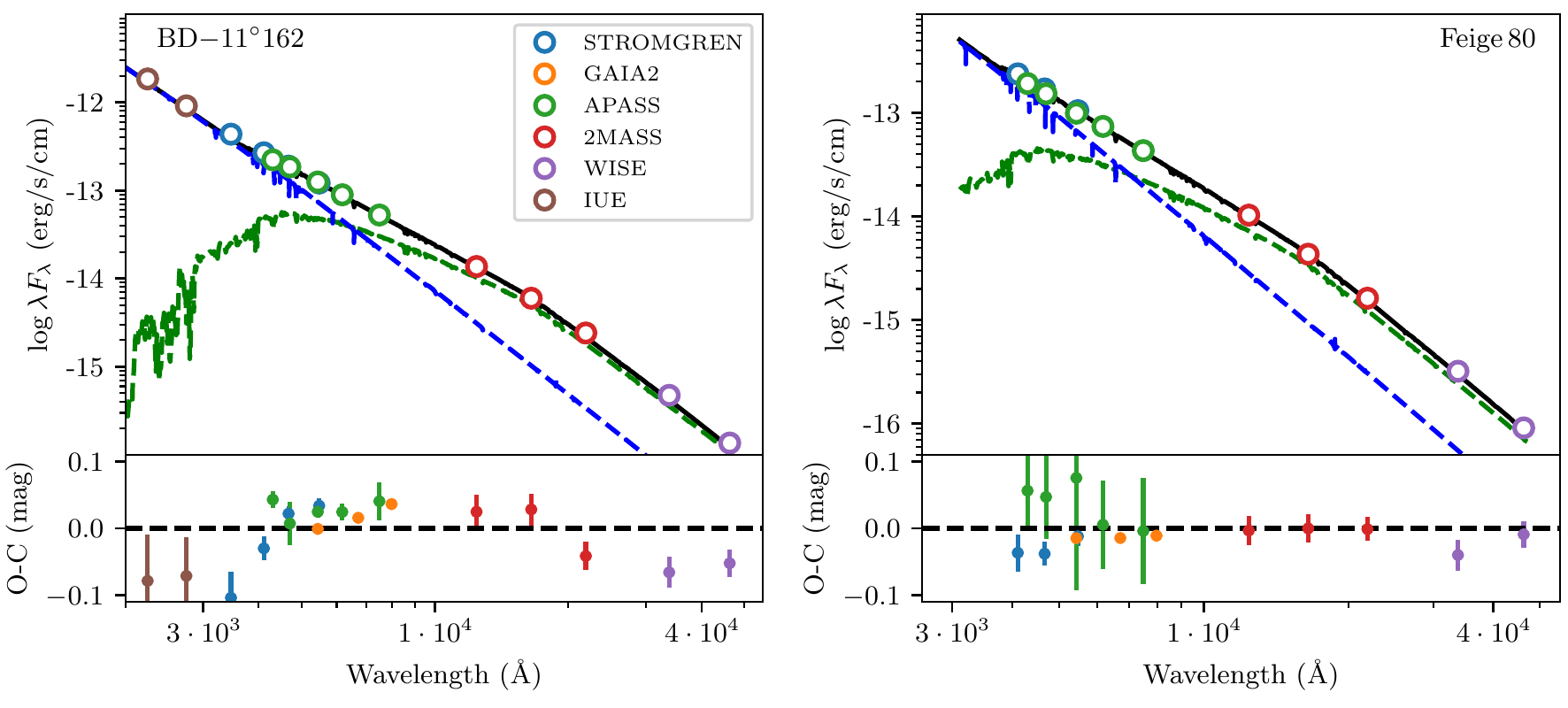}
      \caption{Photometric SEDs for \BD\  on the left and \F\  on the right. The observations are shown in coloured circles, while the best fitting binary model is shown with a solid black  line. The contribution of, respectively, the hot sdO and the cool companion are shown in dashed blue and green lines. The bottom panels show the O-C for the best fitting model.}
\label{fig:sed_fit}
\end{figure*}

The parameters of the best fitting model are shown in Table\,\ref{tb:sed_results} and the observations with the best model fit are shown in Fig.\,\ref{fig:sed_fit}. The obtained radii and luminosities are reasonable for MS stars and subdwarfs.

An interesting observation for \BD\ is that the WISE W4 band shows a slight IR excess. Most WISE W3 and W4 measurements for hot subdwarf stars are upper limits as they are below the detection limit of WISE, but in the case of \BD, the detection is reported as significant, with a S/N of 10.3.

\begin{table}
    \caption{Resulting parameters from the best fitting model to the photometric SED of \BD\  and \F.}
    \centering
    \label{tb:sed_results}
    \begin{tabular}{lr@{ $\pm$ }lr@{ $\pm$ }l}
        \hline\hline
        Parameter & \multicolumn{2}{c}{\BD} & \multicolumn{2}{c}{\F} \\\hline
        \noalign{\smallskip}
        R$_{\rm MS}$ ($R_{\odot}$)  &  1.05 & 0.10  & 1.31  & 0.06 \\
        L$_{\rm MS}$ ($L_{\odot}$)  &  1.45 & 0.23  & 2.55   & 0.5  \\
        \noalign{\smallskip}
        R$_{\rm sdO}$ ($R_{\odot}$) & 0.16  & 0.02  & 0.22  & 0.03  \\
        L$_{\rm sdO}$ ($L_{\odot}$) & 42    & 7     & 91    & 15    \\
        \hline
    \end{tabular}
    \tablefoot{See Sect.\ref{sec:sed} for details.}
\end{table}

\section{Mass determination}\label{s:mass_determinatnion}
The spectroscopic parameters of the cool companions can be used to derive their mass by comparing them to theoretical stellar evolution models. This assumes that the cool companions behave as single stars, and did not accrete enough mass to significantly change their evolutionary state. Based on a study of the rotational velocities of the companions, \citet{Vos2018a} concludes that very little mass ($<$ 0.03 $M_{\odot}$) is accreted during the interaction phase, and that this assumption thus holds true. 
Here the MESA Isochrones \& Stellar Tracks (MIST, \citealt{Choi2016, Dotter2016}) models are used. The fit is  achieved  using  an MCMC approach similar to the one described in \citet{Maxted2015}. The fitting algorithm used here, {\sc emcmass}, is implemented in python and the source code is available on github\footnote{https://github.com/vosjo/emcmass}. More details of the implementation of this package are given in \citet{Vos2018b}, as well as a test against the catalogue of the physical properties of eclipsing binaries (DEBCat, \citealt{Southworth2015} based on \citealt{Andersen1991}). The observables that are used here are the effective temperature, the metallicity, and the radius. It would be possible to include the surface gravity, but the error on that parameter is too large to add constrain to the fit.

We find for \BD\  a mass of 0.78 $\pm$ 0.08 M$_{\odot}$, while the cool companion of \F\  has a mass of 1.15 $\pm$ 0.10 M$_{\odot}$. Using the derived mass ratio, we can estimate the masses of the sdO stars, which results in 0.37 $\pm$ 0.04 and 0.48 $\pm$ 0.05 M$_{\odot}$ for, respectively, \BD\  and \F\  (see Table\,\ref{OP_mass}). 

Assuming that the subdwarf has been formed through the standard post-EHB channel from a progenitor with a mass lower than 2 M\sun, the subdwarf mass-orbital period relation given by \citet{Chen2013} can be used to obtain the sdOs masses. 
We can compare those estimates with the results based on the evolution model fitting of the MS stars.

We consider Z = 0.02 (population I stars) and the inclusion of atmospheric RLOF. 
The relation by \citet{Chen2013} is derived for sdB stars. 
However, the evolution from sdB to sdO does not cause a significant change in mass due to the relatively short period in terms of stellar evolution ($\sim$ 100\,Myr) and stellar winds during the sdB phase being negligible. 
Using this method, we find a mass of 0.41 $\pm$ 0.01 and 0.48 $\pm$ 0.01 M$_{\odot}$, respectively, for \BD\  and \F. 
Because of the assumptions used in \citet{Chen2013}, the mass error intervals depend on the period errors. Therefore, they only evaluate the adjustment to the remnant subdwarf mass – orbital period relation, underestimating the error from other sources. Hence, they should not be compared to those obtained from stellar evolution models.
On the other side, meanwhile the derived \F\  subdwarf mass matches the obtained by stellar evolution models, it is not the case of the \BD\ one. This could suggest that the \BD\ subdwarf is not accomplishing the initial assumption of this method: to be formed through the standard post-EHB channel from a progenitor with a mass lower than 2 M\sun.

Thanks to the minimum masses previously obtained from the orbital solution (Table\,\ref{OP Table}), the orbital inclinations for both systems are derived. This is done using the mass obtained from MIST method. The results are shown in Table\,\ref{OP_mass}.

\begin{table}
\caption{Masses of system components and system inclinations.} 
\label{OP_mass}
\centering
\renewcommand{\arraystretch}{1.4}
\begin{tabular}{lcc}
\hline\hline
\noalign{\smallskip}
Systems & \BD & Feige 80 \\
\hline
\noalign{\smallskip}
MS mass (M\sun)         &   0.78    $\pm$ 0.08       & 1.15        $\pm$ 0.10\\
sdO mass (M\sun)        &   0.37    $\pm$ 0.04       & 0.48        $\pm$ 0.05\\\hline
Inclination ($^o$)      &   70      $\pm$ 4          & 61          $\pm$ 5\\
\hline
\end{tabular}
\tablefoot{Masses obtained by fitting stellar evolution models (MIST) to the MS stars. The system inclination is determined by comparing the minimum mass with the derived mass.}
\end{table}

\section{Galactic orbits}
\begin{figure*}
  \centering
  \includegraphics[width=14cm]{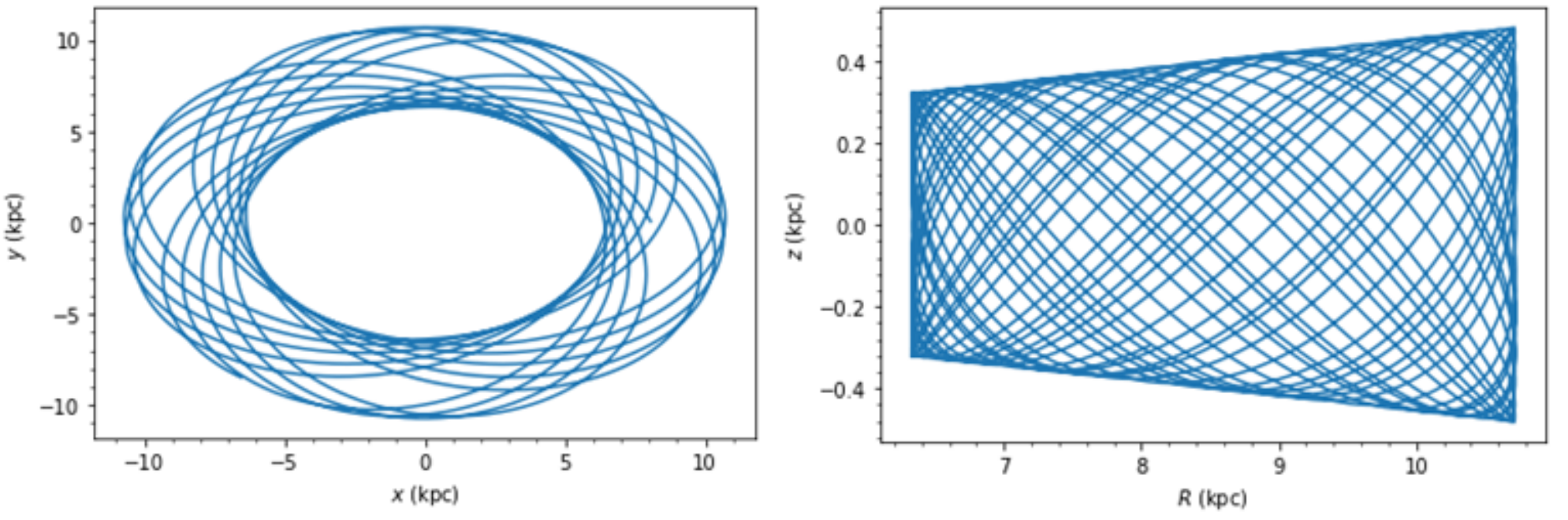}
  \includegraphics[width=14cm]{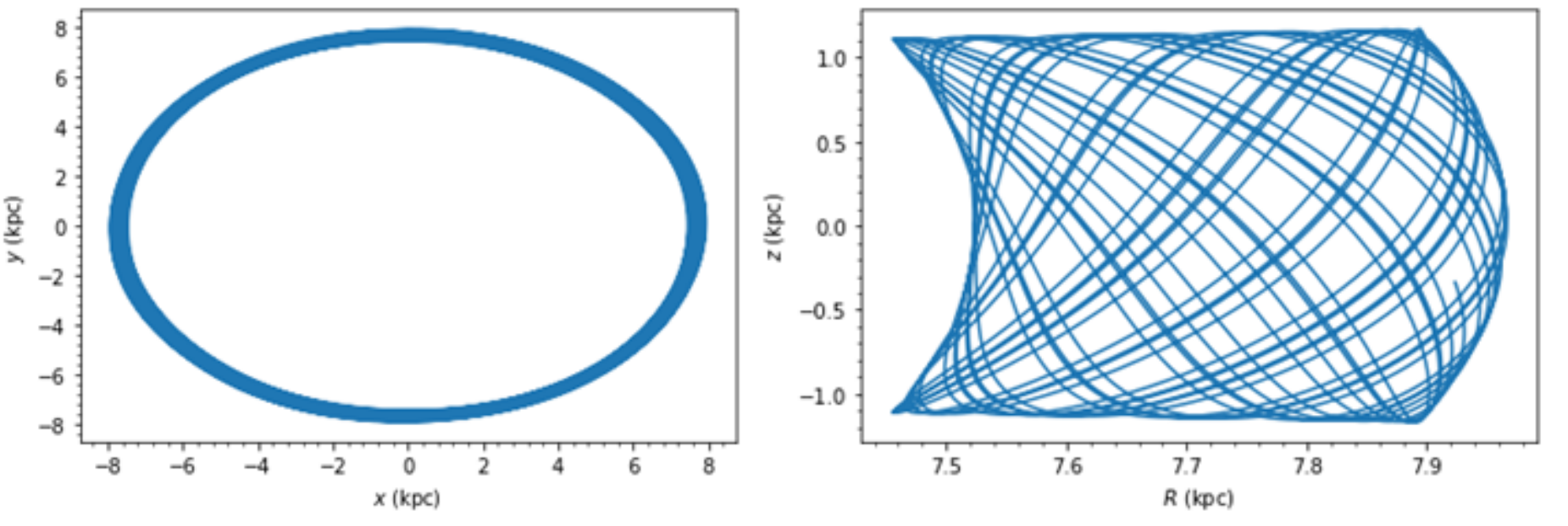}
      \caption{Galactic orbits of \BD\ (top) and \F\ (bottom) projected onto the Galactic plane (left) and the vertical Galactic component against the radius with respect the Galactic centre (right).}
\label{OrbitBDF}
\end{figure*}

The {\sc galpy} software package \citep{Bovy2015ApJS..216...29B} was used to calculate the Galactic orbits of \BD\ and \F. The calculated orbits can be seen in Fig. \ref{OrbitBDF}.
\BD\ shows a moderate eccentricity of 0.26 $\pm$ 0.01 and Galactic vertical component of the angular momentum of 1744 $\pm$ 5 kpc km s$^{-1}$. These values are on the edge between the thin- and thick-disk populations \citep{Pauli2006A&A...447..173P}. 

Regarding \F, its low eccentricity of 0.03 $\pm$ 0.01 and relatively high Galactic vertical component of the angular momentum of 1686 $\pm$ 5 kpc km s$^{-1}$ might indicate thin disk membership. However, it shows an orbit with a maximum vertical height exceeding the Galactic plane of 1.17 $\pm$ 0.02 kpc. Considering that the edge of the thin disk is located at about 0.4-0.5 kpc, this means that it transits the thick disk for a large part of its orbit.

\section{Discussion and conclusions}
\BD\  and \F\  are the first post-stable-RLOF sdO binaries with solved orbits. 
In preceding works on wide sdB binaries, it was shown that many of them have significantly eccentric orbits, even although they are predicted to circularise before the onset of mass loss, a puzzling feature that has also been found in many other evolved binaries systems.
The orbital parameters indicate that \BD\  belongs to group II in the orbital period-eccentricity diagram with an orbital period around 600 days, while \F\  belongs to the majority group. This reinforces the actual bimodal behaviour in the period-eccentricity and period-mass ratios of long-period subdwarf binaries. The addition of \BD\  to the few exceptional systems outside the main period-eccentricity and period-mass ratio relations might provide a better insight into these systems.

Recently, \citet{Vos2019} studied the $P$-$q$ relation and {\bf suggest} that it is linked to the stability of RLOF during the RGB stage. A follow-up study shows that the $P$-$q$ relation is caused by a combination of the Galactic metallicity evolution and the binary interaction mechanism that creates sdB stars \citep{Vos2020}. However, this work does not provide an explanation of the systems in the minority group.

\begin{figure}
\centering
\includegraphics[width=9cm]{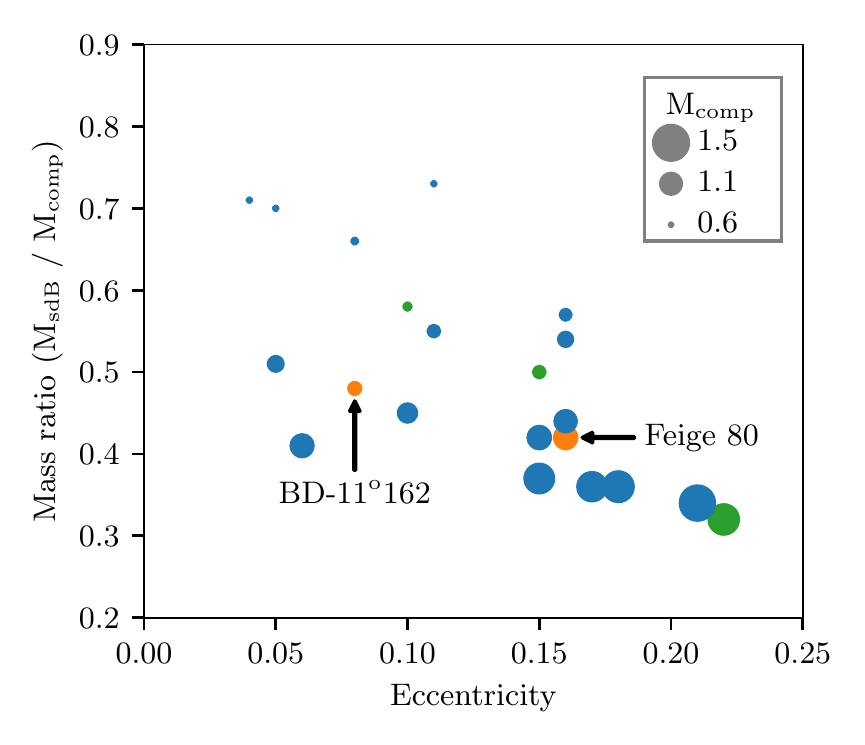}
  \caption{Mass ratio-eccentricity distribution. Group II is shown in green. Bubble size indicates the MS companion mass (to make the differences clear, they are plotted as MS mass minus 0.6 M\sun). For clarity,  error bars are not included. Orbital parameters are obtained from \citet{Vos2019}.}
  \label{q-Ecc}
\end{figure}

Since there is an observed correlation between both $P$-$q$ and $P$-$e$, it is not a big step to infer that there will also be a correlation between the mass ratio and the eccentricity. We show this correlation in Fig.\,\ref{q-Ecc}. An interesting property of this figure is that it does not show the strong bimodal behaviour that is so apparent in the $P$-$q$ and $P$-$e$ correlations. We show that the mass ratio-eccentricity relation indicates a stronger correlation than the traditional orbital period-eccentricity one. 
 
The final mass ratio in wide sdB binaries is strongly dependent on the interaction phase. \citet{Vos2020} shows that, for a given initial donor mass, only a small range in initial mass ratios is capable of producing an sdB star. This range in $q_{\rm i}$ then dictates a narrow range in the final mass ratio. 
Since the eccentricity correlates quite strongly with the mass ratio, independent of the orbital period, this could indicate that the main cause of the eccentricity of the systems is not so much related to the periods, but more to the conditions necessary to create an sdB star during the interaction phase. A more detailed theoretical study is, however, necessary to confirm this.

Mass transfer during the RLOF phase in wide hot subdwarf binary systems causes angular momentum to be gained by the companion stars \citep{Vos2018a}. It is expected that the companion will spin up \citep[e.g.][]{Popham1991ApJ...370..604P} and that its higher-than-average rotational velocity is evidence of the previous interaction.

\citet{Pelisoli2020A&A...642A.180P} studied the light curves of wide hot subdwarf systems using observations from the Transiting  Exoplanet  Survey  Satellite  (TESS).
The authors conclude that a large fraction of the studied sample of the MS companions in wide subdwarf binaries show rotation periods significantly smaller than usual with respect to field MS stars. 
This increased rotation is observed in at least 60\% of the sample. However, when taking the detection limits of TESS into account, evidence for binary interaction is detected in nearly 100\% of the studied systems. 

The authors presented a distribution range of rotation rates from the MS companions that shows evidence of interaction. Their distribution is in contrast with the one obtained from the sample of MS field stars by \citet{ Reinhold2015A&A...583A..65R}.
The rotation period derived from TESS observations of the MS companion in \BD\  is $\log{P} = 0.55$ d (negligible error).
The \citet{Pelisoli2020A&A...642A.180P} distribution is described by $\log{P} = 0.35$ d ($\sigma$=0.27 d). 
This MS star is included in the distribution, pointing to a past binary interaction episode.
Feige\,80 was not included in the study from \citet{Pelisoli2020A&A...642A.180P}.

The atmospheric parameters indicate that the cool companion stars  of \BD\  and \F\  are of type G1V and F7IV/V, respectively. 
The enhancement of $\alpha$ elements in these metal-poor MS stars suggests that they might belong to the Galactic thick disk \citep{Duong2018} or be edge systems between the thin and thick disks \citep{Adibekyan2013}.
This is corroborated by the Galactic orbits of both systems, which show edge features between thin- and thick-disk stars. This is consistent with their moderately poor metallicity and the chemical abundances of the majority of elements in the MS stars except the yttrium and carbon anomalies.

Regarding the chemical anomalies, both MS companions show a depleted Y abundance. As Y has a high condensation temperature, and would be depleted first from CB gas, this could indicate that a CB disk was created during the RLOF phase and that element depletion in re-accreted gas is the cause of the final Y depletion in the MS. This supports the suggestion from \citet{Vos2015} that CB disks are involved in the eccentricity of these systems. However, with only two systems studied, no definitive conclusions can be drawn.

\BD\ has an interesting anomaly in its C abundance. Its cool companion shows a rather strong C enhancement compared to the Galactic abundance trend. The derived mass of the \F\  subdwarf suggests that it was formed following the standard post-EHB channel. However, for the formation of the \BD\  subdwarf, some scenarios are open: non-degenerate He ignition by an sdB’s progenitor with a mass larger than 2 M\sun, the late hot flasher scenario, and the post-RGB scenario due to strong binary interactions and mass loss. In the last case, there is no He ignition taking place, and \BD\  would be a pre-WD instead of a subdwarf. The C anomaly could indicate that the progenitor of the subdwarf of \BD\  might have been a star with a mass larger than 2 M\sun\, that ignited He in a non-degenerate core. The low derived sdO mass (see Sect.\,\ref{s:mass_determinatnion}) also suggests a formation different than the standard post-EHB channel with a degenerate core ignition. Unlike the standard channel, the final mass of the subdwarf obtained by ignition of a non-degenerate helium core does not follow the tight relation between final mass and orbital period studied by \citet{Chen2013}. It could lead to systems with lower orbital periods \citep{Chen2002MNRAS.335..948C,Chen2003MNRAS.341..662C}.

However, even for the most optimistic scenario, the stellar evolution of a star with mass slightly larger than 2 M\sun\, would indicate a younger system than the metallicity of the system suggests.
On the other hand, the derived low mass of the \BD\  sdO is still on the edge, compatible with a core-He flash whenever the sdO's progenitor is close to 2 M\sun. In this case, the MS's C anomaly would remain unexplained and other scenarios should be evaluated.

As \BD\  is a member of the peculiar group II, and the C abundance enhancement is not visible in \F. This might hint at a possible progenitor mass difference between groups I and II. However, with only two systems studied, no conclusions can be drawn.

The abundances of Ba as an s-process prototype element are close to the scaled solar values. 
This supports the exclusion of the post-AGB formation channel for both hot subdwarfs. Although, as commented above, depletion processes linked to the accretion phase could be distorting the abundances of these elements with high condensation temperatures.
The lack of s-process element enrichment is consistent with the assumption that the sdO star of \F\  is the evolved product of a canonical sdB that had its interaction phase near the tip of the RGB (standard post-EHB channel with degenerate He ignition).

The positions of the two sdO stars in the $T_{\rm eff}-\log{g}$ diagram in Fig.\,\ref{fig:Tefflogg} clearly show the evolved nature of both hot subdwarfs. 
The high surface temperature and low gravity places them well beyond the EHB (core He burning), between $\log{L/L_{\rm Edd}=-3}$ and $-2.6$, as well as beyond the post-EHB (shell He burning) stage, between $\log{L/L_{\rm Edd}=-2.6}$ and $-1.6$. 
The Eddington luminosity fractions are close to the maximum value post-EHB stars can reach; therefore, the future evolution of both sdO stars is, most likely, a rapid contraction and heating at a nearly constant bolometric luminosity, before they reach the WD cooling track, that corresponds to their mass. 
The surface helium and metal abundances will further decrease along this evolution.

\begin{acknowledgements}
This work was supported by a fellowship for postdoctoral researchers from the Alexander von Humboldt Foundation awarded to JV.
PN acknowledges support from the Grant Agency of the Czech Republic (GA\v{C}R 18-20083S).
The research leading to these results has (partially) received funding from the KU~Leuven Research Council (grant C16/18/005: PARADISE), from the Research Foundation Flanders (FWO) under grant agreement G0H5416N (ERC Runner Up Project), as well as from the Belgian federal Science Policy Office (BELSPO) through PRODEX grant PLATO.
This research has used the services of \mbox{\url{www.Astroserver.org}} under reference \aref.
Based on observations made with the Mercator Telescope, operated on the island of La Palma by the Flemish Community, at the Spanish Observatorio del Roque de los Muchachos of the Instituto de Astrofísica de Canarias.
Based on observations obtained with the HERMES spectrograph, which is supported by the Research Foundation - Flanders (FWO), Belgium, the Research Council of KU Leuven, Belgium, the Fonds National de la Recherche Scientifique (F.R.S.-FNRS), Belgium, the Royal Observatory of Belgium, the Observatoire de Genève, Switzerland and the Thüringer Landessternwarte Tautenburg, Germany.
This paper makes use of the following ALMA data: ADS/JAO.ALMA\#2018.1.00682.S and ADS/JAO.ALMA\#2017.1.00614.S. ALMA is a partnership of ESO (representing its member states), NSF (USA) and NINS (Japan), together with NRC (Canada), MOST and ASIAA (Taiwan), and KASI (Republic of Korea), in cooperation with the Republic of Chile. The Joint ALMA Observatory is operated by ESO, AUI/NRAO and NAOJ.
\end{acknowledgements}

\bibliographystyle{aa}
\bibliography{bibliography}

\begin{thebibliography}{118}
\expandafter\ifx\csname natexlab\endcsname\relax\def\natexlab#1{#1}\fi

\bibitem[{{Abia} {et~al.}(2003){Abia}, {Dom{\'{\i}}nguez}, {Gallino}, {Busso},
  {Straniero}, {de Laverny}, \& {Wallerstein}}]{Abia2003}
{Abia}, C., {Dom{\'{\i}}nguez}, I., {Gallino}, R., {et~al.} 2003,
  \href{http://dx.doi.org/10.1071/AS03021}{\color{magenta}\pasa},
  \href{https://ui.adsabs.harvard.edu/abs/2003PASA...20..314A}{20, 314}

\bibitem[{{Adibekyan} {et~al.}(2013){Adibekyan}, {Figueira}, {Santos},
  {Hakobyan}, {Sousa}, {Pace}, {Delgado Mena}, {Robin}, {Israelian}, \&
  {Gonz{\'a}lez Hern{\'a}ndez}}]{Adibekyan2013}
{Adibekyan}, V.~Z., {Figueira}, P., {Santos}, N.~C., {et~al.} 2013,
  \href{http://dx.doi.org/10.1051/0004-6361/201321520}{\color{magenta}\aap},
  \href{https://ui.adsabs.harvard.edu/abs/2013A&A...554A..44A}{554, A44}

\bibitem[{Akaike(1974)}]{akaike1974}
Akaike, H. 1974, IEEE Transactions on Automatic Control, 19, 19

\bibitem[{{Amarsi} {et~al.}(2015){Amarsi}, {Asplund}, {Collet}, \&
  {Leenaarts}}]{Amarsi2015}
{Amarsi}, A.~M., {Asplund}, M., {Collet}, R., \& {Leenaarts}, J. 2015,
  \href{http://dx.doi.org/10.1093/mnrasl/slv122}{\color{magenta}\mnras},
  \href{https://ui.adsabs.harvard.edu/abs/2015MNRAS.454L..11A}{454, L11}

\bibitem[{{Amarsi} {et~al.}(2016){Amarsi}, {Asplund}, {Collet}, \&
  {Leenaarts}}]{Amarsi2016}
{Amarsi}, A.~M., {Asplund}, M., {Collet}, R., \& {Leenaarts}, J. 2016,
  \href{http://dx.doi.org/10.1093/mnras/stv2608}{\color{magenta}\mnras},
  \href{https://ui.adsabs.harvard.edu/abs/2016MNRAS.455.3735A}{455, 3735}

\bibitem[{{Amarsi} {et~al.}(2019{\natexlab{a}}){Amarsi}, {Nissen}, {Asplund},
  {Lind}, \& {Barklem}}]{Amarsi2019A&A...622L...4A}
{Amarsi}, A.~M., {Nissen}, P.~E., {Asplund}, M., {Lind}, K., \& {Barklem},
  P.~S. 2019{\natexlab{a}},
  \href{http://dx.doi.org/10.1051/0004-6361/201834480}{\color{magenta}\aap},
  \href{https://ui.adsabs.harvard.edu/abs/2019A&A...622L...4A}{622, L4}

\bibitem[{{Amarsi} {et~al.}(2019{\natexlab{b}}){Amarsi}, {Nissen}, \&
  {Sk{\'u}lad{\'o}ttir}}]{Amarsi2019A&A...630A.104A}
{Amarsi}, A.~M., {Nissen}, P.~E., \& {Sk{\'u}lad{\'o}ttir}, {\'A}.
  2019{\natexlab{b}},
  \href{http://dx.doi.org/10.1051/0004-6361/201936265}{\color{magenta}\aap},
  \href{https://ui.adsabs.harvard.edu/abs/2019A&A...630A.104A}{630, A104}

\bibitem[{{Andersen}(1991)}]{Andersen1991}
{Andersen}, J. 1991,
  \href{http://dx.doi.org/10.1007/BF00873538}{\color{magenta}\aapr},
  \href{http://adsabs.harvard.edu/abs/1991A%26ARv...3...91A}{3, 91}

\bibitem[{{Andrews} \& {Williams}(2007)}]{Andrews2007}
{Andrews}, S.~M. \& {Williams}, J.~P. 2007,
  \href{http://dx.doi.org/10.1086/522885}{\color{magenta}\apj},
  \href{https://ui.adsabs.harvard.edu/abs/2007ApJ...671.1800A}{671, 1800}

\bibitem[{{Asplund} {et~al.}(2009){Asplund}, {Grevesse}, {Sauval}, \&
  {Scott}}]{Asplund09}
{Asplund}, M., {Grevesse}, N., {Sauval}, A.~J., \& {Scott}, P. 2009,
  \href{http://dx.doi.org/10.1146/annurev.astro.46.060407.145222}{\color{magenta}\araa},
  \href{https://ui.adsabs.harvard.edu/abs/2009ARA&A..47..481A}{47, 481}

\bibitem[{{Bensby} {et~al.}(2014){Bensby}, {Feltzing}, \& {Oey}}]{Bensby2014}
{Bensby}, T., {Feltzing}, S., \& {Oey}, M.~S. 2014,
  \href{http://dx.doi.org/10.1051/0004-6361/201322631}{\color{magenta}\aap},
  \href{https://ui.adsabs.harvard.edu/abs/2014A&A...562A..71B}{562, A71}

\bibitem[{{Berger} \& {Fringant}(1980)}]{Berger1980}
{Berger}, J. \& {Fringant}, A.~M. 1980, \aap,
  \href{https://ui.adsabs.harvard.edu/abs/1980A&A....85..367B}{85, 367}

\bibitem[{{Bohlin} {et~al.}(2017){Bohlin}, {M{\'e}sz{\'a}ros}, {Fleming},
  {Gordon}, {Koekemoer}, \& {Kov{\'a}cs}}]{Bohlin17}
{Bohlin}, R.~C., {M{\'e}sz{\'a}ros}, S., {Fleming}, S.~W., {et~al.} 2017,
  \href{http://dx.doi.org/10.3847/1538-3881/aa6ba9}{\color{magenta}\aj},
  \href{https://ui.adsabs.harvard.edu/abs/2017AJ....153..234B}{153, 234}

\bibitem[{{Bovy}(2015)}]{Bovy2015ApJS..216...29B}
{Bovy}, J. 2015,
  \href{http://dx.doi.org/10.1088/0067-0049/216/2/29}{\color{magenta}\apjs},
  \href{https://ui.adsabs.harvard.edu/abs/2015ApJS..216...29B}{216, 29}

\bibitem[{{Brassard} {et~al.}(2001){Brassard}, {Fontaine}, {Bill{\`e}res},
  {Charpinet}, {Liebert}, \& {Saffer}}]{2001ApJ...563.1013B}
{Brassard}, P., {Fontaine}, G., {Bill{\`e}res}, M., {et~al.} 2001,
  \href{http://dx.doi.org/10.1086/323959}{\color{magenta}\apj},
  \href{https://ui.adsabs.harvard.edu/abs/2001ApJ...563.1013B}{563, 1013}

\bibitem[{{Bruntt} {et~al.}(2010){Bruntt}, {Bedding}, {Quirion}, {Lo Curto},
  {Carrier}, {Smalley}, {Dall}, {Arentoft}, {Bazot}, \& {Butler}}]{Bruntt2010}
{Bruntt}, H., {Bedding}, T.~R., {Quirion}, P.~O., {et~al.} 2010,
  \href{http://dx.doi.org/10.1111/j.1365-2966.2010.16575.x}{\color{magenta}\mnras},
  \href{https://ui.adsabs.harvard.edu/abs/2010MNRAS.405.1907B}{405, 1907}

\bibitem[{{Chen} \& {Han}(2002)}]{Chen2002MNRAS.335..948C}
{Chen}, X. \& {Han}, Z. 2002,
  \href{http://dx.doi.org/10.1046/j.1365-8711.2002.05680.x}{\color{magenta}\mnras},
  \href{https://ui.adsabs.harvard.edu/abs/2002MNRAS.335..948C}{335, 948}

\bibitem[{{Chen} \& {Han}(2003)}]{Chen2003MNRAS.341..662C}
{Chen}, X. \& {Han}, Z. 2003,
  \href{http://dx.doi.org/10.1046/j.1365-8711.2003.06449.x}{\color{magenta}\mnras},
  \href{https://ui.adsabs.harvard.edu/abs/2003MNRAS.341..662C}{341, 662}

\bibitem[{{Chen} {et~al.}(2013){Chen}, {Han}, {Deca}, \&
  {Podsiadlowski}}]{Chen2013}
{Chen}, X., {Han}, Z., {Deca}, J., \& {Podsiadlowski}, P. 2013,
  \href{http://dx.doi.org/10.1093/mnras/stt992}{\color{magenta}\mnras},
  \href{https://ui.adsabs.harvard.edu/abs/2013MNRAS.434..186C}{434, 186}

\bibitem[{{Chiappini} {et~al.}(1997){Chiappini}, {Matteucci}, \&
  {Gratton}}]{Chiappini1997ApJ...477..765C}
{Chiappini}, C., {Matteucci}, F., \& {Gratton}, R. 1997,
  \href{http://dx.doi.org/10.1086/303726}{\color{magenta}\apj},
  \href{https://ui.adsabs.harvard.edu/abs/1997ApJ...477..765C}{477, 765}

\bibitem[{{Chiappini} {et~al.}(2001){Chiappini}, {Matteucci}, \&
  {Romano}}]{Chiappini2001ApJ...554.1044C}
{Chiappini}, C., {Matteucci}, F., \& {Romano}, D. 2001,
  \href{http://dx.doi.org/10.1086/321427}{\color{magenta}\apj},
  \href{https://ui.adsabs.harvard.edu/abs/2001ApJ...554.1044C}{554, 1044}

\bibitem[{{Choi} {et~al.}(2016){Choi}, {Dotter}, {Conroy}, {Cantiello},
  {Paxton}, \& {Johnson}}]{Choi2016}
{Choi}, J., {Dotter}, A., {Conroy}, C., {et~al.} 2016,
  \href{http://dx.doi.org/10.3847/0004-637X/823/2/102}{\color{magenta}\apj},
  \href{http://adsabs.harvard.edu/abs/2016ApJ...823..102C}{823, 102}

\bibitem[{{Collet} {et~al.}(2011){Collet}, {Magic}, \& {Asplund}}]{Collet2011}
{Collet}, R., {Magic}, Z., \& {Asplund}, M. 2011, in Journal of Physics
  Conference Series, Vol. 328, Journal of Physics Conference Series,
  \href{https://ui.adsabs.harvard.edu/abs/2011JPhCS.328a2003C}{012003}

\bibitem[{{Cutri} \& {et al.}(2012)}]{Cutri2012}
{Cutri}, R.~M. \& {et al.} 2012, VizieR Online Data Catalog,
  \href{http://cdsads.u-strasbg.fr/abs/2012yCat.2311....0C}{2311}

\bibitem[{{de Ruyter} {et~al.}(2006){de Ruyter}, {van Winckel}, {Maas}, {Lloyd
  Evans}, {Waters}, \& {Dejonghe}}]{deRuyter2006}
{de Ruyter}, S., {van Winckel}, H., {Maas}, T., {et~al.} 2006,
  \href{http://dx.doi.org/10.1051/0004-6361:20054062}{\color{magenta}\aap},
  \href{https://ui.adsabs.harvard.edu/abs/2006A&A...448..641D}{448, 641}

\bibitem[{{De Smedt} {et~al.}(2016){De Smedt}, {Van Winckel}, {Kamath},
  {Siess}, {Goriely}, {Karakas}, \& {Manick}}]{DeSmedt2016}
{De Smedt}, K., {Van Winckel}, H., {Kamath}, D., {et~al.} 2016,
  \href{http://dx.doi.org/10.1051/0004-6361/201527430}{\color{magenta}\aap},
  \href{https://ui.adsabs.harvard.edu/abs/2016A&A...587A...6D}{587, A6}

\bibitem[{{Deca} {et~al.}(2012){Deca}, {Marsh}, {{\O}stensen}, {Morales-Rueda},
  {Copperwheat}, {Wade}, {Stark}, {Maxted}, {Nelemans}, \& {Heber}}]{Deca2012}
{Deca}, J., {Marsh}, T.~R., {{\O}stensen}, R.~H., {et~al.} 2012,
  \href{http://dx.doi.org/10.1111/j.1365-2966.2012.20483.x}{\color{magenta}\mnras},
  \href{https://ui.adsabs.harvard.edu/abs/2012MNRAS.421.2798D}{421, 2798}

\bibitem[{{Delgado Mena} {et~al.}(2017){Delgado Mena}, {Tsantaki}, {Adibekyan},
  {Sousa}, {Santos}, {Gonz{\'a}lez Hern{\'a}ndez}, \&
  {Israelian}}]{Delgado2017}
{Delgado Mena}, E., {Tsantaki}, M., {Adibekyan}, V.~Z., {et~al.} 2017,
  \href{http://dx.doi.org/10.1051/0004-6361/201730535}{\color{magenta}\aap},
  \href{https://ui.adsabs.harvard.edu/abs/2017A&A...606A..94D}{606, A94}

\bibitem[{{Dervi{\textcommabelow s}o{\v{g}}lu}
  {et~al.}(2018){Dervi{\textcommabelow s}o{\v{g}}lu}, {Pavlovski}, {Lehmann},
  {Southworth}, \& {Bewsher}}]{Dervisoglu2018}
{Dervi{\textcommabelow s}o{\v{g}}lu}, A., {Pavlovski}, K., {Lehmann}, H.,
  {Southworth}, J., \& {Bewsher}, D. 2018,
  \href{http://dx.doi.org/10.1093/mnras/sty2684}{\color{magenta}\mnras},
  \href{https://ui.adsabs.harvard.edu/abs/2018MNRAS.481.5660D}{481, 5660}

\bibitem[{{Dorman} {et~al.}(1993){Dorman}, {Rood}, \& {O'Connell}}]{dorman93}
{Dorman}, B., {Rood}, R.~T., \& {O'Connell}, R.~W. 1993,
  \href{http://dx.doi.org/10.1086/173511}{\color{magenta}\apj},
  \href{https://ui.adsabs.harvard.edu/abs/1993ApJ...419..596D}{419, 596}

\bibitem[{{Dotter}(2016)}]{Dotter2016}
{Dotter}, A. 2016,
  \href{http://dx.doi.org/10.3847/0067-0049/222/1/8}{\color{magenta}\apjs},
  \href{http://adsabs.harvard.edu/abs/2016ApJS..222....8D}{222, 8}

\bibitem[{{Doyle} {et~al.}(2014){Doyle}, {Davies}, {Smalley}, {Chaplin}, \&
  {Elsworth}}]{Doyle2014}
{Doyle}, A.~P., {Davies}, G.~R., {Smalley}, B., {Chaplin}, W.~J., \&
  {Elsworth}, Y. 2014,
  \href{http://dx.doi.org/10.1093/mnras/stu1692}{\color{magenta}\mnras},
  \href{https://ui.adsabs.harvard.edu/abs/2014MNRAS.444.3592D}{444, 3592}

\bibitem[{{Driebe} {et~al.}(1998){Driebe}, {Schoenberner}, {Bloecker}, \&
  {Herwig}}]{1998A&A...339..123D}
{Driebe}, T., {Schoenberner}, D., {Bloecker}, T., \& {Herwig}, F. 1998, \aap,
  \href{https://ui.adsabs.harvard.edu/abs/1998A&A...339..123D}{339, 123}

\bibitem[{{Duong} {et~al.}(2018){Duong}, {Freeman}, {Asplund}, {Casagrande},
  {Buder}, {Lind}, {Ness}, {Bland-Hawthorn}, {De Silva}, {D'Orazi}, {Kos},
  {Lewis}, {Lin}, {Martell}, {Schlesinger}, {Sharma}, {Simpson}, {Zucker},
  {Zwitter}, {Anguiano}, {Da Costa}, {Hyde}, {Horner}, {Kafle}, {Nataf},
  {Reid}, {Stello}, {Ting}, \& {Wyse}}]{Duong2018}
{Duong}, L., {Freeman}, K.~C., {Asplund}, M., {et~al.} 2018,
  \href{http://dx.doi.org/10.1093/mnras/sty525}{\color{magenta}\mnras},
  \href{https://ui.adsabs.harvard.edu/abs/2018MNRAS.476.5216D}{476, 5216}

\bibitem[{{Edvardsson} {et~al.}(1993){Edvardsson}, {Andersen}, {Gustafsson},
  {Lambert}, {Nissen}, \& {Tomkin}}]{Edvardsson1993A&A...275..101E}
{Edvardsson}, B., {Andersen}, J., {Gustafsson}, B., {et~al.} 1993, \aap,
  \href{https://ui.adsabs.harvard.edu/abs/1993A&A...275..101E}{500, 391}

\bibitem[{{Escorza} {et~al.}(2019){Escorza}, {Karinkuzhi}, {Jorissen}, {Siess},
  {Van Winckel}, {Pourbaix}, {Johnston}, {Miszalski}, {Oomen}, {Abdul-Masih},
  {Boffin}, {North}, {Manick}, {Shetye}, \& {Miko{\l}ajewska}}]{Escorza2019}
{Escorza}, A., {Karinkuzhi}, D., {Jorissen}, A., {et~al.} 2019,
  \href{http://dx.doi.org/10.1051/0004-6361/201935390}{\color{magenta}\aap},
  \href{https://ui.adsabs.harvard.edu/abs/2019A&A...626A.128E}{626, A128}

\bibitem[{{Evans} {et~al.}(2018){Evans}, {Riello}, {De Angeli}, {Carrasco},
  {Montegriffo}, {Fabricius}, {Jordi}, {Palaversa}, {Diener}, {Busso},
  {Cacciari}, {van Leeuwen}, {Burgess}, {Davidson}, {Harrison}, {Hodgkin},
  {Pancino}, {Richards}, {Altavilla}, {Balaguer-N{\'u}{\~n}ez}, {Barstow},
  {Bellazzini}, {Brown}, {Castellani}, {Cocozza}, {De Luise}, {Delgado},
  {Ducourant}, {Galleti}, {Gilmore}, {Giuffrida}, {Holl}, {Kewley}, {Koposov},
  {Marinoni}, {Marrese}, {Osborne}, {Piersimoni}, {Portell}, {Pulone},
  {Ragaini}, {Sanna}, {Terrett}, {Walton}, {Wevers}, \&
  {Wyrzykowski}}]{Evans2018}
{Evans}, D.~W., {Riello}, M., {De Angeli}, F., {et~al.} 2018,
  \href{http://dx.doi.org/10.1051/0004-6361/201832756}{\color{magenta}\aap},
  \href{https://ui.adsabs.harvard.edu/abs/2018A%26A...616A...4E}{616, A4}

\bibitem[{{Frost} \& {Lattanzio}(1996)}]{Frost1996}
{Frost}, C. \& {Lattanzio}, J. 1996,
  \href{https://ui.adsabs.harvard.edu/abs/1996astro.ph..1017F}{arXiv e-prints,
  astro}

\bibitem[{{Fulbright} \& {Johnson}(2003)}]{Fulbright2003}
{Fulbright}, J.~P. \& {Johnson}, J.~A. 2003,
  \href{http://dx.doi.org/10.1086/377443}{\color{magenta}\apj},
  \href{https://ui.adsabs.harvard.edu/abs/2003ApJ...595.1154F}{595, 1154}

\bibitem[{{Gaia Collaboration} {et~al.}(2018){Gaia Collaboration}, {Brown},
  {Vallenari}, {Prusti}, {de Bruijne}, {Babusiaux}, {Bailer-Jones}, {Biermann},
  {Evans}, {Eyer}, \& et~al.}]{GaiaDR2}
{Gaia Collaboration}, {Brown}, A.~G.~A., {Vallenari}, A., {et~al.} 2018,
  \href{http://dx.doi.org/10.1051/0004-6361/201833051}{\color{magenta}\aap},
  \href{https://ui.adsabs.harvard.edu/abs/2018A%26A...616A...1G}{616, A1}

\bibitem[{{Gorlova} {et~al.}(2013){Gorlova}, {Van Winckel}, {Vos},
  {{\O}stensen}, {Jorissen}, {Van Eck}, \& {Ikonnikova}}]{Gorlova2013}
{Gorlova}, N., {Van Winckel}, H., {Vos}, J., {et~al.} 2013, in EAS Publications
  Series, ed. K.~{Pavlovski}, A.~{Tkachenko}, \& G.~{Torres}, Vol.~64,
  \href{https://ui.adsabs.harvard.edu/abs/2013EAS....64..163G}{163--170}

\bibitem[{{Graham}(1970)}]{Graham1970}
{Graham}, J.~A. 1970,
  \href{http://dx.doi.org/10.1086/129033}{\color{magenta}\pasp},
  \href{https://ui.adsabs.harvard.edu/abs/1970PASP...82.1305G}{82, 1305}

\bibitem[{{Gratton} {et~al.}(2000){Gratton}, {Sneden}, {Carretta}, \&
  {Bragaglia}}]{Gratton2000}
{Gratton}, R.~G., {Sneden}, C., {Carretta}, E., \& {Bragaglia}, A. 2000, \aap,
  \href{https://ui.adsabs.harvard.edu/abs/2000A&A...354..169G}{354, 169}

\bibitem[{{Green} {et~al.}(2001){Green}, {Liebert}, \& {Saffer}}]{Green2001}
{Green}, E.~M., {Liebert}, J., \& {Saffer}, R.~A. 2001, in Astronomical Society
  of the Pacific Conference Series, Vol. 226, 12th European Workshop on White
  Dwarfs, ed. J.~L. {Provencal}, H.~L. {Shipman}, J.~{MacDonald}, \&
  S.~{Goodchild},
  \href{https://ui.adsabs.harvard.edu/abs/2001ASPC..226..192G}{192}

\bibitem[{{Han} {et~al.}(2003){Han}, {Podsiadlowski}, {Maxted}, \&
  {Marsh}}]{Han2003}
{Han}, Z., {Podsiadlowski}, P., {Maxted}, P.~F.~L., \& {Marsh}, T.~R. 2003,
  \href{http://dx.doi.org/10.1046/j.1365-8711.2003.06451.x}{\color{magenta}\mnras},
  \href{https://ui.adsabs.harvard.edu/abs/2003MNRAS.341..669H}{341, 669}

\bibitem[{{Han} {et~al.}(2002){Han}, {Podsiadlowski}, {Maxted}, {Marsh}, \&
  {Ivanova}}]{Han2002}
{Han}, Z., {Podsiadlowski}, P., {Maxted}, P.~F.~L., {Marsh}, T.~R., \&
  {Ivanova}, N. 2002,
  \href{http://dx.doi.org/10.1046/j.1365-8711.2002.05752.x}{\color{magenta}\mnras},
  \href{https://ui.adsabs.harvard.edu/abs/2002MNRAS.336..449H}{336, 449}

\bibitem[{{Han} {et~al.}(2000){Han}, {Tout}, \&
  {Eggleton}}]{2000MNRAS.319..215H}
{Han}, Z., {Tout}, C.~A., \& {Eggleton}, P.~P. 2000,
  \href{http://dx.doi.org/10.1046/j.1365-8711.2000.03839.x}{\color{magenta}\mnras},
  \href{https://ui.adsabs.harvard.edu/abs/2000MNRAS.319..215H}{319, 215}

\bibitem[{{Hardy} {et~al.}(2016){Hardy}, {Schreiber}, {Parsons}, {Caceres},
  {Brinkworth}, {Veras}, {G{\"a}nsicke}, {Marsh}, \& {Cieza}}]{Hardy2016}
{Hardy}, A., {Schreiber}, M.~R., {Parsons}, S.~G., {et~al.} 2016,
  \href{http://dx.doi.org/10.1093/mnras/stw976}{\color{magenta}\mnras},
  \href{https://ui.adsabs.harvard.edu/abs/2016MNRAS.459.4518H}{459, 4518}

\bibitem[{{Heber}(2009)}]{Heber2009}
{Heber}, U. 2009,
  \href{http://dx.doi.org/10.1146/annurev-astro-082708-101836}{\color{magenta}\araa},
  \href{https://ui.adsabs.harvard.edu/abs/2009ARA&A..47..211H}{47, 211}

\bibitem[{{Heber}(2016)}]{Heber2016}
{Heber}, U. 2016,
  \href{http://dx.doi.org/10.1088/1538-3873/128/966/082001}{\color{magenta}\pasp},
  \href{https://ui.adsabs.harvard.edu/abs/2016PASP..128h2001H}{128, 082001}

\bibitem[{{Henden} {et~al.}(2015){Henden}, {Levine}, {Terrell}, \&
  {Welch}}]{Henden2015}
{Henden}, A.~A., {Levine}, S., {Terrell}, D., \& {Welch}, D.~L. 2015, in
  American Astronomical Society Meeting Abstracts, Vol. 225, American
  Astronomical Society Meeting Abstracts \#225,
  \href{https://ui.adsabs.harvard.edu/abs/2015AAS...22533616H}{336.16}

\bibitem[{{Hilditch}(2001)}]{Hilditch2001}
{Hilditch}, R.~W. 2001, {An Introduction to Close Binary Stars}

\bibitem[{{Hillen} {et~al.}(2014){Hillen}, {Menu}, {Van Winckel}, {Min},
  {Gielen}, {Wevers}, {Mulders}, {Regibo}, \& {Verhoelst}}]{Hillen2014}
{Hillen}, M., {Menu}, J., {Van Winckel}, H., {et~al.} 2014,
  \href{http://dx.doi.org/10.1051/0004-6361/201423749}{\color{magenta}\aap},
  \href{https://ui.adsabs.harvard.edu/abs/2014A&A...568A..12H}{568, A12}

\bibitem[{{Hubeny} \& {Lanz}(2017)}]{Hubeny17}
{Hubeny}, I. \& {Lanz}, T. 2017,
  \href{https://ui.adsabs.harvard.edu/abs/2017arXiv170601937H}{arXiv e-prints,
  arXiv:1706.01937}

\bibitem[{{Israelian} {et~al.}(2004){Israelian}, {Ecuvillon}, {Rebolo},
  {Garc{\'\i}a-L{\'o}pez}, {Bonifacio}, \& {Molaro}}]{Israelian2004}
{Israelian}, G., {Ecuvillon}, A., {Rebolo}, R., {et~al.} 2004,
  \href{http://dx.doi.org/10.1051/0004-6361:20047132}{\color{magenta}\aap},
  \href{https://ui.adsabs.harvard.edu/abs/2004A&A...421..649I}{421, 649}

\bibitem[{{Istrate} {et~al.}(2016){Istrate}, {Marchant}, {Tauris}, {Langer},
  {Stancliffe}, \& {Grassitelli}}]{Istrate2016}
{Istrate}, A.~G., {Marchant}, P., {Tauris}, T.~M., {et~al.} 2016,
  \href{http://dx.doi.org/10.1051/0004-6361/201628874}{\color{magenta}\aap},
  \href{https://ui.adsabs.harvard.edu/abs/2016A&A...595A..35I}{595, A35}

\bibitem[{{Kamath}(2015)}]{Kamath2015}
{Kamath}, D. 2015, in EAS Publications Series, Vol. 71-72, EAS Publications
  Series,
  \href{https://ui.adsabs.harvard.edu/abs/2015EAS....71..129K}{129--134}

\bibitem[{{Karakas} \& {Lattanzio}(2014)}]{Karakas2014}
{Karakas}, A.~I. \& {Lattanzio}, J.~C. 2014,
  \href{http://dx.doi.org/10.1017/pasa.2014.21}{\color{magenta}\pasa},
  \href{https://ui.adsabs.harvard.edu/abs/2014PASA...31...30K}{31, e030}

\bibitem[{{Kippenhahn} {et~al.}(1980){Kippenhahn}, {Ruschenplatt}, \&
  {Thomas}}]{Kippenhahn1980}
{Kippenhahn}, R., {Ruschenplatt}, G., \& {Thomas}, H.~C. 1980, \aap,
  \href{https://ui.adsabs.harvard.edu/abs/1980A&A....91..175K}{91, 175}

\bibitem[{{Kurucz}(1979)}]{Kurucz1979}
{Kurucz}, R.~L. 1979,
  \href{http://dx.doi.org/10.1086/190589}{\color{magenta}\apjs},
  \href{http://cdsads.u-strasbg.fr/abs/1979ApJS...40....1K}{40, 1}

\bibitem[{{Lallement} {et~al.}(2019){Lallement}, {Babusiaux}, {Vergely},
  {Katz}, {Arenou}, {Valette}, {Hottier}, \& {Capitanio}}]{Lallement2019}
{Lallement}, R., {Babusiaux}, C., {Vergely}, J.~L., {et~al.} 2019,
  \href{http://dx.doi.org/10.1051/0004-6361/201834695}{\color{magenta}\aap},
  \href{https://ui.adsabs.harvard.edu/abs/2019A&A...625A.135L}{625, A135}

\bibitem[{{Lanz} {et~al.}(2004){Lanz}, {Brown}, {Sweigart}, {Hubeny}, \&
  {Landsman}}]{Lanz2004}
{Lanz}, T., {Brown}, T.~M., {Sweigart}, A.~V., {Hubeny}, I., \& {Landsman},
  W.~B. 2004, \href{http://dx.doi.org/10.1086/380904}{\color{magenta}\apj},
  \href{https://ui.adsabs.harvard.edu/abs/2004ApJ...602..342L}{602, 342}

\bibitem[{{Lindegren} {et~al.}(2018){Lindegren}, {Hern{\'a}ndez}, {Bombrun},
  {Klioner}, {Bastian}, {Ramos-Lerate}, {de Torres}, {Steidelm{\"u}ller},
  {Stephenson}, {Hobbs}, {Lammers}, {Biermann}, {Geyer}, {Hilger}, {Michalik},
  {Stampa}, {McMillan}, {Casta{\~n}eda}, {Clotet}, {Comoretto}, {Davidson},
  {Fabricius}, {Gracia}, {Hambly}, {Hutton}, {Mora}, {Portell}, {van Leeuwen},
  {Abbas}, {Abreu}, {Altmann}, {Andrei}, {Anglada}, {Balaguer-N{\'u}{\~n}ez},
  {Barache}, {Becciani}, {Bertone}, {Bianchi}, {Bouquillon}, {Bourda},
  {Br{\"u}semeister}, {Bucciarelli}, {Busonero}, {Buzzi}, {Cancelliere},
  {Carlucci}, {Charlot}, {Cheek}, {Crosta}, {Crowley}, {de Bruijne}, {de
  Felice}, {Drimmel}, {Esquej}, {Fienga}, {Fraile}, {Gai}, {Garralda},
  {Gonz{\'a}lez-Vidal}, {Guerra}, {Hauser}, {Hofmann}, {Holl}, {Jordan},
  {Lattanzi}, {Lenhardt}, {Liao}, {Licata}, {Lister}, {L{\"o}ffler},
  {Marchant}, {Martin-Fleitas}, {Messineo}, {Mignard}, {Morbidelli}, {Poggio},
  {Riva}, {Rowell}, {Salguero}, {Sarasso}, {Sciacca}, {Siddiqui}, {Smart},
  {Spagna}, {Steele}, {Taris}, {Torra}, {van Elteren}, {van Reeven}, \&
  {Vecchiato}}]{Lindegren2018}
{Lindegren}, L., {Hern{\'a}ndez}, J., {Bombrun}, A., {et~al.} 2018,
  \href{http://dx.doi.org/10.1051/0004-6361/201832727}{\color{magenta}\aap},
  \href{https://ui.adsabs.harvard.edu/abs/2018A%26A...616A...2L}{616, A2}

\bibitem[{{Luri} {et~al.}(2018){Luri}, {Brown}, {Sarro}, {Arenou},
  {Bailer-Jones}, {Castro-Ginard}, {de Bruijne}, {Prusti}, {Babusiaux}, \&
  {Delgado}}]{Luri2018}
{Luri}, X., {Brown}, A.~G.~A., {Sarro}, L.~M., {et~al.} 2018,
  \href{http://dx.doi.org/10.1051/0004-6361/201832964}{\color{magenta}\aap},
  \href{https://ui.adsabs.harvard.edu/abs/2018A%26A...616A...9L}{616, A9}

\bibitem[{{Maas} {et~al.}(2005){Maas}, {Van Winckel}, \& {Lloyd
  Evans}}]{Maas2005}
{Maas}, T., {Van Winckel}, H., \& {Lloyd Evans}, T. 2005,
  \href{http://dx.doi.org/10.1051/0004-6361:20041688}{\color{magenta}\aap},
  \href{https://ui.adsabs.harvard.edu/abs/2005A&A...429..297M}{429, 297}

\bibitem[{{Magic} {et~al.}(2013){Magic}, {Collet}, {Asplund}, {Trampedach},
  {Hayek}, {Chiavassa}, {Stein}, \& {Nordlund}}]{Magic2013}
{Magic}, Z., {Collet}, R., {Asplund}, M., {et~al.} 2013,
  \href{http://dx.doi.org/10.1051/0004-6361/201321274}{\color{magenta}\aap},
  \href{https://ui.adsabs.harvard.edu/abs/2013A&A...557A..26M}{557, A26}

\bibitem[{{Matteucci} \& {Francois}(1989)}]{Matteucci1989MNRAS.239..885M}
{Matteucci}, F. \& {Francois}, P. 1989,
  \href{http://dx.doi.org/10.1093/mnras/239.3.885}{\color{magenta}\mnras},
  \href{https://ui.adsabs.harvard.edu/abs/1989MNRAS.239..885M}{239, 885}

\bibitem[{{Maxted} {et~al.}(2015){Maxted}, {Serenelli}, \&
  {Southworth}}]{Maxted2015}
{Maxted}, P.~F.~L., {Serenelli}, A.~M., \& {Southworth}, J. 2015,
  \href{http://dx.doi.org/10.1051/0004-6361/201425331}{\color{magenta}\aap},
  \href{http://adsabs.harvard.edu/abs/2015A%26A...575A..36M}{575, A36}

\bibitem[{{Mengel} {et~al.}(1975){Mengel}, {Norris}, \&
  {Gross}}]{1975BAAS....7Q.256M}
{Mengel}, J.~G., {Norris}, J., \& {Gross}, P.~G. 1975, in \baas, Vol.~7,
  \href{https://ui.adsabs.harvard.edu/abs/1975BAAS....7Q.256M}{256}

\bibitem[{{Meyer} {et~al.}(2008){Meyer}, {Nittler}, {Nguyen}, \&
  {Messenger}}]{Meyer2008}
{Meyer}, B.~S., {Nittler}, L.~R., {Nguyen}, A.~N., \& {Messenger}, S. 2008,
  \href{http://dx.doi.org/10.2138/rmg.2008.68.4}{\color{magenta}Reviews in
  Mineralogy and Geochemistry},
  \href{https://ui.adsabs.harvard.edu/abs/2008RvMG...68...31M}{68, 31}

\bibitem[{{Min} {et~al.}(2009){Min}, {Dullemond}, {Dominik}, {de Koter}, \&
  {Hovenier}}]{Min2009}
{Min}, M., {Dullemond}, C.~P., {Dominik}, C., {de Koter}, A., \& {Hovenier},
  J.~W. 2009,
  \href{http://dx.doi.org/10.1051/0004-6361/200811470}{\color{magenta}\aap},
  \href{https://ui.adsabs.harvard.edu/abs/2009A&A...497..155M}{497, 155}

\bibitem[{{Moc{\'a}k} {et~al.}(2010){Moc{\'a}k}, {Campbell}, {M{\"u}ller}, \&
  {Kifonidis}}]{Mocak2010A&A...520A.114M}
{Moc{\'a}k}, M., {Campbell}, S.~W., {M{\"u}ller}, E., \& {Kifonidis}, K. 2010,
  \href{http://dx.doi.org/10.1051/0004-6361/201014461}{\color{magenta}\aap},
  \href{https://ui.adsabs.harvard.edu/abs/2010A&A...520A.114M}{520, A114}

\bibitem[{{Murphy} {et~al.}(2018){Murphy}, {Moe}, {Kurtz}, {Bedding},
  {Shibahashi}, \& {Boffin}}]{Murphy2018}
{Murphy}, S.~J., {Moe}, M., {Kurtz}, D.~W., {et~al.} 2018,
  \href{http://dx.doi.org/10.1093/mnras/stx3049}{\color{magenta}\mnras},
  \href{https://ui.adsabs.harvard.edu/abs/2018MNRAS.474.4322M}{474, 4322}

\bibitem[{{Nemeth}(2019)}]{Nemeth2019}
{Nemeth}, P. 2019, in Astronomical Society of the Pacific Conference Series,
  Vol. 519, Astronomical Society of the Pacific Conference Series, ed.
  K.~{Werner}, C.~{Stehle}, T.~{Rauch}, \& T.~{Lanz},
  \href{https://ui.adsabs.harvard.edu/abs/2019ASPC..519..117N}{117}

\bibitem[{{N{\'e}meth} {et~al.}(2012){N{\'e}meth}, {Kawka}, \&
  {Vennes}}]{Nemeth12}
{N{\'e}meth}, P., {Kawka}, A., \& {Vennes}, S. 2012,
  \href{http://dx.doi.org/10.1111/j.1365-2966.2012.22009.x}{\color{magenta}\mnras},
  \href{https://ui.adsabs.harvard.edu/abs/2012MNRAS.427.2180N}{427, 2180}

\bibitem[{{Nissen} {et~al.}(2007){Nissen}, {Akerman}, {Asplund}, {Fabbian},
  {Kerber}, {Kaufl}, \& {Pettini}}]{Nissen2007A&A...469..319N}
{Nissen}, P.~E., {Akerman}, C., {Asplund}, M., {et~al.} 2007,
  \href{http://dx.doi.org/10.1051/0004-6361:20077344}{\color{magenta}\aap},
  \href{https://ui.adsabs.harvard.edu/abs/2007A&A...469..319N}{469, 319}

\bibitem[{{Nissen} {et~al.}(2014){Nissen}, {Chen}, {Carigi}, {Schuster}, \&
  {Zhao}}]{Nissen2014A&A...568A..25N}
{Nissen}, P.~E., {Chen}, Y.~Q., {Carigi}, L., {Schuster}, W.~J., \& {Zhao}, G.
  2014,
  \href{http://dx.doi.org/10.1051/0004-6361/201424184}{\color{magenta}\aap},
  \href{https://ui.adsabs.harvard.edu/abs/2014A&A...568A..25N}{568, A25}

\bibitem[{{Oomen} {et~al.}(2020){Oomen}, {Pols}, {Van Winckel}, \&
  {Nelemans}}]{Oomen2020}
{Oomen}, G.-M., {Pols}, O., {Van Winckel}, H., \& {Nelemans}, G. 2020,
  \href{http://dx.doi.org/10.1051/0004-6361/202038341}{\color{magenta}\aap},
  \href{https://ui.adsabs.harvard.edu/abs/2020A&A...642A.234O}{642, A234}

\bibitem[{{{\O}stensen} \& {Van Winckel}(2012)}]{Ostensen2012}
{{\O}stensen}, R.~H. \& {Van Winckel}, H. 2012, in Astronomical Society of the
  Pacific Conference Series, Vol. 452, Fifth Meeting on Hot Subdwarf Stars and
  Related Objects, ed. D.~{Kilkenny}, C.~S. {Jeffery}, \& C.~{Koen},
  \href{https://ui.adsabs.harvard.edu/abs/2012ASPC..452..163O}{163}

\bibitem[{{O'Toole}(2008)}]{Otoole2008ASPC..392...67O}
{O'Toole}, S.~J. 2008, in Astronomical Society of the Pacific Conference
  Series, Vol. 392, Hot Subdwarf Stars and Related Objects, ed. U.~{Heber},
  C.~S. {Jeffery}, \& R.~{Napiwotzki},
  \href{https://ui.adsabs.harvard.edu/abs/2008ASPC..392...67O}{67}

\bibitem[{{Paczynski}(1976)}]{1976IAUS...73...75P}
{Paczynski}, B. 1976, in IAU Symposium, Vol.~73, Structure and Evolution of
  Close Binary Systems, ed. P.~{Eggleton}, S.~{Mitton}, \& J.~{Whelan},
  \href{https://ui.adsabs.harvard.edu/abs/1976IAUS...73...75P}{75}

\bibitem[{{Pauli} {et~al.}(2006){Pauli}, {Napiwotzki}, {Heber}, {Altmann}, \&
  {Odenkirchen}}]{Pauli2006A&A...447..173P}
{Pauli}, E.~M., {Napiwotzki}, R., {Heber}, U., {Altmann}, M., \& {Odenkirchen},
  M. 2006,
  \href{http://dx.doi.org/10.1051/0004-6361:20052730}{\color{magenta}\aap},
  \href{https://ui.adsabs.harvard.edu/abs/2006A&A...447..173P}{447, 173}

\bibitem[{{Paunzen}(2015)}]{Paunzen2015}
{Paunzen}, E. 2015,
  \href{http://dx.doi.org/10.1051/0004-6361/201526413}{\color{magenta}\aap},
  \href{https://ui.adsabs.harvard.edu/abs/2015A&A...580A..23P}{580, A23}

\bibitem[{{Pelisoli} {et~al.}(2020){Pelisoli}, {Vos}, {Geier}, {Schaffenroth},
  \& {Baran}}]{Pelisoli2020A&A...642A.180P}
{Pelisoli}, I., {Vos}, J., {Geier}, S., {Schaffenroth}, V., \& {Baran}, A.~S.
  2020,
  \href{http://dx.doi.org/10.1051/0004-6361/202038473}{\color{magenta}\aap},
  \href{https://ui.adsabs.harvard.edu/abs/2020A&A...642A.180P}{642, A180}

\bibitem[{{Popham} \& {Narayan}(1991)}]{Popham1991ApJ...370..604P}
{Popham}, R. \& {Narayan}, R. 1991,
  \href{http://dx.doi.org/10.1086/169847}{\color{magenta}\apj},
  \href{https://ui.adsabs.harvard.edu/abs/1991ApJ...370..604P}{370, 604}

\bibitem[{{Raghavan} {et~al.}(2010){Raghavan}, {McAlister}, {Henry}, {Latham},
  {Marcy}, {Mason}, {Gies}, {White}, \& {ten Brummelaar}}]{2010ApJS..190....1R}
{Raghavan}, D., {McAlister}, H.~A., {Henry}, T.~J., {et~al.} 2010,
  \href{http://dx.doi.org/10.1088/0067-0049/190/1/1}{\color{magenta}\apjs},
  \href{https://ui.adsabs.harvard.edu/abs/2010ApJS..190....1R}{190, 1}

\bibitem[{{Ramstedt} \& {Olofsson}(2014)}]{Ramstedt2014}
{Ramstedt}, S. \& {Olofsson}, H. 2014,
  \href{http://dx.doi.org/10.1051/0004-6361/201423721}{\color{magenta}\aap},
  \href{https://ui.adsabs.harvard.edu/abs/2014A&A...566A.145R}{566, A145}

\bibitem[{{Raskin} {et~al.}(2011){Raskin}, {van Winckel}, {Hensberge},
  {Jorissen}, {Lehmann}, {Waelkens}, {Avila}, {de Cuyper}, {Degroote},
  {Dubosson}, {Dumortier}, {Fr{\'e}mat}, {Laux}, {Michaud}, {Morren}, {Perez
  Padilla}, {Pessemier}, {Prins}, {Smolders}, {van Eck}, \&
  {Winkler}}]{Raskin2011}
{Raskin}, G., {van Winckel}, H., {Hensberge}, H., {et~al.} 2011,
  \href{http://dx.doi.org/10.1051/0004-6361/201015435}{\color{magenta}\aap},
  \href{https://ui.adsabs.harvard.edu/abs/2011A&A...526A..69R}{526, A69}

\bibitem[{{Reinhold} \& {Gizon}(2015)}]{Reinhold2015A&A...583A..65R}
{Reinhold}, T. \& {Gizon}, L. 2015,
  \href{http://dx.doi.org/10.1051/0004-6361/201526216}{\color{magenta}\aap},
  \href{https://ui.adsabs.harvard.edu/abs/2015A&A...583A..65R}{583, A65}

\bibitem[{{Riello} {et~al.}(2018){Riello}, {De Angeli}, {Evans}, {Busso},
  {Hambly}, {Davidson}, {Burgess}, {Montegriffo}, {Osborne}, {Kewley},
  {Carrasco}, {Fabricius}, {Jordi}, {Cacciari}, {van Leeuwen}, \&
  {Holland}}]{Riello2018}
{Riello}, M., {De Angeli}, F., {Evans}, D.~W., {et~al.} 2018,
  \href{http://dx.doi.org/10.1051/0004-6361/201832712}{\color{magenta}\aap},
  \href{https://ui.adsabs.harvard.edu/abs/2018A%26A...616A...3R}{616, A3}

\bibitem[{{Romano} {et~al.}(2020){Romano}, {Franchini}, {Grisoni}, {Spitoni},
  {Matteucci}, \& {Morossi}}]{Romano2020A&A...639A..37R}
{Romano}, D., {Franchini}, M., {Grisoni}, V., {et~al.} 2020,
  \href{http://dx.doi.org/10.1051/0004-6361/202037972}{\color{magenta}\aap},
  \href{https://ui.adsabs.harvard.edu/abs/2020A&A...639A..37R}{639, A37}

\bibitem[{{Romano} {et~al.}(2000){Romano}, {Matteucci}, {Salucci}, \&
  {Chiappini}}]{Romano2000ApJ...539..235R}
{Romano}, D., {Matteucci}, F., {Salucci}, P., \& {Chiappini}, C. 2000,
  \href{http://dx.doi.org/10.1086/309223}{\color{magenta}\apj},
  \href{https://ui.adsabs.harvard.edu/abs/2000ApJ...539..235R}{539, 235}

\bibitem[{{Saffer} {et~al.}(1994){Saffer}, {Bergeron}, {Koester}, \&
  {Liebert}}]{1994ApJ...432..351S}
{Saffer}, R.~A., {Bergeron}, P., {Koester}, D., \& {Liebert}, J. 1994,
  \href{http://dx.doi.org/10.1086/174573}{\color{magenta}\apj},
  \href{https://ui.adsabs.harvard.edu/abs/1994ApJ...432..351S}{432, 351}

\bibitem[{{Sarna} \& {De Greve}(1996)}]{Sarna1996}
{Sarna}, M.~J. \& {De Greve}, J.~P. 1996, \qjras,
  \href{https://ui.adsabs.harvard.edu/abs/1996QJRAS..37...11S}{37, 11}

\bibitem[{Schwarz(1978)}]{schwarz1978}
Schwarz, G. 1978, The Annals of Statistics, 6, 6

\bibitem[{{Shulyak} {et~al.}(2004){Shulyak}, {Tsymbal}, {Ryabchikova},
  {St{\"u}tz}, \& {Weiss}}]{Shulyak2004}
{Shulyak}, D., {Tsymbal}, V., {Ryabchikova}, T., {St{\"u}tz}, C., \& {Weiss},
  W.~W. 2004,
  \href{http://dx.doi.org/10.1051/0004-6361:20034169}{\color{magenta}\aap},
  \href{https://ui.adsabs.harvard.edu/abs/2004A&A...428..993S}{428, 993}

\bibitem[{{Simon} \& {Sturm}(1994)}]{Simon94}
{Simon}, K.~P. \& {Sturm}, E. 1994, \aap,
  \href{https://ui.adsabs.harvard.edu/abs/1994A&A...281..286S}{281, 286}

\bibitem[{{Skrutskie} {et~al.}(2006){Skrutskie}, {Cutri}, {Stiening},
  {Weinberg}, {Schneider}, {Carpenter}, {Beichman}, {Capps}, {Chester},
  {Elias}, {Huchra}, {Liebert}, {Lonsdale}, {Monet}, {Price}, {Seitzer},
  {Jarrett}, {Kirkpatrick}, {Gizis}, {Howard}, {Evans}, {Fowler}, {Fullmer},
  {Hurt}, {Light}, {Kopan}, {Marsh}, {McCallon}, {Tam}, {Van Dyk}, \&
  {Wheelock}}]{Skrutskie2006}
{Skrutskie}, M.~F., {Cutri}, R.~M., {Stiening}, R., {et~al.} 2006,
  \href{http://dx.doi.org/10.1086/498708}{\color{magenta}\aj},
  \href{http://esoads.eso.org/abs/2006AJ....131.1163S}{131, 1163}

\bibitem[{{Southworth}(2015)}]{Southworth2015}
{Southworth}, J. 2015, in ASPCS, Vol. 496, Living Together: Planets, Host Stars
  and Binaries, ed. S.~M. {Rucinski}, G.~{Torres}, \& M.~{Zejda},
  \href{http://adsabs.harvard.edu/abs/2015ASPC..496..164S}{164}

\bibitem[{{Takeda} \& {Honda}(2005)}]{Takeda2005}
{Takeda}, Y. \& {Honda}, S. 2005,
  \href{http://dx.doi.org/10.1093/pasj/57.1.65}{\color{magenta}\pasj},
  \href{https://ui.adsabs.harvard.edu/abs/2005PASJ...57...65T}{57, 65}

\bibitem[{{Tkachenko}(2015)}]{Tkachenko2015}
{Tkachenko}, A. 2015,
  \href{http://dx.doi.org/10.1051/0004-6361/201526513}{\color{magenta}\aap},
  \href{https://ui.adsabs.harvard.edu/abs/2015A&A...581A.129T}{581, A129}

\bibitem[{{Travaglio} {et~al.}(2004){Travaglio}, {Gallino}, {Arnone}, {Cowan},
  {Jordan}, \& {Sneden}}]{Travaglio2004ApJ...601..864T}
{Travaglio}, C., {Gallino}, R., {Arnone}, E., {et~al.} 2004,
  \href{http://dx.doi.org/10.1086/380507}{\color{magenta}\apj},
  \href{https://ui.adsabs.harvard.edu/abs/2004ApJ...601..864T}{601, 864}

\bibitem[{{Tsymbal}(1996)}]{Tsymbal1996}
{Tsymbal}, V. 1996, Astronomical Society of the Pacific Conference Series, Vol.
  108, {STARSP: A Software System For the Analysis of the Spectra of Normal
  Stars}, ed. S.~J. {Adelman}, F.~{Kupka}, \& W.~W. {Weiss}, 198

\bibitem[{{Ulla} \& {Thejll}(1998)}]{Ulla1998A&AS..132....1U}
{Ulla}, A. \& {Thejll}, P. 1998,
  \href{http://dx.doi.org/10.1051/aas:1998439}{\color{magenta}\aaps},
  \href{https://ui.adsabs.harvard.edu/abs/1998A&AS..132....1U}{132, 1}

\bibitem[{{Vos} {et~al.}(2020){Vos}, {Bobrick}, \&
  {Vu{\v{c}}kovi{\'c}}}]{Vos2020}
{Vos}, J., {Bobrick}, A., \& {Vu{\v{c}}kovi{\'c}}, M. 2020,
  \href{http://dx.doi.org/10.1051/0004-6361/201937195}{\color{magenta}\aap},
  \href{https://ui.adsabs.harvard.edu/abs/2020A&A...641A.163V}{641, A163}

\bibitem[{{Vos} {et~al.}(2018{\natexlab{a}}){Vos}, {N{\'e}meth},
  {Vu{\v{c}}kovi{\'c}}, {{\O}stensen}, \& {Parsons}}]{Vos2018a}
{Vos}, J., {N{\'e}meth}, P., {Vu{\v{c}}kovi{\'c}}, M., {{\O}stensen}, R., \&
  {Parsons}, S. 2018{\natexlab{a}},
  \href{http://dx.doi.org/10.1093/mnras/stx2198}{\color{magenta}\mnras},
  \href{https://ui.adsabs.harvard.edu/abs/2018MNRAS.473..693V}{473, 693}

\bibitem[{{Vos} {et~al.}(2012){Vos}, {{\O}stensen}, {Degroote}, {De Smedt},
  {Green}, {Heber}, {Van Winckel}, {Acke}, {Bloemen}, {De Cat}, {Exter},
  {Lampens}, {Lombaert}, {Masseron}, {Menu}, {Neyskens}, {Raskin}, {Ringat},
  {Rauch}, {Smolders}, \& {Tkachenko}}]{Vos2012}
{Vos}, J., {{\O}stensen}, R.~H., {Degroote}, P., {et~al.} 2012,
  \href{http://dx.doi.org/10.1051/0004-6361/201219723}{\color{magenta}\aap},
  \href{https://ui.adsabs.harvard.edu/abs/2012A&A...548A...6V}{548, A6}

\bibitem[{{Vos} {et~al.}(2015){Vos}, {{\O}stensen}, {Marchant}, \& {Van
  Winckel}}]{Vos2015}
{Vos}, J., {{\O}stensen}, R.~H., {Marchant}, P., \& {Van Winckel}, H. 2015,
  \href{http://dx.doi.org/10.1051/0004-6361/201526019}{\color{magenta}\aap},
  \href{https://ui.adsabs.harvard.edu/abs/2015A&A...579A..49V}{579, A49}

\bibitem[{{Vos} {et~al.}(2013){Vos}, {{\O}stensen}, {N{\'e}meth}, {Green},
  {Heber}, \& {Van Winckel}}]{Vos2013}
{Vos}, J., {{\O}stensen}, R.~H., {N{\'e}meth}, P., {et~al.} 2013,
  \href{http://dx.doi.org/10.1051/0004-6361/201322200}{\color{magenta}\aap},
  \href{https://ui.adsabs.harvard.edu/abs/2013A&A...559A..54V}{559, A54}

\bibitem[{{Vos} {et~al.}(2017){Vos}, {{\O}stensen}, {Vu{\v{c}}kovi{\'c}}, \&
  {Van Winckel}}]{Vos2017}
{Vos}, J., {{\O}stensen}, R.~H., {Vu{\v{c}}kovi{\'c}}, M., \& {Van Winckel}, H.
  2017,
  \href{http://dx.doi.org/10.1051/0004-6361/201730958}{\color{magenta}\aap},
  \href{https://ui.adsabs.harvard.edu/abs/2017A&A...605A.109V}{605, A109}

\bibitem[{{Vos} {et~al.}(2019){Vos}, {Vu{\v{c}}kovi{\'c}}, {Chen}, {Han},
  {Boudreaux}, {Barlow}, {{\O}stensen}, \& {N{\'e}meth}}]{Vos2019}
{Vos}, J., {Vu{\v{c}}kovi{\'c}}, M., {Chen}, X., {et~al.} 2019,
  \href{http://dx.doi.org/10.1093/mnras/sty3017}{\color{magenta}\mnras},
  \href{https://ui.adsabs.harvard.edu/abs/2019MNRAS.482.4592V}{482, 4592}

\bibitem[{{Vos} {et~al.}(2018{\natexlab{b}}){Vos}, {Zorotovic},
  {Vu{\v{c}}kovi{\'c}}, {Schreiber}, \& {{\O}stensen}}]{Vos2018b}
{Vos}, J., {Zorotovic}, M., {Vu{\v{c}}kovi{\'c}}, M., {Schreiber}, M.~R., \&
  {{\O}stensen}, R. 2018{\natexlab{b}},
  \href{http://dx.doi.org/10.1093/mnrasl/sly050}{\color{magenta}\mnras},
  \href{https://ui.adsabs.harvard.edu/abs/2018MNRAS.477L..40V}{477, L40}

\bibitem[{{Webbink}(1984)}]{1984ApJ...277..355W}
{Webbink}, R.~F. 1984,
  \href{http://dx.doi.org/10.1086/161701}{\color{magenta}\apj},
  \href{https://ui.adsabs.harvard.edu/abs/1984ApJ...277..355W}{277, 355}

\bibitem[{{Werner} {et~al.}(2003){Werner}, {Deetjen}, {Dreizler}, {Nagel},
  {Rauch}, \& {Schuh}}]{Werner2003}
{Werner}, K., {Deetjen}, J.~L., {Dreizler}, S., {et~al.} 2003, in ASPCS, Vol.
  288, Stellar Atmosphere Modeling, ed. I.~{Hubeny}, D.~{Mihalas}, \&
  K.~{Werner}, \href{http://adsabs.harvard.edu/abs/2003ASPC..288...31W}{31}

\bibitem[{{Woitke} {et~al.}(2016){Woitke}, {Min}, {Pinte}, {Thi}, {Kamp},
  {Rab}, {Anthonioz}, {Antonellini}, {Baldovin-Saavedra}, {Carmona}, {Dominik},
  {Dionatos}, {Greaves}, {G{\"u}del}, {Ilee}, {Liebhart}, {M{\'e}nard},
  {Rigon}, {Waters}, {Aresu}, {Meijerink}, \& {Spaans}}]{Woitke2016}
{Woitke}, P., {Min}, M., {Pinte}, C., {et~al.} 2016,
  \href{http://dx.doi.org/10.1051/0004-6361/201526538}{\color{magenta}\aap},
  \href{https://ui.adsabs.harvard.edu/abs/2016A&A...586A.103W}{586, A103}

\bibitem[{{Wyatt}(2008)}]{Wyatt2008}
{Wyatt}, M.~C. 2008,
  \href{http://dx.doi.org/10.1146/annurev.astro.45.051806.110525}{\color{magenta}\araa},
  \href{https://ui.adsabs.harvard.edu/abs/2008ARA&A..46..339W}{46, 339}

\bibitem[{{Zverko} {et~al.}(2007){Zverko}, {{\v{Z}}i{\v{z}}{\v{n}}ovsk{\'y}},
  {Mikul{\'a}{\v{s}}ek}, \& {Iliev}}]{Zverko2007}
{Zverko}, J., {{\v{Z}}i{\v{z}}{\v{n}}ovsk{\'y}}, J., {Mikul{\'a}{\v{s}}ek}, Z.,
  \& {Iliev}, I.~K. 2007, Contributions of the Astronomical Observatory
  Skalnate Pleso,
  \href{https://ui.adsabs.harvard.edu/abs/2007CoSka..37...49Z}{37, 49}

\bibitem[{{Zwicky}(1957)}]{Zwicky1957moas.book.....Z}
{Zwicky}, F. 1957, {Morphological astronomy}

\end{thebibliography}

%
%

\begin{appendix} 

\onecolumn
\section{Observational data and radial velocities}

\begin{longtable}{lccccc}
\caption{\label{A1}\BD: Observational data and RVs of MS and sdO stars.}\\
\hline
\hline
 &  & \multicolumn{2}{c}{MS} & \multicolumn{2}{c}{sdO} \\
\cline{3-6}
BJD-2455053   & Exp.time (s) &  RV (km s$^{-1}$) & error ($\pm$) &   RV (km s$^{-1}$) & error ($\pm$)\\
\hline
\noalign{\smallskip}
\endfirsthead
\caption{Continued.} \\
\hline
\endhead
\hline
\endfoot
\hline
\endlastfoot
 0.72040   & 1230  &  -4.27 &   0.09 &  15.37  &   1.86   \\
73.53177   & 1230  &   0.95 &   0.12 &   8.58  &   2.05   \\
73.54662   & 1230  &   1.00 &   0.10 &  11.13  &   2.12   \\
365.67960  & 2700  &  -0.13 &   0.08 &  11.89  &   1.49   \\
444.60462  & 2700  &  -4.96 &   0.07 &  21.14  &   1.96   \\
454.44219  & 2700  &  -5.10 &   0.08 &  24.64  &   5.42   \\
521.32650  & 2700  &  -2.92 &   0.09 &  17.23  &   1.96   \\
525.32419  & 1800  &  -2.43 &   0.08 &  19.73  &   1.83   \\
525.34560  & 1800  &  -2.44 &   0.10 &  16.91  &   2.56   \\
717.69416  & 1800  &  12.43 &   0.09 &  21.31  &   2.16   \\
726.69127  & 1800  &  12.77 &   0.07 &  15.32  &   0.80   \\
746.62395  & 1800  &  12.16 &   0.07 &  13.15  &   1.54   \\
756.66505  & 1800  &  11.81 &   0.07 &  12.42  &   2.23   \\
775.61453  & 1800  &  10.36 &   0.06 &  10.27  &   1.15   \\
784.61870  & 1800  &  9.53  &   0.07 &  -8.23  &   2.81   \\
787.51668  & 1800  &  8.85  &   0.08 &  -6.52  &   2.17   \\
790.51022  & 1800  &  8.08  &   0.08 &  -6.18  &   2.06   \\
807.57059  & 1800  &  6.61  &   0.06 &  -1.65  &   1.60   \\
818.52215  & 1800  &  5.21  &   0.06 &  -4.73  &   1.42   \\
820.54345  & 1800  &  4.83  &   0.07 &   2.88  &   1.41   \\
827.46841  & 1800  &  4.00  &   0.05 &   3.12  &   1.98   \\
830.48514  & 1800  &  3.62  &   0.06 &   6.48  &   1.29   \\
833.47654  & 1600  &  2.90  &   0.08 &   7.29  &   1.41   \\
857.37153  & 1400  &  0.43  &   0.10 &  11.47  &   2.32   \\
861.41670  & 1800  & -0.05      &   0.10 &   9.98  &   1.47   \\
885.35460  & 1800  & -2.57      &   0.08 &  12.74  &   3.06   \\
902.34084  & 1800  & -3.68      &   0.10 &  16.45  &   2.71   \\
1082.67041 & 900   &  2.36      &   0.12 &   9.26  &   1.83   \\
1082.68141 & 900   &  2.74      &   0.10 &   4.19  &   2.17   \\
1086.68673 & 2000  &  2.89      &   0.08 &   5.03  &   1.08   \\
1092.71389 & 1800  &  3.76      &   0.06 &   4.92  &   1.64   \\
1106.67753 & 1800  &  4.82      &   0.06 &   2.73  &   1.14   \\
1157.55971 & 1800  &  9.81      &   0.09 &  -8.21  &   1.15   \\
1270.33560 & 1800  & 10.31      &   0.08 &  -5.04  &   1.21   \\
1430.69455 & 1250  & -4.79      &   0.10 &  18.79  &   2.60   \\
1482.63187 & 1800  & -4.57      &   0.06 &  21.78  &   1.58   \\
1547.44805 & 1800  & -0.55      &   0.09 &  18.90  &   1.94   \\
1610.33796 & 1800  &  5.43      &   0.08 &  -0.13  &   1.52   \\
1839.65250 & 1800  &  1.64      &   0.06 &   8.46  &   2.06   \\
1851.63462 & 1200  &  0.16      &   0.10 &   9.07  &   1.26   \\
1919.49692 & 1300  & -4.65      &   0.07 &  20.59  &   1.68   \\
1984.36401 & 1800  & -4.36      &   0.11 &  24.78  &   2.57   \\
2168.63948 & 1800  & 10.92      &   0.09 & -15.38  &   1.95   \\
2289.49920 & 1800  &  7.86      &   0.12 &  -9.58  &   2.52   \\
2344.34644 & 1800  &  0.67      &   0.12 &   6.86  &   1.76   \\
2571.57778 & 1800  &  1.85      &   0.13 &   8.50  &   3.52   \\
2658.51723 & 2100  & 10.71      &   0.15 &   1.15  &   4.03   \\
2709.33454 & 1800  & 12.99      &   0.15 & -16.56  &   2.99   \\
\end{longtable}

\begin{longtable}{lccccc}
\caption{\F: Observational data and RVs of MS and sdO stars.}\\
\hline
\hline
 &  & \multicolumn{2}{c}{MS} & \multicolumn{2}{c}{sdO} \\
\cline{3-6}
BJD-2455030   & Exp.time (s) &  RV ( km s$^{-1}$) & error ($\pm$) &   RV (km s$^{-1}$) & error ($\pm$)\\
\hline
\noalign{\smallskip}
\endfirsthead
\caption{Continued.} \\
\hline
\endhead
\hline
\endfoot
\hline
\endlastfoot
 0.42431    &   2250   &   36.35        &   0.87  & 54.04       &   3.31   \\
 0.45097    &   2250   &   33.95        &   0.35  & 50.74       &   2.85   \\
 187.76108  &   900    &   38.18        &   0.58  & 51.34       &   2.47   \\
 187.77210  &   900    &   37.39        &   0.51  & **      &   **   \\
 187.78314  &   900    &   35.38        &   0.78  & 48.34       &   4.61   \\
 201.71473  &   900    &   36.74        &   0.72  & 46.34       &   4.93   \\
 201.72574  &   900    &   36.18        &   0.90  & 51.10       &   3.03   \\
 201.73677  &   900    &   36.55        &   0.61  & 48.03       &   3.04   \\
 264.54985  &   2700   &   39.19        &   0.55  & 44.19       &   4.11   \\
 311.50146  &   2700   &   40.26        &   0.23  & 38.91       &   3.37   \\
 323.40135  &   2700   &   40.43        &   0.29  & 38.30       &   2.88   \\
 343.41275  &   2700   &   41.23        &   0.32  & 37.33       &   3.54   \\
 536.74478  &   2700   &   47.44        &   0.48  & 21.56       &   0.70   \\
 586.61438  &   2100   &   47.72        &   0.67  & 24.78       &   1.38   \\
 610.61157  &   2700   &   46.90        &   1.27  & 23.07       &   1.05   \\
 623.60091  &   1780   &   45.92        &   0.80  & 27.48       &   2.00   \\
 626.58436  &   1268   &   45.98        &   1.18  & 24.42       &   1.39   \\
 627.58453  &   1619   &   45.98        &   0.74  & 29.12       &   2.47   \\
 628.57055  &   1650   &   45.65        &   0.92  & 28.40       &   2.22   \\
 630.59930  &   1400   &   45.72        &   0.79  & 26.51       &   1.67   \\
 631.63454  &   1550   &   45.40        &   0.82  & 26.02       &   1.58   \\
 692.45431  &   1800   &   44.06        &   1.26  & 29.10       &   2.49   \\
 907.77202  &   1800   &   37.89        &   0.23  & 46.02       &   4.18   \\
 912.71709  &   2400   &   38.43        &   0.29  & 46.79       &   3.53   \\
 936.65947  &   2600   &   35.75        &   0.33  & 47.15       &   3.22    \\
 956.61946  &   1800   &   39.12        &   0.68  & 49.70       &   3.94    \\
 998.57192  &   1800   &   35.40        &   0.36  & 47.33       &   3.17    \\
 1011.53217 &   1800   &   35.36        &   0.40  & 50.65       &   3.14    \\
 1024.46467 &   1800   &   35.81        &   0.38  & 49.60       &   3.66    \\
 1028.42657 &   1800   &   35.67        &   0.59  & 46.84       &   3.86    \\
 1033.41583 &   1800   &   35.24        &   0.84  & 46.89       &   4.58    \\
 1033.43725 &   1800   &   34.06        &   0.75  & 46.06       &   4.44    \\
 1034.54128 &   1800   &   34.43        &   0.43  & 49.23       &   3.00    \\
 1036.44989 &   974    &   35.77        &   0.53  & 52.09       &   2.19    \\
 1038.43905 &   1488   &   35.85        &   0.39  & 50.14       &   3.18    \\
 1108.41471 &   1800   &   34.91        &   0.40  & 55.09       &   2.91    \\
 1111.37819 &   1800   &   34.48        &   0.62  & 50.89       &   2.89    \\
 1269.78883 &   1800   &   35.95        &   0.39  & 53.63       &   2.20    \\
 1290.66486 &   1800   &   35.56        &   0.24  & 50.65       &   3.07    \\
 1292.71528 &   2500   &   36.26        &   0.28  & 50.81       &   2.12    \\
 1310.70938 &   1800   &   37.93        &   0.79  & 45.48       &   3.72    \\
 1361.53271 &   2000   &   37.29        &   0.40  & 47.21       &   3.02    \\
 1402.48208 &   1800   &   37.87        &   0.56  & 43.18       &   4.58    \\
 1410.45867 &   1800   &   37.13        &   0.37  & 45.24       &   3.17    \\
 1593.76397 &   2200   &   46.36        &   3.70  & 26.67       &   1.92    \\
 1649.68690 &   1800   &   46.94        &   0.70  & 22.55       &   1.04    \\
 1677.74508 &   2000   &   47.38        &   0.95  & 23.40       &   1.30    \\
 1723.56597 &   1800   &   46.92        &   0.94  & 24.06       &   1.04    \\
 1763.47919 &   1800   &   46.94        &   0.75  & 23.77       &   1.06    \\
 1780.40388 &   1800   &   45.93        &   0.90  & 26.43       &   1.33    \\
 1979.75380 &   1800   &   40.12        &   0.86  & 39.30       &   3.82    \\
 2052.74894 &   1800   &   40.70        &   1.36  & 43.51       &   4.67    \\
 2083.53034 &   1800   &   37.82        &   0.69  & 46.26       &   4.70    \\
 2086.57898 &   1800   &   36.92        &   0.38  & 46.95       &   4.28    \\
 2088.58872 &   1800   &   37.93        &   0.37  & 47.62       &   4.10    \\
 2100.59817 &   1800   &   37.23        &   0.50  & 47.86       &   2.20    \\
 2159.41844 &   1800   &   35.79        &   0.28  & 49.81       &   1.78    \\
 2434.65722 &   1800   &   37.19        &   0.58  & 52.69       &   3.47    \\
 2435.63450 &   1800   &   36.36        &   0.54  & 51.64       &   4.61    \\
 2457.64088 &   1800   &   36.15        &   0.67  & 49.70       &   2.28    \\
 2752.69744 &   1800   &   46.65        &   0.44  & 25.65       &   1.35    \\
 2779.59285 &   1800   &   47.38        &   1.17  & 24.10       &   1.39    \\
 2780.61971 &   2300   &   47.63        &   2.07  & 24.03       &   2.05    \\
 2818.61864 &   1800   &   47.62        &   0.94  & 21.85       &   1.25    \\
\end{longtable}
\clearpage

\begin{table}
\centering
\caption{\BD: RVs and errors from binned spectra within 5\% of phase.}
\begin{tabular}{lr@{ $\pm$ }lr@{ $\pm$ }l}
\hline\hline
\noalign{\smallskip}
Phase bin  &  \multicolumn{2}{c}{RV MS (km s$^{-1}$)} & \multicolumn{2}{c}{RV sdO (km s$^{-1}$)} \\
\hline
\noalign{\smallskip}
0.025  &     8.50     & 0.05        &    -7.49    &   1.13    \\
0.075  &     5.15     & 0.04        &    -0.30    &   0.73    \\
0.125  &     2.65     & 0.04        &     7.46    &   1.22    \\
0.175  &     0.09     & 0.04        &    10.51    &   0.78    \\
0.216  &    -2.57     & 0.09        &    12.74    &   2.95    \\
0.275  &    -4.21     & 0.06        &    18.88    &   1.66    \\
0.325  &    -4.96     & 0.04        &    21.97    &   2.19    \\
0.425  &    -4.44     & 0.06        &    21.11    &   1.57    \\
0.475  &    -2.62     & 0.05        &    18.03    &   1.42    \\
0.548  &    -0.55     & 0.10        &    18.90    &   2.18    \\
0.575  &     0.98     & 0.07        &     9.73    &   1.25    \\
0.625  &     2.87     & 0.04        &     6.05    &   1.06    \\
0.675  &     5.07     & 0.05        &     1.49    &   1.14    \\
0.775  &    10.33     & 0.05        &    -9.14    &   1.11    \\
0.875  &    12.71     & 0.06        &   -17.21    &   1.13    \\
0.937  &    12.16     & 0.07        &   -13.15    &   1.46    \\
0.975  &    10.82     & 0.05        &    -9.54    &   0.75    \\
\hline
\end{tabular}
\tablefoot{The phase of an interval made up of a single observation is declared with its determined phase. In the rest, in which several spectra are binned, the nominal phases correspond to the central value of phase of the 5\% interval.}
\end{table}

\begin{table}
\centering
\caption{\F: RVs and errors from binned spectra within 5\% of phase.}
\begin{tabular}{lr@{ $\pm$ }lr@{ $\pm$ }l}
\hline\hline
\noalign{\smallskip}
Phase bin  &  \multicolumn{2}{c}{RV MS (km s$^{-1}$)} & \multicolumn{2}{c}{RV sdO (km s$^{-1}$)} \\
\hline
\noalign{\smallskip}
0.025  &    47.36    &  0.23        &   22.43   &  0.81    \\
0.075  &    47.15    &  0.47        &   23.97   &  1.01    \\
0.125  &    45.97    &  0.27        &   26.32  &   1.27    \\
0.167  &    44.06    &  1.00        &   29.10   &  1.86    \\
0.289  &    40.12     & 0.79        &   39.30   &  4.05    \\
0.375  &    37.74     & 0.16        &   46.61   &  3.38    \\
0.425  &    35.55     & 0.19        &   49.30   &  3.32    \\
0.475  &    35.44     & 0.19        &   49.33   &  3.63    \\
0.525  &    34.67     & 0.36        &   52.63   &  3.43    \\
0.575  &    35.08     & 0.24            &   52.67   &  2.43    \\
0.675  &    36.18    &  0.16            &   51.76  &   2.18    \\
0.725  &    36.72    &  0.21            &   48.58  &   5.01    \\
0.775  &    37.84    &  0.24            &   45.05  &   4.53    \\
0.825  &    40.32     & 0.18            &   38.22   &  3.64    \\
0.863  &    41.23     & 0.26        &   37.33    & 3.59    \\
0.975  &    47.04     & 0.58        &   25.03    & 1.19    \\
\hline
\end{tabular}
\tablefoot{The phase of an interval made up of a single observation is declared with its determined phase. In the rest in which several spectra are binned, the nominal phases correspond to the central value of phase of the 5\% interval.}
\end{table}
\clearpage

\twocolumn
\section{Alma observations of \BD\  and BD$-$7$^{\rm{o}}$5977} \label{sec:alma_observations}
In an attempt to search for CB matter around sdB or sdO binaries, two such long-period sd+MS binaries have been observed with ALMA. The goal was to search for the possible signature of matter expelled during the interaction phase that still remained in the system. Gas would be removed early on from a potential CB disk due to photo-evaporation. However, models using the Robertson-Poynting drag based on the method outlined in \citet{Wyatt2008} showed that large dust grains ($\ge$ 60 $\mu$m) at a distance of 100 AU or more from the binary could have survived for the duration of the He core and shell burning phase \citep[see e.g.][]{Hardy2016}.

ALMA observations were taken in Band 7 (950 $\mu$m) as it offered the best compromise between weather dependence and highest flux expectations. The observations were set up to be able to detect a signal of 0.1 mJy with a S/N of 5. These limits were determined based on the minimum expected flux of a CB disk with a mass of 1 earth mass at a distance of 100 AU from the system. The flux estimates were obtained using the MC dust radiative transfer code MCMAX \citep{Min2009}, while varying system and disk parameters corresponding to estimates from \citet{Andrews2007}, \citet{Vos2015}, and \citet{Woitke2016}.

\begin{figure}
    \centering
    \includegraphics[width=9cm]{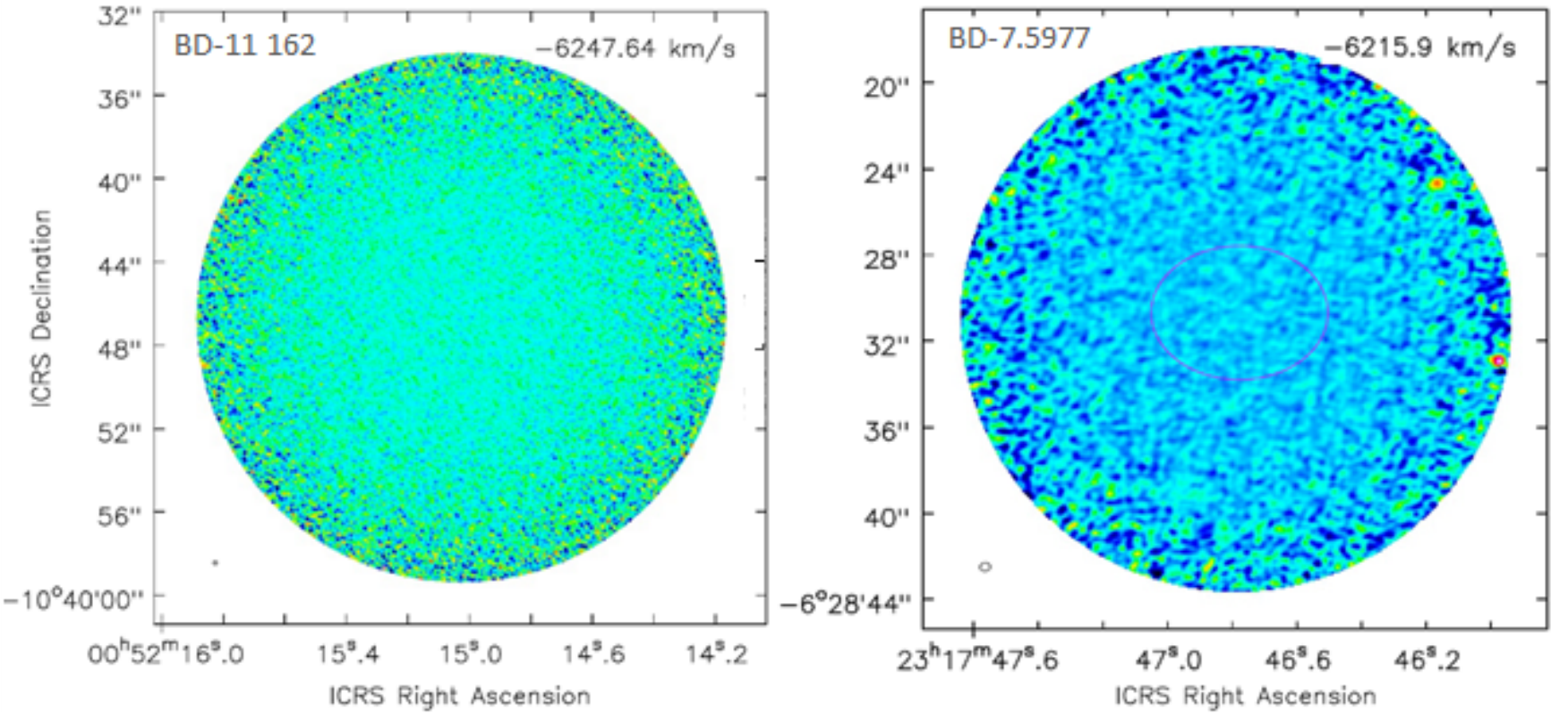}
    \caption{Radio ALMA observations on the \BD\ and BD-7$^{\rm o}$5977 systems. There is no signal visible higher than the noise level that could indicate circumbinary material.}
    \label{fig:alma_results}
\end{figure}

For both systems no signal of CB matter is detected in the observations. This indicates that no dust is present any more in the systems, or at least that the amount is less than the detection limit of the observations. It does however not exclude the presence of a CB disk in the past of the system. The reduced images are shown in Fig.\,\ref{fig:alma_results}, on which it can be seen that there is no signal visible higher than the noise level.

\clearpage

\onecolumn
\section{Intervals and lines used in abundance calculations}

\begin{longtable}{p{0.20\textwidth}p{0.20\textwidth}p{0.40\textwidth}}

\caption{\label{BDInt}\BD: Used intervals (Int) in constrained abundance results from GSSP.}\\
\hline
Element  &    Int (\AA) &   Line (Int) (\AA)\\
\hline
\noalign{\smallskip}
\endfirsthead
\caption{Continued.}\\
\hline
\endhead
\hline
\endfoot
\hline
\endlastfoot
C   &                               &  7115.19 (7078-7151) \\
    &                               &  7113.18 \& 7115.19 \& 7116.99 (7078-7151) \\
    &                               &  8335.14 (8326-8336) \\
N   &                               &  8683.40 (8673-8744  \\
O   &                               &  7771.94 \& 7774.17 \& 7775.39 (7078-7151) \\
    &                               &  8446.25 \& 8446.36 \& 8446.76 (8432-8448) \\
Na  &   5657-5713                   &  5688.21 (5657-5713) \\
    &                               &  6154.22 \& 6160.75 (5657-5713) \\
Mg  &   5117-5210                   &  5528.40 (5496-5545)\\
    &                               &  8717.82 (8673-8745) \\
    &                               &  8736.02 (8673-8745) \\
Al  &                               &  7835.31 \& 7836.13 (7778-7838) \\
    &                               &  8772.87 \& 8773.90 (8674-8755$^{a}$) \\
Si  &   5904-6273 \& 6334-6510      &  5772.15 (5752-5818)    \\
    &                               &  6155.13 (5904-6273 \& 6334-6510)    \\
    &                               &  6347.10 (5904-6273 \& 6334-6510)    \\
    &                               &  6371.36 (5904-6273 \& 6334-6510)    \\
S   &                               &  6748.79 (6731-6789) \\
    &                               &  6757.16 (6731-6789)  \\
K   &                               &  7698.97 (7698-7744$^{a}$) \\
Ca  &    4519-4608                  &  4581.40 \& 4581.47 (4519-4608) \\
    &    5199-5354                  &  6717.70 (6642-6728) \\
    &    5904-6273 \& 6334-6510     &                      \\
Sc  &                               &  4246.82 (4198-4252) \\
    &                               &  6604.58 (6590-6682) \\
Ti  &    5002-5052                  &                       \\
    &    5974-6273 \& 6334-6510     &                       \\
V   &    4378-4445                  &                       \\
Cr  &    4563-4608                  &                       \\
    &    5199-5354                  &                       \\
Mn  &    4749-4769                  &                       \\
    &    4782-4824                  &                       \\
    &    5980-6273 \& 6334-6510     &                       \\
Fe  &    4519-4608                  &                       \\
    &    5199-5354                  &                       \\
    &    5904-6273 \& 6334-6510     &                       \\
Co  &    4110-4124                  &                       \\
    &    5199-5354                  &                   \\
Ni  &    5904-6273 \& 6334-6510     &  6767.77 (6731-6789) \\
    &                               &  6772.32 (6731-6789) \\
Cu  &                               &  5218.20 (5199-5354) \\
    &                               &  5700.24 (5694-5716) \\
    &                               &  5782.13 (5752-5818) \\
Zn  &                               &  4722.16 (4698-4752) \\
    &                               &  4810.53 (4809-4839) \\
Sr  &                               &  4077.71 (4069-4087)\\
    &                               &  4215.52 (4198-4252) \\
Y   &                               &  4883.68 (4877-4926) \\
    &                               &  4900.12 (4877-4926) \\
Zr  &                               &  4149.20 (4145-4197) \\
Ba  &                               &  4554.03 (4548-4593) \\
    &                               &  6141.71 (6006-6197) \\
La  &                               &  4123.22 (4117-4160)   \\
    &                               &  4333.74 (4378-4445$^{a}$) \\
Ce  &                               &  4562.36 (4548-4593)   \\
    &                               &  4628.16 (4627-4630)   \\
\end{longtable}
\tiny{a: not possible to constraint dilution from interval containing the line due to telluric bands or G-band. Dilution fixed using a clean nearby interval}

\begin{longtable}{p{0.20\textwidth}p{0.25\textwidth}p{0.4\textwidth}}
\caption{\label{FGInt}\F: Used intervals (Int) in constrained abundance results from GSSP.}\\
\hline
Element  &    Int (\AA) &   Line (Int) (\AA)\\
\hline
\noalign{\smallskip}
\endfirsthead
\caption{Continued.} \\
\hline
\endhead
\hline
\endfoot
\hline
\endlastfoot
C   &                               &  7115.19 (7080-7152) \\
    &                               &  7111.47 \& 7113.18 \& 7115.19 \& 7116.99 \& 7119.67 (7080-7152) \\
N   &                               &  8683.40 (8673-8740) \\
O   &                               &  7771.94 \& 7774.17 \& 7775.39 (7770-7848) \\
    &                               &  8446.25 \& 8446.36 \& 8446.76 (8673-8740$^{a}$) \\
Na  &    5657-5713                  &  5688.21 (5657-5713) \\
    &                               &  6154.22 \& 6160.75 (6133-6199) \\
Mg  &    5117-5210                  &  4702.99 (4698-4724)   \\
    &                               &  5528.40 (5496-5545) \\
    &                               &  8717.82 (8673-8739) \\
    &                               &  8736.02 (8673-8739) \\
Al  &                               &  8673-8740 (d 8772.87 \& 8773.9$^{a}$)\\
Si  &   5904-6267 \& 6333-6510      &  5772.15 (5751-5818)    \\
    &                               &  6155.13 (5904-6267 \& 6333-6510)    \\
    &                               &  6347.10 (5904-6267 \& 6333-6510)    \\
    &                               &  8728.01 (8673-8740)                 \\    
S   &                               &  6748.79 (6694-6812) \\
    &                               &  6757.16 (6694-6812)  \\
K   &                               &  7698.97 (7697-7791$^{a}$) \\
Ca  &    5904-6267 \& 6333-6510     &  4581.40 \& 4581.47 (4498-4609) \\
    &    4498-4609                  &  6717.70 (6694-6812)  \\
    &    5200-5355                  &                       \\
Ti  &    4987-5054                  &                       \\
    &    5974-6267 \& 6333-6510     &                       \\
V   &    4378-4446                  &                       \\
Cr  &    4563-4608                  &                       \\
    &    5200-5354                  &                       \\
Mn  &    5974-6267                  &                       \\
    &    4749-4769                  &                       \\
    &    4782-4824                  &                       \\
Fe  &    5904-6267 \& 6333-6510     &                       \\
    &    4519-4608                  &                       \\
    &    5200-5354                  &                       \\
Co  &    4117-4162                  &                       \\
Ni  &    5904-6267 \& 6333-6510     &  6767.77 (6694-6812)  \\
Cu  &                               &  5782.13 (5751-5818)  \\
Zn  &                               &  4722.16 (4694-4738)  \\
    &                               &  4810.53 (4782-4830)  \\
Sr  &                               &  4077.71 (4043-4086)  \\
    &                               &  4215.52 (4197-4252)  \\
Y   &                               &  4883.68 (4877-4927)  \\
    &                               &  4900.12 (4877-4927)  \\
Zr  &                               &  4149.20 (44117-4162) \\
    &                               &  4211.88 (4197-4252)  \\
Ba  &                               &  4554.03 (4497-4609)  \\
    &                               &  6141.71 (6006-6197)  \\
\end{longtable}
\tiny{a: not possible to constraint dilution from interval containing the line due to telluric bands. Dilution fixed using a clean nearby interval}
\tablefoot{When abundance is derived from specific line/s, the used procedure is to cut and normalise an interval containing the studied line (Line(Int)), firstly developing an abundance calculation with GSSP, based on that interval. From this preliminary study, the dilution value of the interval is obtained. Eventually, a new abundance calculation is developed on the specific line/s (short interval around) meanwhile the value of dilution, previously obtained, is fixed.}
\clearpage

\begin{longtable}{p{0.1\textwidth}p{0.10\textwidth}}
\caption{\label{Lines}Lines used for specific line studies of abundance.}\\
\hline
Element  &   Wavelength (\AA) \\
\hline
\noalign{\smallskip}
\endfirsthead
\caption{Continued.} \\
\hline
\endhead
\hline
\endfoot
\hline
\endlastfoot
\ion{C}{I}   &  7111.47    \\
\ion{C}{I}   &  7113.18    \\
\ion{C}{I}   &  7115.19    \\
\ion{C}{I}   &  7116.99    \\
\ion{C}{I}   &  7119.67    \\
\ion{C}{I}   &  8335.14    \\
\ion{N}{I}   &  8683.40    \\
\ion{O}{I}   &  7771.94    \\
\ion{O}{I}   &  7774.17    \\
\ion{O}{I}   &  7775.39    \\
\ion{O}{I}   &  8446.25    \\
\ion{O}{I}   &  8446.36    \\
\ion{O}{I}   &  8446.76    \\
\ion{Na}{I}  &  5688.21    \\
\ion{Na}{I}  &  6154.22    \\
\ion{Na}{I}  &  6160.70    \\
\ion{Mg}{I}  &  4702.99    \\
\ion{Mg}{I}  &  5528.40    \\
\ion{Mg}{I}  &  8717.82    \\
\ion{Mg}{I}  &  8736.02    \\
\ion{Al}{I}  &  7835.31    \\
\ion{Al}{I}  &  7836.13    \\
\ion{Al}{I}  &  8772.87    \\
\ion{Al}{I}  &  8773.9     \\
\ion{Si}{I}  &  5772.15    \\
\ion{Si}{I}  &  6155.13    \\
\ion{Si}{II} &  6347.11    \\
\ion{Si}{II} &  6371.36    \\
\ion{Si}{I}  &  8728.01    \\
\ion{S}{I}   &  6748.79    \\
\ion{S}{I}   &  6757.16    \\
\ion{K}{I}   &  7698.97     \\
\ion{Ca}{I}  &  4581.40     \\
\ion{Ca}{I}  &  4581.47     \\
\ion{Ca}{I}  &  6717.70     \\
\ion{Sc}{II} &  4246.82     \\
\ion{Sc}{II} &  6604.58     \\
\ion{Ni}{I}  &  6767.77     \\
\ion{Ni}{I}  &  6772.32     \\
\ion{Cu}{I}  &  5218.20     \\
\ion{Cu}{I}  &  5700.24     \\
\ion{Cu}{I}  &  5782.13     \\
\ion{Zn}{I}  &  4722.16     \\
\ion{Zn}{I}  &  4810.53     \\
\ion{Sr}{II} &  4077.71     \\
\ion{Sr}{II} &  4215.52     \\
\ion{Y}{II}  &  4883.68     \\
\ion{Y}{II}  &  4900.12     \\
\ion{Zr}{II} &  4149.20     \\
\ion{Zr}{II} &  4211.88     \\
\ion{Ba}{II} &  4554.03     \\
\ion{Ba}{II} &  6141.71     \\
\ion{La}{II} &  4123.22     \\
\ion{La}{II} &  4333.74     \\
\ion{Ce}{II} &  4562.36     \\
\ion{Ce}{II} &  4628.16     \\
\end{longtable}
\tablefoot{To obtain additional information: \newline
\url{https://fys.kuleuven.be/ster/meetings/binary-2015/gssp-software-package}}

\begin{figure*}
  \centering
  \includegraphics[width=18.2cm,height=11cm,keepaspectratio]{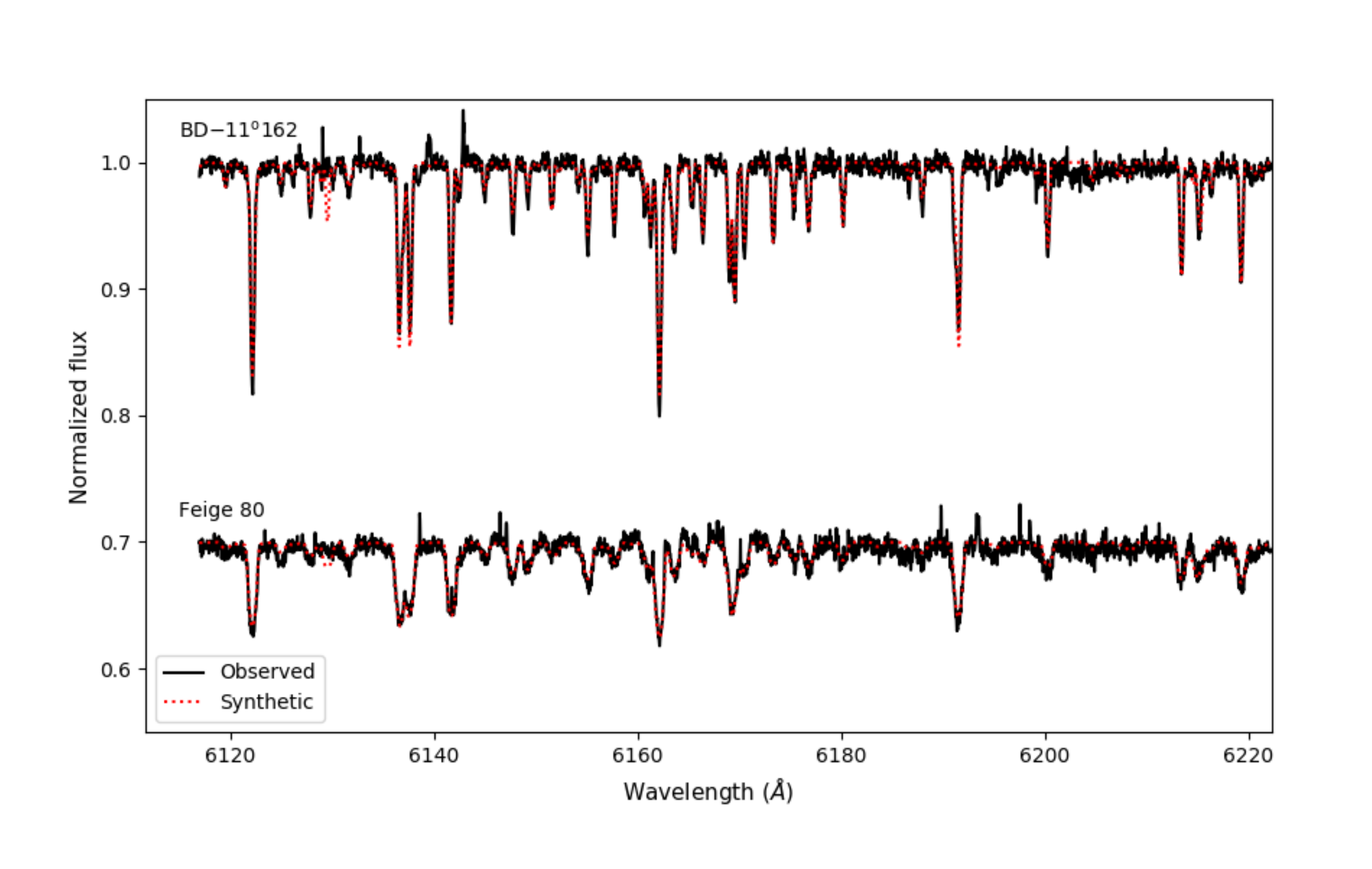}
      \caption{Observed normalised spectrum (solid black line) and the best fitting GSSP model (dotted red line) for the wavelength range (6120-6220 \AA) used to determine the atmospheric parameters with the GSSP code.}
\label{fitAP}
\end{figure*}

\begin{figure*}
  \centering
  \includegraphics[width=18.2cm, height=11cm,keepaspectratio]{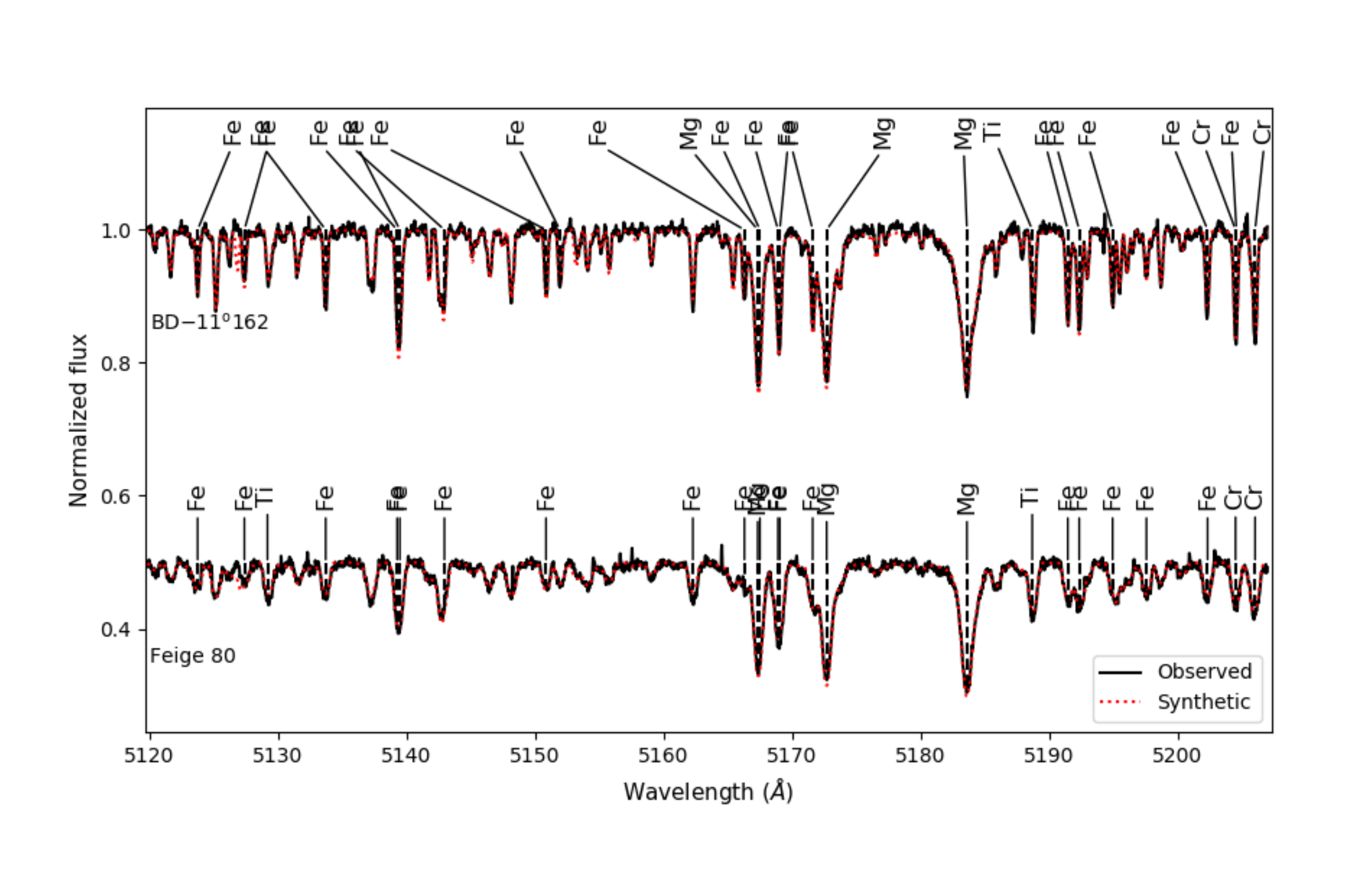}
      \caption{Observed normalised spectrum (solid black line) and the best fitting GSSP model (dotted red line) for the wavelength range 5120-5210 \AA. It includes the strong \ion{Mg}{i} triplet at 5167, 5172, and 5183 \AA,\, which is used to determine the Mg abundance. To prevent label overlapping, an intensity threshold was applied for the line labelling.}
\label{fitMg}
\end{figure*}

\end{appendix}
\end{document}